\PassOptionsToPackage{unicode}{hyperref}
\PassOptionsToPackage{hyphens}{url}
\PassOptionsToPackage{dvipsnames,svgnames,x11names}{xcolor}
\documentclass[
  12pt]{article}

\usepackage{amsmath,amssymb}
\usepackage{iftex}
\usepackage{comment}
\usepackage{graphicx} 
\usepackage{mathptmx}
\usepackage{amsthm}
\usepackage{amsmath}
\usepackage{mathtools}
\usepackage{amssymb}
\usepackage{xcolor}
\usepackage{algorithm}
\usepackage{algpseudocode}
\usepackage{etoolbox}
\newtheorem{remark}{Remark}
\newtheorem{theorem}{Theorem}
\newtheorem{definition}{Definition}
\newtheorem{corollary}{Corollary}

\newtheorem{proposition}{Proposition}

\newtheorem{assumption}{Assumption}
\usepackage{multirow}
\usepackage{booktabs}
\usepackage{bm}

\newcommand{\thb}{\bm{\theta}}

\renewcommand{\thb}{\bm{\theta}}
\newcommand{\SW}{\text{SW}}
\newcommand{\SWh}{\widehat{\text{SW}}}
\newcommand{\W}{\text{W}}

\newcommand{\Gbar}{\bar{G}}

\newcommand{\Sbb}{\mathbb{S}}

\newcommand{\BB}{\mathcal{B}}

\newcommand{\XX}{\mathcal{X}}

\newcommand{\EEb}{\mathbb{E}}
\newcommand{\SSS}{\mathcal{S}}

\newcommand{\Ga}{\text{Ga}}

\newcommand{\dDPP}{\text{dDPP}}

\renewcommand{\Re}{\mathbb{R}}
\newcommand{\PP}{\mathcal{P}}

\newcommand{\thi}[1]{\theta_{#1}}                 
\newcommand{\Gmix}{G}                             
\newcommand{\Gmixbar}{\bar{G}}                 
\newcommand{\Gmixbarp}{\bar{G'}}            
\newcommand{\Gtrue}{G_0}           
\newcommand{\thtilde}[1]{\widetilde\theta_{#1}}   
\newcommand{\thbase}[1]{\nu_{#1}}            
\newcommand{\thbasenew}{\theta_{\mathrm{new}}}   
\newcommand{\thbaseprop}{\theta_{\mathrm{prop}}} 
\newcommand{\thstar}[1]{\theta^*_{#1}}           
\newcommand{\thstarvec}{\thb^*}            

\newcommand{\argmax}{\text{arg\,max}}
\newcommand{\argmin}{\text{arg\,min}}
\ifpdftex
  \usepackage[T1]{fontenc}
  \usepackage[utf8]{inputenc}
  \usepackage{textcomp} 
\else 
  \usepackage{unicode-math}
  \defaultfontfeatures{scale=matchlowercase}
  \defaultfontfeatures[\rmfamily]{ligatures=tex,scale=1}
\fi
\usepackage{lmodern}
\ifpdftex\else  
\fi
\IfFileExists{upquote.sty}{\usepackage{upquote}}{}
\IfFileExists{microtype.sty}{
  \usepackage[]{microtype}
  \UseMicrotypeSet[protrusion]{basicmath} 
}{}
\makeatletter
\@ifundefined{komaclassname}{
  \IfFileExists{parskip.sty}{%
    \usepackage{parskip}
  }{
    \setlength{\parindent}{0pt}
    \setlength{\parskip}{6pt plus 2pt minus 1pt}}
}{
  \KOMAoptions{parskip=half}}
\makeatother
\usepackage{xcolor}
\makeatletter
\ifx\paragraph\undefined\else
  \let\oldparagraph\paragraph
  \renewcommand{\paragraph}{
    \@ifstar
      \xxxparagraphstar
      \xxxparagraphnostar
  }
  \newcommand{\xxxparagraphstar}[1]{\oldparagraph*{#1}\mbox{}}
  \newcommand{\xxxparagraphnostar}[1]{\oldparagraph{#1}\mbox{}}
\fi
\ifx\subparagraph\undefined\else
  \let\oldsubparagraph\subparagraph
  \renewcommand{\subparagraph}{
    \@ifstar
      \xxxsubparagraphstar
      \xxxsubparagraphnostar
  }
  \newcommand{\xxxsubparagraphstar}[1]{\oldsubparagraph*{#1}\mbox{}}
  \newcommand{\xxxsubparagraphnostar}[1]{\oldsubparagraph{#1}\mbox{}}
\fi
\makeatother

\usepackage{longtable,booktabs,array}
\usepackage{calc} 
\usepackage{etoolbox}
\makeatletter
\patchcmd\longtable{\par}{\if@noskipsec\mbox{}\fi\par}{}{}
\makeatother
\IfFileExists{footnotehyper.sty}{\usepackage{footnotehyper}}{\usepackage{footnote}}
\makesavenoteenv{longtable}
\usepackage{graphicx}
\makeatletter
\def\maxwidth{\ifdim\Gin@nat@width>\linewidth\linewidth\else\Gin@nat@width\fi}
\def\maxheight{\ifdim\Gin@nat@height>\textheight\textheight\else\Gin@nat@height\fi}
\makeatother
\setkeys{Gin}{width=\maxwidth,height=\maxheight,keepaspectratio}
\makeatletter
\def\fps@figure{htbp}
\makeatother

\makeatletter
\@ifpackageloaded{caption}{}{\usepackage{caption}}
\AtBeginDocument{%
\ifdefined\contentsname
  \renewcommand*\contentsname{Table of contents}
\else
  \newcommand\contentsname{Table of contents}
\fi
\ifdefined\listfigurename
  \renewcommand*\listfigurename{List of Figures}
\else
  \newcommand\listfigurename{List of Figures}
\fi
\ifdefined\listtablename
  \renewcommand*\listtablename{List of Tables}
\else
  \newcommand\listtablename{List of Tables}
\fi
\ifdefined\figurename
  \renewcommand*\figurename{Figure}
\else
  \newcommand\figurename{Figure}
\fi
\ifdefined\tablename
  \renewcommand*\tablename{Table}
\else
  \newcommand\tablename{Table}
\fi
}
\@ifpackageloaded{float}{}{\usepackage{float}}
\floatstyle{ruled}
\@ifundefined{c@chapter}{\newfloat{codelisting}{h}{lop}}{\newfloat{codelisting}{h}{lop}[chapter]}
\floatname{codelisting}{listing}

\makeatother
\makeatletter
\@ifpackageloaded{caption}{}{\usepackage{caption}}
\@ifpackageloaded{subcaption}{}{\usepackage{subcaption}}
\makeatother

\ifluatex
  \usepackage{selnolig}  
\fi
\usepackage[]{natbib}
\usepackage{bookmark}

\IfFileExists{xurl.sty}{\usepackage{xurl}}{} 
\hypersetup{
  pdftitle={title},
  pdfauthor={author 1; author 2},
  pdfkeywords={3 to 6 keywords, that do not appear in the title},
  colorlinks=true,
  linkcolor={blue},
  filecolor={maroon},
  citecolor={blue},
  urlcolor={blue},
  pdfcreator={latex via pandoc}}

\newcommand{\anon}{1}

\begin{document}

\def\spacingset#1{\renewcommand{\baselinestretch}%
{#1}\small\normalsize} \spacingset{1}


\if1\anon
{
  \title{\bf Distributional Determinantal Point Process  for
    Repulsive Clustering of Distributions}
  \author{Khai Nguyen$^1$, Yang Ni$^2$, Elizabeth Juarez-Colunga$^4$, and Peter M\"uller$^{2,3}$ \\
   $^1$Department of Statistics, Texas A\&M University\\
    $^2$Department of Statistics and Data Sciences, University of
    Texas at Austin \\
    $^3$Department of Mathematics, University of Texas at Austin \\
    $^4$Department of Biostatistics and Informatics,
University of Colorado Anschutz}
  \maketitle
} \fi

\if0\anon
{
  \bigskip
  \bigskip
  \bigskip
  \begin{center}
    {\large\bf Distributional Determinantal Point Process  for
    Repulsive Clustering of Distributions}
\end{center}
  \medskip
} \fi

\bigskip
\begin{abstract}
We introduce the distributional determinantal point process (dDPP) as
a novel repulsive point process whose atoms are probability
distributions rather than points in a real space. The dDPP is
constructed via an L-ensemble with a sliced Wasserstein (SW)
kernel between distributions. We show its validity as a well-defined
point process. In the discrete setting, we
derive concentration results for plug-in estimators of the L-ensemble,
the correlation kernel, and their determinants given i.i.d. samples
from the distributional atoms. Leveraging this framework, we propose a
distribution-valued random partition model by way of a repulsive
generalized Bayesian mixture model. The model places a dDPP prior over
the atoms of the mixing measure and defines a generalized likelihood
based on SW distance. 
To summarize posterior inference,
we develop a decision-theoretic approach to report a point estimate
of the mixing measure as a Bayes rule under a hierarchical optimal
transport utility function.
The latter is a natural choice given that the mixing measure is
itself a distribution over distributions.  We use
the proposed framework for inference with
single-cell gene expression data and human epilepsy data,
producing interpretable and well-separated clusters that reflect
meaningful structure in the data. 
\end{abstract}

\noindent%
\vfill

\newpage
\spacingset{1.8} 

\section{Introduction}
\label{sec:introduction}

We propose the distributional determinantal point process (dDPP) as a
novel repulsive point process  with distribution-valued atoms. The
construction leverages a sliced Wasserstein (SW)
kernel~\citep{kolouri2016sliced,carriere2017sliced} between
distributions to build an L-ensemble.
 We use the proposed dDPP to 
formulate a distribution-valued random partition model by way of
setting up a mixture model with (i) a dDPP prior over  the atoms of the
mixing measure and, (ii) a generalized
likelihood~\citep{bissiri2016general,chakraborty2025robust} based on
SW distance~\citep{rabin2012wasserstein,bonneel2015sliced,nguyen2025introduction}.

Probabilistic clustering (random partition) 
by discrete mixture models uses latent allocation
variables to associate each observation with a specific mixture
component. Interpreting these latent variables as cluster membership
indicators defines the desired clusters. Specifying a prior over the
partition in this setup reduces to placing a prior over the mixing
measure that governs the mixture. Under symmetry conditions known as
exchangeability, every random partition admits such a characterization
based on a prior over the discrete mixing
measure~\citep{kingman1978representation}. Prior distributions on
random probability measures are known as  nonparametric Bayesian
models. Prominent examples are the Ferguson-Dirichlet process (DP)
\citep{ferguson1973bayesian},
the Pitman--Yor process \citep{pitman1997two}, and the wider family of
normalized completely random measures~\citep{lijoi2005hierarchical,
  lijoi2007controlling}. 

A common problem of the mentioned models is the tendency to produce
poorly separated mixture components, which can
undermine the interpretability of the resulting clusters.
The problem has been studied
for large samples \citep{rousseau2011asymptotic}, and
can be avoided by repulsive priors
\citep{beraha2022mcmc,cremaschi2025repulsion,petralia2012repulsive,xie2020bayesian,pedroso2026bayesian}. 
Placing a repulsive point process prior on the
atoms of the mixing measure, i.e., the mixture locations,
effectively pushes components of the mixture model apart.
In particular, the determinantal point process
(DPP)~\citep{hough2009zeros,macchi1975coincidence,kulesza2012determinantal,lavancier2015determinantal}
has been widely used as a prior in repulsive mixture
models~\citep{bianchini2020determinantal,xu2016bayesian}. 
Recently,
~\citet{beraha2025bayesian} and \citet{song2025repulsive} proposed a unified
framework for analyzing the associated mixing measure and
characterizing the distribution of the DPP prior both a priori and a
posteriori.
 In this paper, we follow a similar strategy to develop clustering for
distribution-valued data by way of a novel distribution-valued dDPP. 

Alternatively, multilevel clustering~\citep{ho2017multilevel} can also
be viewed as a method for finding partitions of distributions. 
K-means and  K-centers algorithms in the Wasserstein space
are proposed in ~\citet{zhuang2022wasserstein}
and~\citet{okano2025wasserstein}, respectively. For Bayesian
approaches, nested DP mixture models~\citep{rodriguez2008nested}
provide two levels of clustering, including clustering at the
distributional level. A common challenge in this literature is the
high computational complexity arising from the rich structure of the
atoms (distributions). In particular, Wasserstein distance is
computationally expensive, with super-cubic
complexity~\citep{peyre2019computational}.  Bayesian nonparametric
models such as the nested DP require a full specification of the
generative model for tractable posterior inference,   tend to produce poorly separated clusters, and
involve computationally expensive implementation of posterior
inference. 
Adding repulsiveness into the prior poses
both theoretical and computational challenges.

To the best of our knowledge, no prior work has considered repulsive
priors with distributions as atoms or derived associated mixture
models. Despite the inherent challenges, repulsive mixture models for
distributional data are essential in practice when interpretable
inference is required. We consider two typical examples as motivating
applications. In one example, we aim to cluster donors
based on single-cell gene expression data,
which can be treated as
(empirical) distributions (over cells) of gene expression. Another example is
clustering of epilepsy patients based on their history of seizures,
which can be cast as a 
distribution over windows of observed binary repeat measurements.
Inference in both examples involves
clustering of distributions.
We implement the desired model-based clustering by the proposed
extension of DPP mixture models to distribution-valued data.

We set up dDPP using a SW kernel for distributions.
The choice of the SW kernel is
particularly appealing for this construction because,
first,  it
allows construction of a positive definite kernel
that guarantees the dDPP to be a well-defined point
process (we discuss details later).
This is a unique property that many alternative distance functions for
distributions such as the Wasserstein distance do not possess. Second,
in the discrete setting  where computation is more tractable,
concentration results for 
plug-in estimators of both the L-ensemble kernel and the correlation kernel
of the DPP construction allow for the use of empirical distributions
as practical proxies for distributional atoms.
By placing a
dDPP prior
over the atoms of the mixing measure and adopting SW distance to
construct a generalized likelihood~\citep{bissiri2016general}, the
resulting repulsive mixture model favors well-separated clusters
without requiring a fully specified generative model, substantially
broadening its applicability. 

The remainder of the article is organized as
follows. Section~\ref{sec:background} reviews SW distance and SW
kernel that underpin our construction. In section~\ref{seq:D2PP}, we
introduce the dDPP, establish its well-definedness under a compactness
assumption, and derive concentration results for plug-in estimates
(using empirical distributions) of
the L-ensemble and correlation kernels in the discrete
setting. Section~\ref{sec:mixture} introduces a generalized repulsive
Bayesian mixture model and its posterior consistency. Section~\ref{sec:posterior_inference} includes
results on posterior characterization, marginal Markov chain Monte Carlo
inference, and a decision-theoretic framework to summarize
the posterior random partition.
In
section~\ref{sec:experiments}, we demonstrate the proposed framework
with inference for single-cell gene expression data and human epilepsy
data, showing that the 
repulsive prior yields interpretable and well-separated clusters. We
conclude with a final discussion in
section~\ref{sec:conclusion}.
Technical proofs and additional experimental results are provided in
the Supplementary Materials.

A brief note on notation.
The Dirac delta measure concentrated at a point $x$ is written as
$\delta_x$.
For any integer $d \ge 2$, the set
$\Sbb^{d-1} = \{\theta \in \Re^d : \|\theta\|_2 = 1\}$
denotes the unit hypersphere in $\Re^d$.
When comparing two sequences $a_n$
and $b_n$, notation $a_n = \mathcal{O}(b_n)$ means that there is
an absolute constant $c > 0$ satisfying $a_n \le c\, b_n$ for every $n
\ge 1$. Let $(\mathcal{X}_1, \Sigma_1)$ and $(\mathcal{X}_2,
\Sigma_2)$ be measurable spaces, and suppose $f : \XX_1 \to
\XX_2$ is a measurable map.  For a measure $\mu$ on
$(\XX_1, \Sigma_1)$, the push-forward measure $f \sharp \mu$
on $(\XX_2, \Sigma_2)$ is given by 
$
    f \sharp \mu(B) = \mu\left(f^{-1}(B)\right), \quad \forall B \in \Sigma_2.
$
The standard $(Q-1)$-simplex is denoted $\Delta^{Q-1}=\left\{ \pi \in \Re^Q : \pi_q \geq 0 \ \forall\, q,
    \quad \sum_{q=1}^Q \pi_q = 1\right\}$. 
Finally, 
$\PP_2(\Re^d)$ denotes the set of all distributions supported on $\Re^d$
with finite second moment.
Further notation is introduced as needed.

\section{Sliced Wasserstein Distance and Kernel}
\label{sec:background}

By way of a brief review of Wasserstein distance, sliced Wasserstein
distance, and sliced Wasserstein kernel, we introduce some notation
and definitions.
For $\mu, \nu \in \PP_2(\Re^d)$,
Wasserstein
distance~\citep{villani2009optimal} between  $\mu$ and $\nu$ is
defined as:
\begin{align}
   \W_2^2(\mu, \nu) = \inf_{\pi \in \Pi(\mu, \nu)} \int_{\Re^d \times
     \Re^d} \|x- y\|_2^2 \, \mathrm{d}\pi(x, y),
   \nonumber
\end{align}
where $\Pi(\mu, \nu) = \left\{\pi \in \PP(\Re^d \times \Re^d) \mid
\pi(A\times  \Re^d) = \mu(A), \ \pi(\Re^d \times  B) = \nu(B)
\ \forall A, B \subset \Re^d \right\}$ is the set of all
transportation plans/couplings.  While being geometrically meaningful,
Wasserstein distance is computationally expensive.
For continuous cases, Wasserstein distance is usually
intractable, except for some special cases like Gaussians
 and the univariate setting. 
For
discrete cases, the time complexity  of Wasserstein distance is
$\mathcal{O}(m^3 \log m)$~\citep{peyre2019computational} with $m$
being the maximum number of atoms of two distributions. Therefore, it
is burdensome to use it  for large discrete  distributions. 

One of the alternatives to avoid the computational issue of  Wasserstein
distance is  
sliced Wasserstein (SW)
distance~\citep{rabin2014adaptive,nguyen2025introduction}. SW distance
between two distributions $\mu,\nu \in
\mathcal{P}_2(\Re^d)$ is defined as:
\begin{align*}
\SW_2^2(\mu, \nu) = \mathbb{E}_{v \sim
  \mathcal{U}(\Sbb^{d-1})}[W_2^2(P_v \sharp \mu, P_v
\sharp \nu)],
\end{align*}
where  $\mathcal{U}(\Sbb^{d-1})$ is the uniform distribution over the unit hypersphere in $d$ dimension, and $P_v \sharp \mu$
and $P_v \sharp \nu$ denote the pushforward distribution of $\mu$ and
$\nu$ through the function $P_v(x) =  v^\top x$.
Utilizing closed-form expression for one-dimensional Wasserstein distance~\citep{dall1956sugli},
we rewrite SW as:
\begin{align}
\label{eq:SW_closed_form}
    \SW_2^2(\mu, \nu) = \mathbb{E}_{v \sim
  \mathcal{U}(\Sbb^{d-1})}\left[ \int_0^1 (F_{P_v\sharp \mu}^{-1}(t) - F_{P_v\sharp \nu}^{-1}(t))^2 \mathrm{d} t\right],
\end{align}
where $F_{P_v\sharp \mu}^{-1}(t)$ and $F_{P_v\sharp \nu}^{-1}(t)$ are
quantile functions of $P_v\sharp \mu$ and $P_v\sharp \nu$,
respectively. When $\mu$ and $\nu$ are discrete distributions with at
most $m$ atoms, evaluating  $\int_0^1 (F_{P_v\sharp \mu}^{-1}(t) -
F_{P_v\sharp \nu}^{-1}(t))^2 \mathrm{d} t$ has the time
complexity of only $\mathcal{O}(m\log m)$. The expectation
in~\eqref{eq:SW_closed_form} is often intractable, and hence,
numerical approximation is
needed~\citep{nguyen2024quasimonte,leluc2024slicedwasserstein},
for example via Monte Carlo integral~\citep{bonneel2015sliced} as:
\begin{align*}
    \SWh_2^2(\mu, \nu;V) = \frac{1}{V}\sum_{l=1}^V \W_2^2(P_{v_l} \sharp \mu, P_{v_l} \sharp \nu),
\end{align*}
where $v_1,\ldots,v_V \overset{i.i.d.}{\sim} \mathcal{U}(\Sbb^{d-1})$ with $V$ being the number of Monte Carlo samples or the number of projections. The overall time complexity of this approximation is $\mathcal{O}(V m \log m)$.

Beyond the computational benefit, SW has an unique property compared
to Wasserstein distance. In particular, SW is \emph{Hilbertian} while
Wasserstein distance is not,
 implying in particular that we can use SW to build a positive
definite kernel  needed for DPP. 
In more detail, there exists a mapping  $\Psi: \PP_2(\Re^d) \to \mathbb{L}_2(\Sbb^{d-1} \times [0,1])$, such that:
\begin{align*}
    \SW_2^2(\mu, \nu) = \|\Psi(\mu)-\Psi(\nu)\|_{\mathbb{L}_2(\Sbb^{d-1} \times \Re)}^2,
\end{align*}
where $\|\cdot\|_{\mathbb{L}_2(\Sbb^{d-1} \times \Re)}$ is the
functional norm. From~\eqref{eq:SW_closed_form},  one example of the
mapping is $\Psi[\mu](v,t) = F_{P_v\sharp \mu}^{-1}(t)$
(there are alternative mappings --  see examples in~\citet{kolouri2016sliced}). With the Hilbertian property, SW distance results in a valid kernel between distributions, which is defined as follows:
\begin{align}
\label{eq:SWkernel}
         \text{SWK}(\mu,\nu)
        = \exp\bigl[-\gamma\,\SW_2^2(\mu,\nu)\bigr],
\end{align}
where $\gamma > 0$ is a scale (bandwidth) parameter. The above SW
kernel is positive
definite~\citep{kolouri2016sliced,carriere2017sliced,meunier2022distribution}.
 Importantly, replacing $\SW_2^2(\mu,\nu)$ by its numerical
approximation $\SWh_2^2(\mu, \nu;V)$ still defines a positive definite
kernel. 
Finally, we note that there are also other alternative definitions of
SW-based kernels~\citep{kolouri2016sliced,luong2025unbiased}.
However, we will focus on \eqref{eq:SWkernel}.

\section{Distributional Determinantal Point Processes}
\label{seq:D2PP}

\subsection{A Distribution-Valued Determinantal Point Processes (dDPP)}

We introduce the distributional determinantal point process (dDPP) as
a DPP on a set of distributions $\Theta \subset
\mathcal{P}_2(\Re^d)$. We equip 
$\Theta$ with SW topology and its Borel $\sigma$-algebra
$\BB(\Theta)$.
In the construction, we use a base measure $P_0$ on
$(\Theta,\mathcal{B}(\Theta))$. As an example, $\Theta$ can be the
subset of finitely supported probability measures (distributions): 
\begin{equation*}
\Theta =
\left\{
  \nu = \sum_{j=1}^m {\alpha}_j\delta_{x_j} :
  m\in\mathbb{N},
  \alpha_j > 0,
  x_j\in\Re^d,
  \sum_{j=1}^m \alpha_j = 1
\right\},
\end{equation*}
and $P_0$ may be taken as a normalized
random measure (NRM)~\citep{regazzini2003distributional}.
As another example, $\Theta$ can be a finite set of distributions
\begin{equation*}
    \Theta = \{\nu_1 \ldots, \nu_Q\} \subset \mathcal{P}_2(\Re^d),
\end{equation*}
where each $\nu_i$ is a fixed probability measure on $\Re^d$. In this discrete setting, $P_0$ is naturally taken as a probability measure on the finite set $\Theta$, e.g., the uniform measure
$
    P_0 = \frac{1}{Q}\sum_{i=1}^{Q} \delta_{\nu_i}.
$ With a base measure $P_0$, we now define dDPP through a L-ensemble kernel.
\begin{definition}[dDPP]
\label{def:D2PP}
A \emph{distributional determinantal point process} (dDPP) on $ \Theta $
with L-ensemble kernel $L$ between distributions and base probability measure $P_0$
is a point process $\Phi$  whose J\'anossy density with respect to
$P_0^{\otimes k}$ is: 
\begin{align}
\label{eq:janossy}
j_k(\nu_1,\dots,\nu_k)
\propto \det(L(\nu_i,\nu_j))_{i,j=1}^k
 \end{align}
for every $k\ge 1$ and every unordered finite
configuration $\{\nu_1,\dots,\nu_k\}\subset \Theta $ with
normalizing constant
\begin{align*}
    \sum_{k=0}^{\infty} \frac{1}{k!}
\int_{\Theta^k}
\det\big( L(\nu_i, \nu_j) \big)_{i,j=1}^k
\, \mathrm{d} P_0(\nu_1)\ldots \mathrm{d} P_0(\nu_k).
\end{align*}
\end{definition}

Next, we prove that
under a compactness assumption on $\Theta$, the dDPP in Definition~\ref{def:D2PP}
with SW kernel~\eqref{eq:SWkernel} for $L(\nu_i,\nu_j)$
is a well-defined DPP.
In particular, beyond showing that  $L
(\cdot,\cdot)$ is symmetric positive definite, we show that the
normalization constant is finite so that~\eqref{eq:janossy} is a
well-defined density.

\begin{assumption}[Compactness of $\Theta$ under $\SW_2$]
\label{assumption:compact}
We assume that $\Theta \subset \mathcal{P}_2(\Re^d)$ is compact under the  $\SW_2$ metric. This holds if the following two conditions are satisfied:
\begin{enumerate}
    \item \textit{Closedness:} $\Theta$ is closed in $(\mathcal{P}_2(\Re^d), \W_2)$.
    \item \textit{Uniform integrability of second moments:}
    $
        \lim_{R \to \infty} \sup_{\nu \in \Theta}
        \int_{\|x\| \geq R} \|x\|^2 \, \mathrm{d}\nu(x) = 0 .
    $
\end{enumerate}
\end{assumption}

Condition~2 implies uniform tightness of $\Theta$ i.e., it gives $M := \sup_{\nu \in \Theta} \int \|x\|^2 \mathrm{d}\nu(x) < \infty$, and hence $\sup_{\nu \in \Theta} \nu\big(\{\|x\| \geq r\}\big) \leq M / r^2$ by Markov's
inequality. By Prokhorov's theorem, every sequence in $\Theta$ therefore admits a weakly convergent subsequence. Since $\W_2$-convergence is equivalent to weak convergence together with convergence of second moments~\citep[Theorem~6.9]{villani2009optimal}, Condition~2 upgrades this convergence to
$\W_2$. Thus $\Theta$ is relatively compact in $(\mathcal{P}_2(\Re^d), \W_2)$, and Condition~1 makes it compact. Finally, since $x \mapsto \langle v, x\rangle$ is $1$-Lipschitz for every $v\in \mathbb{S}^{d-1}$, pushing forward any coupling of
$\mu$ and $\nu$ yields $\W_2(P_v \sharp \mu, P_v\sharp \nu) \leq \W_2(\mu,\nu)$ and
therefore $\SW_2(\mu,\nu) \leq \W_2(\mu,\nu)$ for any $\mu,\nu \in \mathcal{P}_2(\mathbb{R}^d)$. The identity map
$(\mathcal{P}_2(\Re^d), \W_2) \to (\mathcal{P}_2(\Re^d), \SW_2)$ is consequently
$1$-Lipschitz, and continuous images of compact sets are compact. Therefore, $\Theta$ is
compact under $\SW_2$. A simple example satisfying Assumption~\ref{assumption:compact} is
\begin{equation*}
    \Theta = \Big\{ \nu \in \mathcal{P}_2(\Re^d) :
    \mathrm{supp}(\nu) \subset \bar{B}(0,R) \Big\},
\end{equation*}
for a fixed $R < \infty$, where $\bar{B}(0,R)$ denotes the closed Euclidean ball of radius $R$ centred at the origin. Uniform integrability of second moments: $\nu \in \Theta$, so $\sup_{\nu \in \Theta} \int_{\|x\| \geq r} \|x\|^2 \mathrm{d}\nu(x) = 0$, and the limit as $r \to \infty$ is zero. Closedness also holds: if $\nu_n \in \Theta$ and $\W_2(\nu_n, \nu) \to 0$, then $\nu_n \to \nu$ weakly, and since $\bar{B}(0,R)^c$ is open then $\nu\big(\bar{B}(0,R)^c\big) \leq \liminf_n \nu_n\big(\bar{B}(0,R)^c\big) = 0$, so $\nu \in \Theta$.  With this assumption, we now show that dDPP  with a SW kernel  is a
well-defined point process. 

\begin{theorem}[Well-definedness]
\label{theorem:well_defined}
Under Assumption~\ref{assumption:compact} on $\Theta$,
the dDPP with sliced Wasserstein
kernel~\eqref{eq:SWkernel} as the L-ensemble kernel is a well-defined
DPP on $\Theta$.
We write $\dDPP(\Theta,P_0,\gamma)$.
\end{theorem}
The proof of Theorem~\ref{theorem:well_defined} is given in
Section A.1 of the Supplementary Materials.
For the positive definiteness of the SW
kernel, we provide a direct proof rather than relying on the
conditional negative definiteness of SW distance as in existing
works~\citep{kolouri2016sliced}.  Let 
$T_L: \mathbb{L}_2(\Theta,P_0)\to \mathbb{L}_2(\Theta,P_0)$ be the
integral operator of $L$ (compare \eqref{TL} in the proof):
\begin{align*}
(T_L f)(\nu) = \int_{\Theta} L(\nu,\nu')\,f(\nu')\mathrm{d}P_0(\nu').
\end{align*}
We prove that $ \sum_{k=0}^{\infty} \frac{1}{k!}
\int_{\Theta^k}
\det\big( L(\nu_i, \nu_j) \big)_{i,j=1}^k
\, \mathrm{d} P_0(\nu_1)\ldots \mathrm{d} P_0(\nu_k):=\det(I+T_L)$ is finite by proving $T_L$ is trace-class using
Mercer's theorem.    While the proof focuses on the SW kernel
in~\eqref{eq:SWkernel}, it might be adaptable to other SW
kernels~\citep{kolouri2016sliced} and other probability kernels as
long as they are symmetric positive definite and the associated
integral operator is trace-class. 

\begin{remark}[Mercer's theorem]
As shown in the proof of Theorem~\ref{theorem:well_defined}, the SW kernel $L$ is symmetric positive definite and continuous. Under the compactness assumption of $\Theta$ (Assumption~\ref{assumption:compact}), Mercer's theorem for general metric spaces~\citep{steinwart2012mercer} applies since $P_0$ is a finite measure on $\Theta$. There exist non-negative eigenvalues $\{\lambda_\ell\}_{\ell \ge 1}$ and an orthonormal basis of eigenfunctions $\{\psi_\ell\}_{\ell \ge 1}$ of $\mathbb{L}_2(\Theta,P_0)$ such that
\begin{align*}
L(\nu,\nu') = \sum_{\ell=1}^\infty \lambda_\ell \psi_\ell(\nu)\psi_\ell(\nu'),
\end{align*}
with uniform convergence on $\Theta \times \Theta$.
\end{remark}
With Mercer's theorem, we can define the correlation kernel of the dDPP, which will be later used for posterior inference. 
\begin{definition}[Correlation kernel]
Let $T_L$ be the integral operator associated with the SW kernel $L$ on $\mathbb{L}_2(\Theta,P_0)$ and $K$ denote the correlation operator associated with $L$ and $P_0$.
For every $k \ge 1$, the $k$-th order correlation function (or $k$-point intensity function) is defined as
\begin{align*}
\rho_k(\nu_1,\dots,\nu_k)
=
\det\bigl(K(\nu_i,\nu_j)\bigr)_{i,j=1}^k,
\qquad \text{for }\nu_1,\dots,\nu_k \in \Theta.
\end{align*}
 Assume that $L$ admits the spectral decomposition $\{(\psi_\ell)\}_{\ell\ge 1}$. The \emph{correlation kernel} $K$ is  the integral kernel of the operator
\begin{align*}
T_K := T_L (I + T_L)^{-1}.
\end{align*}
Equivalently, $K$ admits the following expansion:
\begin{align*}
K(\nu,\nu')
=
\sum_{\ell=1}^{\infty}
\frac{\lambda_\ell}{1+\lambda_\ell}\,\psi_\ell(\nu)\psi_\ell(\nu'),
\qquad \text{for }\nu,\nu'\in\Theta.
\end{align*}
with $\lambda'_\ell = \lambda_\ell/(1+\lambda_\ell) < 1$.
\end{definition}

The relation $T_K = T_L(I+T_L)^{-1}$ converts the L-ensemble weights into inclusion intensities. The kernel $L$ enters through the J\'anossy density of Definition~\ref{def:D2PP}: $j_k(\nu_1,\dots,\nu_k) =
\det(L(\nu_i,\nu_j))_{i,j=1}^k / \det(I+T_L)$ is the intensity (with respect to $P_0^{\otimes k}$) that a realization $\Phi$ consists of exactly the $k$ atoms $\nu_1,\dots,\nu_k$, normalized by the
partition function $\det(I+T_L)$. The correlation function $\rho_k$ instead measures the intensity that $\Phi$ contains atoms at $\nu_1,\dots,\nu_k$ regardless of its remaining atoms, and is recovered from the J\'anossy densities by summing over and integrating out the unobserved atoms,
\begin{align*}
    \rho_k(\nu_1,\dots,\nu_k)
    = \sum_{n=0}^\infty \frac{1}{n!}
    \int_{\Theta^n} j_{k+n}(\nu_1,\dots,\nu_k,\eta_1,\dots,\eta_n)
    \prod_{r=1}^n \mathrm dP_0(\eta_r).
\end{align*}
Carrying out this marginalization on the determinantal J\'anossy density
cancels the partition function $\det(I+T_L)$ and yields $\rho_k=\det(K(\nu_i,\nu_j))_{i,j=1}^k$, where $K$ is the kernel of $T_K=T_L(I+T_L)^{-1}$. The map $T_L\mapsto T_L(I+T_L)^{-1}$ is precisely
this integrating-out of the rest of the process~\citep{shirai2003random}[Theorem 1.2 and Theorem ~6.17]. In the finite setting
$\Theta=\{\nu_1,\dots,\nu_Q\}$, where $T_L$ and $T_K$ are represented by
the matrices $\mathbf L$ and $\mathbf K=\mathbf L(I+\mathbf L)^{-1}$.

\subsection{Using Empirical Distributions}

We next discuss the approximation quality of the plug-in estimators of the L-ensemble kernel and the correlation kernel. In particular, we derive a concentration bound on the approximation error when we observe i.i.d samples from the underlying distributions.  We would like to note that in practice the distributions might be fully observed (e.g., in a discrete form), and therefore plug-in estimation might not be needed.

\begin{theorem}
\label{thm:sample_complexity}
Let $\Theta = \{\thbase{1}, \ldots, \thbase{Q}\} \subset \mathcal{P}_2(\Re^d)$ be a finite set of
probability measures and $P_0 = \frac{1}{Q}\sum_{i=1}^Q \delta_{\thbase{i}}$. For each
$i \in [Q]$, let
$\hat{\nu}^{(m)}_i = \frac{1}{m}\sum_{t=1}^m \delta_{x_t^{(i)}}$
be the empirical measure formed from $m$ i.i.d.\ samples
$\{x_t^{(i)}\}_{t=1}^m \sim \thbase{i}$, drawn independently across $i \in [Q]$.
Suppose $\mathrm{supp}(\thbase{i}) \subset B(0,R)$ for all $i \in [Q]$ and some fixed
$R < \infty$.
Define the true and estimated L-ensemble and correlation kernel matrices:
\begin{align*}
    L_{ij} 
    &= \exp\left(-\gamma  \SW_2^2(\thbase{i},\thbase{j})\right), 
    &\hat{L}_{ij} 
    &= \exp\left(-\gamma  \SW_2^2\left(\hat{\nu}^{(m)}_i,\hat{\nu}^{(m)}_j\right)\right),\\
    \mathbf{K} 
    &= \mathbf{L}(I + \mathbf{L})^{-1}, 
    &\hat{\mathbf{K}} 
    &= \hat{\mathbf{L}}(I + \hat{\mathbf{L}})^{-1}.
\end{align*}
Let $\lambda_{\min}$ denote the smallest eigenvalue of (p.d.)
$L$, and 
$\kappa := (1 + \lambda_{\min}(\mathbf{L}))^{-1}<1$. Then for any $\delta \in (0,1)$, with probability at least $1 - \delta$,
the following hold simultaneously:
    \begin{align*}
        &\|\hat{\mathbf{L}} - \mathbf{L}\|_{2}
        \leq
        \frac{16\gamma R^2 Q}{\sqrt{m}}
        \sqrt{\pi + \log(2Q^2/\delta)}, 
        \\
        &\|\hat{\mathbf{K}} - \mathbf{K}\|_{2}
        \leq
        \frac{16\kappa\gamma R^2 Q}{\sqrt{m}}
        \sqrt{\pi + \log(2Q^2/\delta)}.
    \end{align*}
\end{theorem}

The proof of Theorem~\ref{thm:sample_complexity} is given in Section A.2 of the Supplementary Materials. The bounds exhibit the parametric $\mathcal{O}(m^{-1/2})$ rate in the number of samples, which is notably dimension-free. The dependence on the number of distributions $Q$ is mild, 
inside the square root, aside from the linear factor arising from bounding the operator norm of a $Q \times Q$ matrix entrywise. Finally, the correlation kernel enjoys a strictly sharper bound than the L-ensemble kernel by the contraction factor $\kappa < 1$, reflecting the smoothing effect of the map $\mathbf{L} \to \mathbf{L}(I+\mathbf{L})^{-1}$.

\begin{corollary}
\label{cor:determinant}
Under the setting of Theorem~\ref{thm:sample_complexity}, let $S =\{i_1,\ldots,i_k\} \subset [Q]$ be any index set of size $k \leq Q$,
and let $\mathbf{L}_S$ and $\hat{\mathbf{L}}_S$ denote the $k \times k$ principal submatrices of $\mathbf{L}$ and $\hat{\mathbf{L}}$,
respectively, and similarly $\mathbf{K}_S$ and $\hat{\mathbf{K}}_S$ for $\mathbf{K}$ and $\hat{\mathbf{K}}$. Then for any $\delta \in
(0,1)$, with probability at least $1 - \delta$:
\begin{align}
    \left|\det(\hat{\mathbf{L}}_S) - \det(\mathbf{L}_S)\right|
    &\leq
    \frac{16\, k(3k)^{k-1} \gamma R^2 Q}{\sqrt{m}}
    \sqrt{\pi + \log(2Q^2/\delta)},
    \label{eq:det_L_bound}
    \\
    \left|\det(\hat{\mathbf{K}}_S) - \det(\mathbf{K}_S)\right|
    &\leq
    \frac{16\, k\,3^{k-1} \kappa\gamma R^2 Q}{\sqrt{m}}
    \sqrt{\pi + \log(2Q^2/\delta)},
    \label{eq:det_K_bound}
\end{align}
where $\kappa := (1 + \lambda_{\min}(\mathbf{L}))^{-1} < 1$.
\end{corollary}

The proof of Corollary~\ref{cor:determinant} is given in Section A.3 of the Supplementary Materials. This corollary transfers the guarantees of Theorem~\ref{thm:sample_complexity} to the determinants of principal submatrices, which govern the DPP probabilities where $\det(\mathbf{L}_S)$ is proportional to the probability that the dDPP configuration 
is exactly the subset $S=\{\thbase{i_1},\ldots,\thbase{i_k}\}$ (i.e., that these and only 
these atoms are jointly present in a realization of dDPP), while $\det(\mathbf{K}_S)$ 
gives the probability that $S$ is contained in the (random) realized configuration.
These probabilities are estimated consistently at the parametric $\mathcal{O}(m^{-1/2})$ rate. The prefactors $(3k)^{k-1}$ and $3^{k-1}$ arise from propagating an entrywise perturbation through a $k \times k$ determinant
and are benign for the small subset sizes of interest. The sharper factor in \eqref{eq:det_K_bound} is due to the fact that eigenvalues of $K$ is bounded by 1.

\section{
  Clustering Distribution-Valued Data}
\label{sec:mixture}
Consider observed distribution-valued data $\SSS = \{F_1, \ldots, F_N\}$ with
$F_i \in \PP_2(\Re^d)$, and assume we aim to infer a random
partition of $\SSS$, or equivalently, to infer the latent mixing
measure in a mixture model for $F_i$.
In this section, we introduce such a mixture model using the discrete
dDPP.

While the dDPP is well-defined for general measures, we consider a
dDPP mixture model under a simple choice of discrete base space
$\Theta$.
We assume $\Theta = \{\thbase{1}, \ldots, \thbase{Q}\}$ (with $0 < Q < \infty$) and
the uniform measure
\begin{equation*}
    P_0 = \frac{1}{Q}\sum_{i=1}^{Q} \delta_{\thbase{i}}
\end{equation*}
as the associated base measure. We focus on this finite discrete case for two reasons. First, it admits more tractable posterior inference, yielding a fast practical mixing rate for  Markov chain Monte Carlo (MCMC) posterior simulation. Second, it allows us to
establish posterior consistency.

Letting $\mathrm{dDPP}(\Theta,P_0,\gamma)$ denote the 
dDPP on $\Theta$ with the base measure $P_0$, we propose the following
hierarchical generalized Bayesian model: 
\begin{align}
\label{eq:likelihood}
    \ell(F_i \mid \thi{i})
    &= \exp\left(-w\SW_2^2(F_i, \thi{i})\right), \\
    \thi{i} \mid \Gmix  &\overset{i.i.d}{\sim} \Gmix, \quad i = 1,\ldots,N, \nonumber
\end{align}
where $\Gmix$ is a discrete random probability measure
with distribution-valued atoms which
define the clusters by way of sampling latent distribution-valued variables 
$\thi{i}$ from $\Gmix$, $w > 0$ is a bandwidth parameter controlling the concentration
of the likelihood around $\thi{i}$, and $\Gmix$ is a normalized random
measure over $\Theta$. Equation~\eqref{eq:likelihood} links $F_i$ with the
clusters. The atoms of the mixing measure $\Gmix$ are generated by a
DPP, as follows: 
\begin{align}
\label{eq:mixing_measure}
    &\Gmix =  \sum_{h=1}^{m'} \frac{s_h}{\sum_{h'=1}^{m'} s_{h'}} \delta_{\thtilde{h}},  \\
    &\{\thtilde{1},\ldots,\thtilde{m}\} \sim \dDPP(\Theta, P_0,\gamma), \nonumber\\
    &s_1,\ldots,s_{m'} \overset{i.i.d}{\sim} \text{Gamma}(a,1), \nonumber
\end{align}
 We say $F_i$ belongs to cluster $h$ if $\thi{i}=\thtilde{h}$.   Here $\thi{i}$ and $\thtilde{h}$ are themselves
distributions (elements of $\Theta$)
(adopting conventional notation from the Bayesian nonparametric
literature to highlight the hierarchical structure). In
summary, we define a random partition of $F_1,\ldots,F_N$ by setting
up a mixture model using a generalized Bayes kernel based on SW
distance. 

 Since the random mixing measure $\Gmix$ is supported on the finite set $\Theta = \{\thbase{1},\ldots,\thbase{Q}\}$, it is
fully characterized by the weight it places on
each atom. Concretely, identifying
\begin{align*}
    \pi_q = \Gmix(\{\thbase{q}\}), \qquad q = 1,\ldots,Q,
\end{align*}
the measure $\Gmix$ corresponds one-to-one to a vector
$\pi = (\pi_1, \ldots, \pi_Q) \in \Delta^{Q-1}$.
For later reference, we note that \eqref{eq:likelihood} and using
$\pi$ to represent a mixing measure $G$ implies
marginal (general) likelihood
$\ell(F_i \mid G) = \sum_q \pi_q \exp\left\{-w
  \SW_2^2(F,\nu_q)\right\}$.

The posterior on $\Gmix$ is characterized by a probability
model  $\Pi_N(\pi \mid F_1,\ldots,F_N)$ as a distribution over
$\Delta^{Q-1}$, and we establish the following posterior consistency
result. 

\begin{theorem}[Posterior Consistency]
\label{theorem:consistency}
Let $\Gtrue \in \mathcal{P}(\mathcal{P}_2(\Re^d))$ be the true data-generating distribution of $F_1, \ldots,F_N  \overset{i.i.d.}{\sim}\Gtrue $,
and assume that
\begin{align}
\label{eq:moment_assumption}
    \mathbb{E}_{F\sim \Gtrue}\left[\exp\left(w\sum_{q=1}^Q \SW_2^2(F,\thbase{q})\right)\right]
    < \infty
\end{align}
(for any $\thbase{q} \in \Theta$). Then for every $\varepsilon > 0$,  we have
\begin{align*}
    \Pi_N\left(\tilde{d}(\pi,\mathcal{M}^*) > \varepsilon
    \Big| F_1,\ldots,F_N\right)
    \xrightarrow{a.s.} 0
    \quad \text{as } N\to\infty,
\end{align*}
where $\tilde{d}(\pi,\mathcal{M}^*) = \min_{\pi'\in\mathcal{M}^*}\|\pi-\pi'\|_1$
and $\mathcal{M}^* := \argmin_{\pi \in \Delta^{Q-1}} R(\pi)$
is the set of minimizers of the population risk
$R(\pi) := \mathbb{E}_{F \sim \Gtrue}[-\log\sum_{q=1}^Q \pi_q\,
\exp(-w\,\SW_2^2(F,\thbase{q}))]$.
\end{theorem}

The proof of Theorem~\ref{theorem:consistency} is given in Section A.4 of the Supplementary Materials. As $N \to \infty$, the posterior concentrates on the set $\mathcal{M}^*$ of
mixing weights minimizing the population risk $R(\pi)$, and hence the procedure recovers
the best mixture representation within $\Theta$.


\begin{remark}[Consistency under empirical approximation]
In the case where each $F_i$ is observed only through an empirical measure $\hat F_i^{(m)}$ formed from $m$ i.i.d. samples, the conclusion of Theorem~\ref{theorem:consistency} continues to hold (in probability) as
$N\to\infty$ and $m=m(N)\to\infty$ jointly, with no condition on their relative rates. Indeed, replacing each $F_i$ by $\hat F_i^{(m)}$ perturbs the generalized likelihood only through the discrepancies $|\SW_2^2(F_i,\thbase{q})-\SW_2^2(\hat F_i^{(m)},\thbase{q})|$, which vanish as $m\to\infty$ because empirical measures of a $\mathcal{P}_2(\Re^d)$ law converge in $W_2$, hence in $\SW_2$ \citep{nadjahi2020statistical}. Assumption~\eqref{eq:moment_assumption} makes these discrepancies uniformly integrable, and hence their average over the $N$ observations tends to $0$.
\end{remark}

\section{Posterior Inference}
\label{sec:posterior_inference}
\subsection{Posterior Characterization}

We adapt Theorems 1--3 of~\citet{beraha2025bayesian} for the posterior
characterization
under model \eqref{eq:likelihood}-\eqref{eq:mixing_measure},
working here with the
correlation kernel $K$. While one could sample directly via a
reversible-jump scheme operating on 
the L-ensemble~\citep{xu2016bayesian,green1995reversible}, we instead adapt the sampler of \citet{beraha2025bayesian},
which yields a P\'olya-urn-like scheme.
 Below we write $H(s)$  generically for the prior
distribution of the marks $s_h$ in
\eqref{eq:mixing_measure}, and $h(s)$ for the corresponding density,
keeping in mind the example of the $\text{Gamma}(a,1)$ prior. 
 The model augmentation with $U_N$ in the following proposition is introduced in anticipation
of a simplification in the conditional posterior for $\Gbar$.
The factor $\{\Gbar(\Theta)\}^N$ in the gamma normalization constant of $p(U_N \mid G)$ cancels against the $1/\Gbar(\Theta)$ in the (augmented)
joint posterior that arises from $p(\theta_i \mid
\Gbar)=\Gbar(\theta_i)/\Gbar(\Theta)$ in \eqref{eq:likelihood}, greatly
simplifying the following posterior conditionals. 

\begin{proposition}[Posterior Characterization of dDPP Mixture]
\label{proposition:posterior}
Since $(\Theta, \SW_2)$ is a compact space and the dDPP
$\dDPP(\Theta,P_0,\gamma)$  is a
well-defined finite point process on $\Theta$
(Theorem~\ref{theorem:well_defined}), the general posterior
characterization of~\citet{beraha2025bayesian}[Theorem 1 and Example
3] applies. Let $\thstarvec = (\thstar{1},\ldots,\thstar{k})$
denote the distinct elements among $\{\thi{1},\ldots,\thi{N}\}$ with
multiplicities $n_1,\ldots,n_k$, and let $U_N$ be a latent variable with
$U_N \mid \Gmix \sim \mathrm{Gamma}(N, \Gmixbar(\Theta))$, where $
\Gmixbar = \sum_{h=1}^k s_h\,\delta_{\thtilde{h}}$ is the unnormalized
random measure.
The posterior of $ \Gmixbar$ given $\thstarvec$ and $U_N = u$ equals
in distribution:
\begin{align*}
 (\Gmixbar \mid \thstarvec, U_N = u)
\overset{d}{=}
\sum_{j=1}^k S^*_j\,\delta_{\thstar{j}} +  \Gmixbarp,
\end{align*}
where \emph{(i)} $S^*_1,\ldots,S^*_k$ are independent with densities $f_{S^*_j}(s) \propto e^{-us}s^{n_j}h(s)$, \emph{(ii)} $ \Gmixbarp$ is an independent residual random measure whose unmarked point process is a
dDPP on $\Theta$ with correlation kernel
\begin{align*}
    K_{\thstarvec}'(\thbasenew,\theta')
= K(\thbasenew,\theta')
- \sum_{i,j=1}^k (K^{-1}_{\thstarvec})_{ij}\,
K(\thbasenew,\thstar{i})\,K(\theta',\thstar{j}),
\end{align*}
where $K_{\thstarvec}$ is the $k \times k$ matrix with entries
$K(\thstar{i},\thstar{j})$, and marks drawn i.i.d. from
$f_{S'}(s;u) \propto e^{-su}h(s)$, and \emph{(iii)} the conditional
density of $U_N$ given $\thstarvec$ satisfies
\begin{align*}
f_{U_N \mid \thstarvec}(u)
\propto u^{N-1}
\prod_{j \geq 1}
\bigl(1 - \lambda^{\thstarvec}_j + \lambda^{\thstarvec}_j\,\psi(u)\bigr)
\prod_{j=1}^k \kappa(u, n_j),
\end{align*}
where $\{\lambda^{\thstarvec}_j\}_{j\geq 1}$ are the eigenvalues of
$K_{\thstarvec}'$,
$\psi(u) = \int_{\Re_+} e^{-us}h(s)\,ds$ is the Laplace
transform of the mark distribution $H$,
and $\kappa(u,n) = \int_{\Re_+} e^{-us}s^n h(s)\,ds$,
 with closed form expressions under the $\text{Gamma}(a,1)$ prior in
\eqref{eq:mixing_measure}. 
\end{proposition}
The residual kernel
$K_{\thstarvec}'$ is the Schur complement of $K_{\thstarvec}$ in the 
augmented kernel matrix. In particular, it is small when $\thbasenew$
is close to the observed atoms $\thstar{j}$ in $\SW_2$, encoding that
the posterior residual process avoids regions of $\Theta$ already
occupied by observed atoms. 
Also note that here, following
\citet{beraha2025bayesian}
we have characterized the posterior dDPP $\Gmixbarp$ by way of the
correlation kernel $K_{\thstarvec}'$ (rather than as L-ensemble).
\begin{proposition}[Marginal and Predictive Distributions of dDPP Mixture]
\label{proposition:predictive}
Under the same conditions as Proposition~\ref{proposition:posterior}, the marginal distribution of $\thi{1},\ldots,\thi{N}$ with $k$ distinct values $\thstarvec$ and counts $n_1,\ldots,n_k$~\citep{beraha2025bayesian}[Theorem 2 and Example 5] is:
\begin{align*}
P(\thi{1},\ldots,\thi{N} \in d\thb)
&= \int_{\Re_+} \frac{u^{N-1}}{\Gamma(N)}
\prod_{j \geq 1}\left(1-\lambda^{\thstarvec}_j
+\lambda^{\thstarvec}_j\psi(u)\right)
\prod_{j=1}^k \kappa(u,n_j)\,du \nonumber
\\ & \quad \times \det\bigl(K(\thstar{i},\thstar{j})\bigr)_{i,j=1}^k
\prod_{j=1}^k P_0(d\thstar{j}).
\end{align*}
The determinant factor $\det(K(\thstar{i},\thstar{j}))_{i,j=1}^k$ is the dDPP factorial moment measure density with respect to $P_0^k$, which is
large when the distinct atoms $\thstar{1},\ldots,\thstar{k}$ are
well-separated under $\SW_2$, directly reflecting the repulsive prior.
Conditionally on $\thstarvec$ and $U_N = u$, the predictive
distribution of $\thi{N+1}$~\citep{beraha2025bayesian}[Theorem 3 and
  Example 7] is: 
\begin{align}\label{CondPred}
&P(\thi{N+1} \in \mathrm d\thbasenew \mid \thstarvec, U_N = u)
\propto
\sum_{j=1}^k \frac{\kappa(u,n_j+1)}{\kappa(u,n_j)}
\delta_{\thstar{j}}(d\thbasenew)\nonumber\\&\qquad
+
\kappa(u,1)
\frac{
\prod_{j\geq 1}(1-\lambda^{\thstarvec,\thbasenew}_j
+\lambda^{\thstarvec,\thbasenew}_j\psi(u))
}{
\prod_{j\geq 1}(1-\lambda^{\thstarvec}_j
+\lambda^{\thstarvec}_j\psi(u))
}
\Delta_K(\thstarvec,\thbasenew)
P_0(\mathrm d\thbasenew),
\end{align}
where $\lambda^{\thstarvec,\thbasenew}_j$ are the eigenvalues of $K_{\thstarvec,\thbasenew}'$ evaluated at the augmented atom set $(\thstarvec,\thbasenew)$, and
\begin{align}\label{DeltaK}
\Delta_K(\thstarvec,\thbasenew)
= K(\thbasenew,\thbasenew) - K^\top_{\thstarvec,\thbasenew}\,K^{-1}_{\thstarvec}\,K_{\thstarvec,\thbasenew},
\end{align}
with $K_{\thstarvec,\thbasenew} = (K(\thbasenew,\thstar{1}),\ldots,K(\thbasenew,\thstar{k}))^\top$, is the SW repulsion term. We have $\Delta_K(\thstarvec,\thbasenew) \in [0,1]$, it approaches $0$ as $\thbasenew \to \thstar{j}$ for any $j$ (the new atom is
$\SW_2$-close to an existing one) and is maximized when $\thbasenew$ is $\SW_2$-distant from all current atoms (maximal distributional diversity).
This gives a precise geometric meaning to repulsion in $\mathcal{P}_2(\Re^d)$ i.e., new mixture components are penalized
proportionally to their sliced Wasserstein proximity to existing ones.
\end{proposition}

\subsection{Posterior Simulation}

We adapt the marginal MCMC sampler  of~\citet{beraha2025bayesian} to
the discrete dDPP mixture. Since $\Theta =
\{\thbase{1},\ldots,\thbase{Q}\}$  
is finite,
 it becomes straightforward to use \eqref{DeltaK} to
analytically integrate out the random measure $\Gmix$. 
Posterior inference proceeds over the latent assignments 
$\mathbf{c} = (c_1,\ldots,c_N)$ with $c_i \in \{1,\ldots,Q\}$ such
that  $\thi{i} = \thbase{c_i}$, together with the auxiliary variable
$U_N$.  As before, we write $\thstarvec =
(\thstar{1},\ldots,\thstar{k})$ for the distinct  occupied atoms among
$\{\thi{j}\}_{j=1}^N$, $n_j$ for the multiplicities of  observations assigned
to $\thstar{j}$, $K_{\thstarvec}$ for the $k \times k$ submatrix  of
$\mathbf{K}$ with entries $K(\thstar{i},\thstar{j})$, and  
$K_{\thstarvec,\thbasenew} =
(K(\thbasenew,\thstar{1}),\ldots,K(\thbasenew,\thstar{k}))^\top$ for
the vector of kernel  evaluations between a candidate atom
$\thbasenew$ and the currently occupied atoms.  The algorithm iterates
over three steps. 

\textbf{Update $U_N$.} 
Given the current assignments $\mathbf{c}$ with $k$ distinct atoms $\thstarvec$,  we sample $U_N$ via Metropolis-Hastings targeting the density in  Proposition~\ref{proposition:posterior}(iii):
\begin{align*}
f_{U_N \mid \thstarvec}(u) \propto u^{N-1}
\prod_{j \geq 1}
\bigl(1 - \lambda^{\thstarvec}_j + \lambda^{\thstarvec}_j\,\psi(u)\bigr)
\prod_{j=1}^k \kappa(u, n_j),
\end{align*}
where $\{\lambda^{\thstarvec}_j\}_{j \geq 1}$ are the eigenvalues of
the residual  correlation kernel $K_{\thstarvec}'$
defined in
Proposition~\ref{proposition:posterior}, 
$\psi(u) = (1+u)^{-a}$, and  $\kappa(u,n) =
\Gamma(n+a)/\Gamma(a)(1+u)^{-(n+a)}$ for  $\Ga(a,1)$ marks. We use a
log-normal Metropolis-Hastings proposal $U_N' = U_N \exp(\epsilon)$
with $\epsilon \sim \mathcal{N}(0,\sigma^2)$. 

\textbf{Update assignments.}
For $i = 1,\ldots,N$, we remove observation $i$ from the current  configuration and let $\thstarvec_{(-i)}$ denote the remaining distinct atoms with counts $n^{(-i)}_j$. We then sample the new assignment $c_i$ from a categorical distribution  with unnormalized  log-weights:
\begin{align*}
\log p_q \propto
\begin{cases}
\log\dfrac{\kappa(u,\,n^{(-i)}_q+1)}{\kappa(u,\,n^{(-i)}_q)}
+ \log \ell(F_i \mid \thbase{q})
& \text{if } \thbase{q} \in \thstarvec_{(-i)}, \\[12pt]
\log \kappa(u,1)
+ \log \widetilde{r}\left(\thstarvec_{(-i)},\thbase{q}; u\right)  + \log \ell(F_i \mid \thbase{q})
- \log n_{\mathrm{aux}}
& \text{if } \thbase{q} \notin \thstarvec_{(-i)},
\end{cases}
\end{align*}
where $\ell(F_i \mid \thbase{q}) = \exp(-w\,\SW_2^2(F_i, \thbase{q}))$ is the SW 
generalized likelihood and
\begin{align*}
\log \widetilde{r}\left(\thstarvec_{(-i)},\thbase{q}; u\right)
\approx
\bigl(\mathrm{Tr}(K'_{(\thstarvec_{(-i)},\thbase{q})})
- \mathrm{Tr}(K'_{\thstarvec_{(-i)}})\bigr)\,(\psi(u) - 1)
\end{align*}
is the Le~Cam approximation~\citep{beraha2025bayesian} to the
log-ratio  of Laplace functionals
in Proposition~\ref{proposition:predictive}
(second line of \eqref{CondPred}), with
$\mathrm{Tr}(K'_{\thstarvec_{(-i)}})$ the trace of the residual kernel
after  
removing observation $i$ and $\mathrm{Tr}(K'_{(\thstarvec_{(-i)},\thbase{q})})$ the trace after further adding $\thbase{q}$. For new atoms, we propose 
$n_{\mathrm{aux}}$ (is later set to 1) candidates drawn proportionally to $\Delta_K(\thstarvec_{(-i)}, \cdot)$ from the unoccupied atoms in $\Theta$, following the auxiliary variable scheme of~\citet{neal2000markov}.

\textbf{Update occupied atoms.}
Given the current assignments $\mathbf{c}$ and occupied atoms
$\thstarvec$, we update each occupied atom $\thstar{h}$ via a
Metropolis-Hastings transition probability. We propose swapping
$\thstar{h}$ with a 
uniformly drawn unoccupied atom  $\thbaseprop \in \Theta \setminus
\thstarvec$ and accept  using log acceptance ratio: 
\begin{align*}
\log \alpha 
= \bigl(\log\det K_{\thstarvec_{\mathrm{prop}}} 
- \log\det K_{\thstarvec}\bigr)
+ \sum_{\{i:\,c_i = h\}}
\bigl(\log\ell(F_i \mid \thbaseprop) 
- \log\ell(F_i \mid \thstar{h})\bigr),
\end{align*}
where $\thstarvec_{\mathrm{prop}}$ is $\thstarvec$ with $\thstar{h}$
replaced by $\thbaseprop$. The first term favors configurations where
the occupied atoms are well-separated under $\SW_2$ via the dDPP
prior, and the second term arises from the SW likelihood of the
observations assigned to cluster $h$. 

\begin{remark} The $Q \times Q$ kernel matrices $\mathbf{L}$ and $\mathbf{K}$ are 
precomputed once from the pairwise $\SW_2$ distances among $\thbase{1},\ldots,\thbase{Q}$, costing $\mathcal{O}(Q^2)$ kernel evaluations. The SW repulsion term $\Delta_K(\thstarvec_{(-i)},\thbase{q})$ is computed via a rank-one Cholesky update of $K_{\thstarvec_{(-i)}}$, costing $\mathcal{O}(k^2)$ per candidate atom rather than $\mathcal{O}(k^3)$ from scratch. The SW generalized likelihoods $\ell(F_i \mid \thbase{q})$ for all $q = 1,\ldots,Q$ 
are also precomputed from the pairwise $\SW_2$ distances between observations $F_i$ and atoms $\thbase{q}$, which are fixed throughout the MCMC. 
\end{remark}
\subsection{Posterior Summarization}
\label{subsec:summarization}

For posterior summarization, we extend the approach
of~\citet{nguyen2026summarizing} to report a point estimate of the
mixing measure $\Gmix$, which in turn implies the random
partition.
We discuss the point estimate $\hat{\Gmix}$ for $\Gmix$ below.
After obtaining
$\hat{\Gmix}:=\sum_{j=1}^k \pi_j \delta_{\thstar{j}}$, a point
estimate of the cluster label $c$ for any data point $F$
is then readily obtained using maximum conditional probabilities: 
\begin{align*}
    c =  \argmax_{j\in \{1,\ldots,k\}} \log  p (c=j \mid \hat{\Gmix},F) =  \argmax_{j\in \{1,\ldots,k\}} \log \pi_j - w \, \SW_2^2(F,\thstar{j}).
\end{align*}

Back to the point estimate for $\Gmix$.
Although the marginal MCMC integrates out $\Gmix$
analytically and targets the posterior over the assignments
$\mathbf{c} = (c_1,\ldots,c_N)$ and the auxiliary variable $U_N$,
samples of the mixing measure itself are recovered at no additional
cost by exploiting the conditional in
Proposition~\ref{proposition:posterior}. At the end of each iteration,
the sampler yields $k$ distinct occupied atoms $\thstarvec =
(\thstar{1},\ldots,\thstar{k})$, counts $n_1,\ldots,n_k$, and $U_N =
u$; the unnormalized weights $S^*_1,\ldots,S^*_k$ are then
conditionally independent with densities $f_{S^*_j}(s) \propto
e^{-us}\, s^{n_j} h(s)$, which for $H = \mathrm{Gamma}(a,1)$ 
reduces to $S^*_j \mid \thstarvec, U_N = u \sim \mathrm{Gamma}(n_j + a,\, 1+u)$ independently across $j$. Normalizing yields $\pi_j = s_j / \sum_{j'} s_{j'}$. We record  the mixing measure  as
$
    \Gmix = \sum_{j=1}^k \pi_j\,\delta_{\thstar{j}}.
$
The full posterior also includes a residual dDPP component on $\Theta
\setminus \thstarvec$ (Proposition~\ref{proposition:posterior}), but
since residual atoms carry no observational support, their posterior
weights are negligible and this component is omitted in
practice.

The key challenge of the default proposed in ~\citet{nguyen2026summarizing} is that $\Gmix$ is a measure over distributions, which
calls for a different utility function. Let $W_{\SW_2^2}(\hat{\Gmix}, \Gmix)$ denote Wasserstein
distance with $\SW_2^2$ as the ground cost. We report a point estimate of $\Gmix$ based on the loss $\ell (\hat{\Gmix}, \Gmix) = W_{\SW_2^2}(\hat{\Gmix}, \Gmix)$, and therefore posterior expected loss
\begin{align*}
    \mathbb{L}(\hat{\Gmix})=\mathbb{E}\left[
    W_{\SW_2^2}\big(\hat{\Gmix}, \Gmix\big)
    \,\Big|\, F_1,\ldots,F_N\right].
\end{align*}
We then report a point estimate $\hat{\Gmix}$ as the Bayes rule by
minimizing posterior expected loss $ \mathbb{L}(\hat{\Gmix})$. Using
imputed posterior MCMC samples $\Gmix_t$ across iterations
$t=1,\ldots,T$, we approximate $ \mathbb{L}$ as a Monte Carlo average,
reporting the point estimate: 
\begin{align}\label{Ghat}
  \hat{\Gmix} = \arg
    \min_{G \in \{\Gmix_1,\ldots,\Gmix_T\}}
    \frac{1}{T}\sum_{t=1}^T
    W_{\SW_2^2}\big(G, \Gmix_t\big).
\end{align}
The resulting point estimate $\hat{\Gmix}$ is thus selected among the
MCMC samples $\{\Gmix_1,\ldots,\Gmix_T\}$ as the Bayes rule
\eqref{Ghat}
minimizing the average $W_{\SW_2^2}$ distance to all other samples. Selecting the
estimate requires the pairwise distance matrix among the $T$ samples,
giving $\mathcal{O}(T^2)$ evaluations of $W_{\SW_2^2}$. Each
evaluation in turn solves an optimal transport problem over the at
most $k$ mixture components, whose ground-cost entries are SW
distances between distributions with at most $m$ atoms. The overall
cost is therefore $\mathcal{O}\big[T^2 (k^3 \log k + k^2\, m\log
m)\big]$, where the $k^3 \log k $ term 
is the optimal transport solve and $k^2\, m\log m$ accounts for forming the
$\SW_2^2$ ground-cost matrix.

\section{ Examples}
\label{sec:experiments}

In the following two applications, we used the dDPP for inference on
random partitions of distribution-valued data.  We compare 
against inference under a Ferguson-Dirichlet process (DP) prior for $\Gmix$
using the same base measure and concentration parameter. Both are used as priors in the
generalized Bayes framework
of~\eqref{eq:likelihood}-\eqref{eq:mixing_measure}. For the DP, we
implement Neal's Algorithm 8~\citep{neal2000markov}, augmented with an
update step for occupied atoms analogous to that used for the dDPP. Since
the models differ only in their choice of prior on $\Gmix$, the comparison is
controlled and fair. We conduct experiments on population-scale
single-cell 
dataset OneK1K~\citep{yazar2022single} in
Section~\ref{subsec:single_cell} and on
data from the Human Epilepsy Project
(HEP)~\citep{French_HumanEpilepsyProject} data in Section~\ref{subsec:hep}. In
both cases, we use the set of data points $\mathcal{S}$ as the atom
set $\Theta$
 with the  base measure $P_0$ being a uniform distribution over
$\Theta$, 
for both DP and dDPP. For the generalized-likelihood
hyperparameter $w$, we consider values chosen relative to the scale of
the pairwise $\SW_2^2$ distances between data points in $\mathcal{S}$,
using the inverse of their median as a reference scale. For
evaluation, we report the posterior expectations of the number of
clusters, mean cluster size, log generalized likelihood, and repulsion
(measured by the log unnormalized density of the dDPP prior). We also
compute these quantities for the point estimates obtained via the
Bayes rule \eqref{Ghat}. In all experiments, we use 1000
projections to approximate SW distance and SW kernels. Those
projections are kept fixed during posterior inference without
resampling. We run 2000 MCMC iterations with 1000 iterations for
burn-in.

\subsection{Single-Cell Data}
\label{subsec:single_cell}

Population-scale single-cell RNA-sequencing count matrices (with thousands of cells and $10{,}786$ genes) were obtained for four immune cell populations (B cells, T cells, natural killer (NK) cells, and monocytes) for each donor. Donors with available data across all four cell types were retained, yielding a common cohort of 961 donors. For each donor, cells from all four cell types were pooled and preprocessed
using per-cell total-count normalization followed by a $\log(1+x)$
transformation $
    \tilde{x}_{ij} = \log\left(1 + \frac{x_{ij}}{\sum_{j'} x_{ij'}}\right),
$
where $x_{ij}$ denotes the raw count of gene $j$ in cell $i$. A donor-level mean expression matrix
$M \in \Re^{961 \times 10{,}786}$ was constructed by averaging the normalized, log-transformed expression profiles of all cells belonging to each donor. We applied principal component analysis (PCA) to $M$ and retained the top 17 principal components 
(cumulative explained variance about $95\%$). The fitted PCA projection was then applied to every
individual cell of each donor, producing for each donor $p$ an
empirical distribution of cells in the $17$-dimensional PCA space.

\begin{table}[!t]
\centering
\caption{
  Single-cell data.
The table reports inference for the number of clusters $k$, average
cluster size, log generalized-likelihood, and repulsion (log
unnormalized dDPP density):  
$\log\det (\exp(-\SW_2^2(\cdot,\cdot)))$). ``Point" is
the Bayes rule \eqref{Ghat};
$\EEb$ and HCI refer to posterior expectation and the 95\%
highest-credible interval (HCI) based on the last $T$ samples. } 
\label{tab:combined_all}
\resizebox{\textwidth}{!}{%
\begin{tabular}{ll cc cc cc cc}
\toprule
 & & \multicolumn{2}{c}{\textbf{Clusters }$k$} & \multicolumn{2}{c}{\textbf{Average  Cluster Size}} & \multicolumn{2}{c}{\textbf{Log-generalized likelihood}} & \multicolumn{2}{c}{\textbf{Repulsion}} \\
\cmidrule(lr){3-4}\cmidrule(lr){5-6}\cmidrule(lr){7-8}\cmidrule(lr){9-10}
\textbf{Model} & $w$ & \textbf{Point} & $\mathbb{E}$ and [HCI]& \textbf{Point} & $\mathbb{E}$ and [HCI]& \textbf{Point} & $\mathbb{E}$ and [HCI]& \textbf{Point} & $\mathbb{E}$ and [HCI]\\
\midrule
\multirow{3}{*}{DP}
  & $5{\times}10^5$ & 14 & 14.45 [11.00, 17.00] & 68.64 & 67.46 [50.58, 80.08] & $-4.17$ & $-4.17$ [$-4.18, -4.16$] & $-183.58$ & $-196.71$ [$-254.93, -146.09$] \\
  & $7{\times}10^5$ & 24 & 23.86 [21.00, 27.00] & 40.04 & 40.51 [35.59, 45.76] & $-5.16$ & $-5.16$ [$-5.18, -5.15$] & $-341.23$ & $-361.61$ [$-430.48, -300.96$] \\
  & $10^6$          & 37 & 34.77 [31.00, 38.00] & 25.97 & 27.73 [25.29, 31.00] & $-6.35$ & $-6.36$ [$-6.38, -6.34$] & $-612.27$ & $-575.06$ [$-639.70, -509.13$] \\
\midrule
\multirow{3}{*}{dDPP ($\gamma=10^3$)}
  & $5{\times}10^5$ & 11 & 11.77 [11.00, 13.00] & 87.36 & 82.06 [73.92, 87.36] & $-4.21$ & $-4.22$ [$-4.23, -4.21$] & $-133.79$ & $-149.21$ [$-172.34, -132.02$] \\
  & $7{\times}10^5$ & 21 & 22.23 [20.00, 24.00] & 45.76 & 43.34 [38.44, 45.76] & $-5.26$ & $-5.26$ [$-5.27, -5.24$] & $-313.32$ & $-336.60$ [$-364.19, -300.58$] \\
  & $10^6$          & 32 & 33.04 [31.00, 36.00] & 30.03 & 29.15 [26.69, 31.00] & $-6.42$ & $-6.45$ [$-6.49, -6.42$] & $-523.72$ & $-542.00$ [$-597.89, -508.47$] \\
\midrule
\multirow{3}{*}{dDPP ($\gamma=10^4$)}
  & $5{\times}10^5$ & 11 & 11.68 [11.00, 13.00] & 87.36 & 82.48 [73.92, 87.36] & $-4.24$ & $-4.25$ [$-4.25, -4.24$] & $-134.30$ & $-146.04$ [$-170.94, -131.94$] \\
  & $7{\times}10^5$ & 20 & 19.76 [18.00, 22.00] & 48.05 & 48.83 [43.68, 53.39] & $-5.23$ & $-5.25$ [$-5.28, -5.23$] & $-287.61$ & $-285.09$ [$-325.09, -261.32$] \\
  & $10^6$          & 33 & 33.08 [30.00, 36.00] & 29.12 & 29.14 [26.69, 32.03] & $-6.49$ & $-6.49$ [$-6.52, -6.47$] & $-519.52$ & $-527.78$ [$-585.97, -470.31$] \\
\midrule
\multirow{3}{*}{dDPP ($\gamma=10^5$)}
  & $5{\times}10^5$ & 12 & 11.88 [10.00, 13.00] & 80.08 & 81.35 [73.92, 96.10] & $-4.19$ & $-4.20$ [$-4.22, -4.19$] & $-158.23$ & $-152.19$ [$-173.01, -120.30$] \\
  & $7{\times}10^5$ & 20 & 21.79 [20.00, 25.00] & 48.05 & 44.31 [38.44, 48.05] & $-5.26$ & $-5.26$ [$-5.29, -5.24$] & $-301.51$ & $-327.03$ [$-379.44, -287.03$] \\
  & $10^6$          & 34 & 35.98 [32.00, 39.00] & 28.26 & 26.79 [24.03, 29.12] & $-6.45$ & $-6.47$ [$-6.53, -6.43$] & $-558.91$ & $-594.74$ [$-650.37, -518.95$] \\
\bottomrule
\end{tabular}%
}
\end{table}

\begin{figure}[!t]
    \centering
    \begin{tabular}{cc}
         \includegraphics[width=0.45\linewidth]{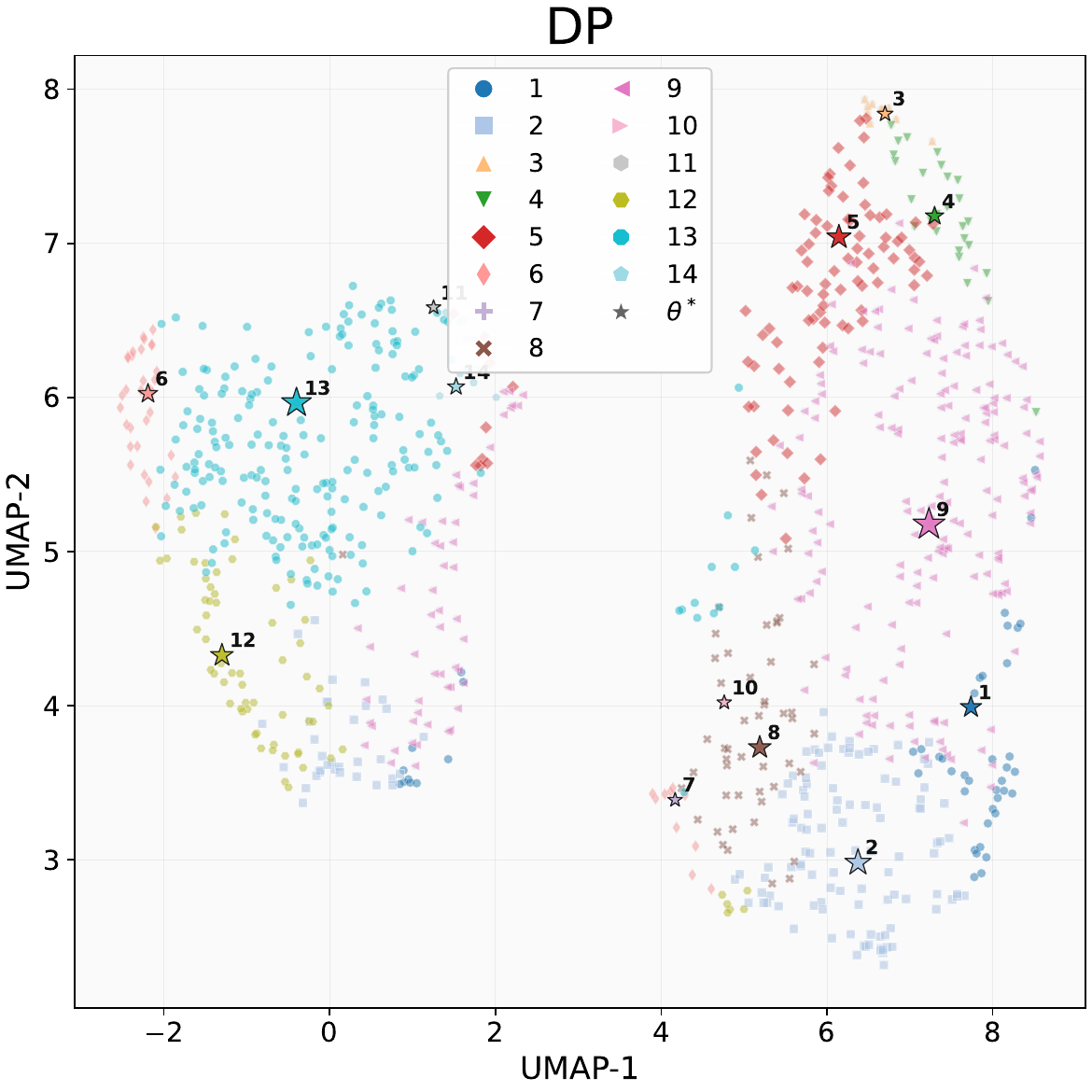}
         &
         \includegraphics[width=0.45\linewidth]{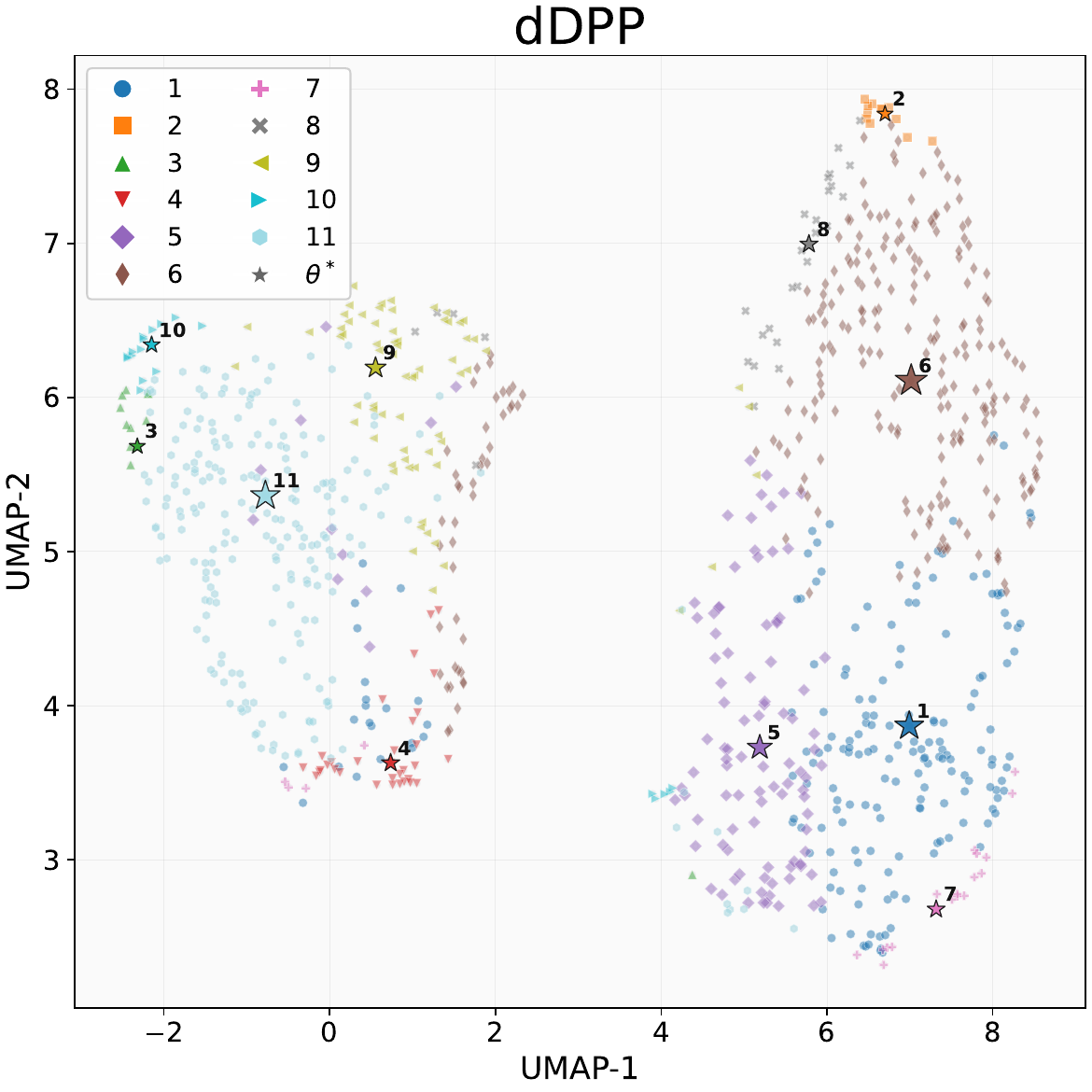} 
    \end{tabular}
    \caption{Single-cell data.
      UMAP visualization based on point estimates \eqref{Ghat}
      of the mixing measures under the DP  and dDPP models
      ($w=5\times10^5,\gamma=10^3$). Colors indicate the corresponding
      partitions. Atoms of the mixing measures are shown as stars,
      with size proportional to their weights.} 
    \label{fig:mixing_500000}
\end{figure}

\begin{figure}[!t]
    \centering
    \includegraphics[width=1\linewidth]{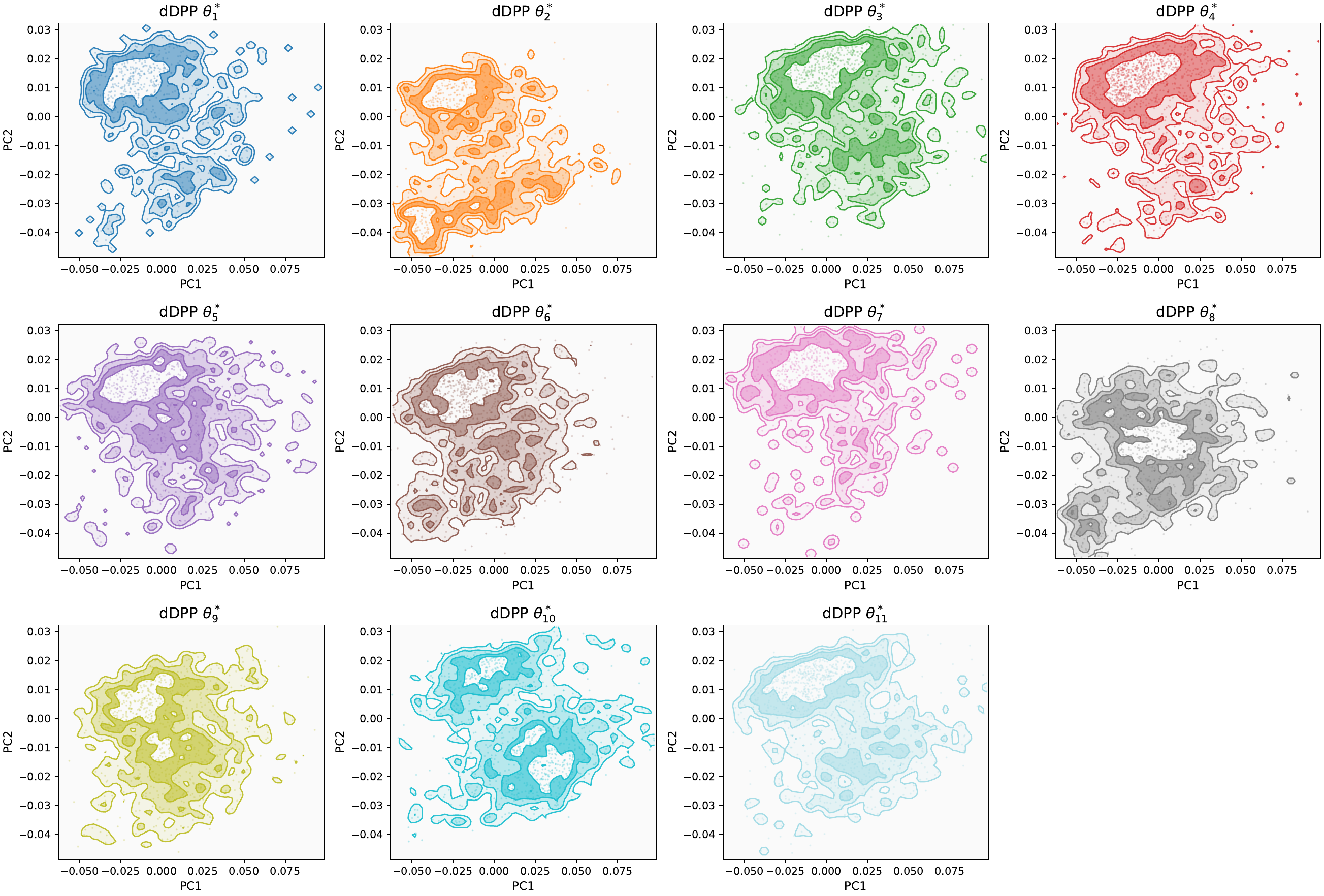}
    \caption{Singe cell data -
      PCA visualization of the atoms of the mixing measure.
      The figure shows kernel density estimates of $\hat{G}$ under
      \eqref{Ghat}
      ($w=5\times10^5,\gamma=10^3$).} 
    \label{fig:mixing_measure_summary}
\end{figure}

The results in Table~\ref{tab:combined_all} report
more parsimonious and better-separated partitions for inference under
the dDPP than under the DP baseline.
Across a range of values for the  likelihood hyperparameter
$w$ dDPP reports fewer clusters, with
correspondingly larger average cluster sizes.
Repulsion is {\em a priori} determined by the repulsion
hyperparameter $\gamma$ of the SW kernel.  Smaller 
values of $\gamma$ induce stronger prior repulsion between
atoms, which typically yields fewer, more separated clusters.
However, practically, inference on the number of clusters remains
robust with respect to choice of $\gamma$ in a wide range, confirming
appropriate dominance of the likelihood.
Similar behavior appears in the repulsion column in Table
\ref{tab:combined_all}.
Inference under the dDPP attains 
substantially higher (less negative) log unnormalized density at every
weight, confirming that its atoms occupy more distinct regions of the
embedding. Importantly, this improved separation  of clusters 
costs almost nothing in log generalized likelihood. Inference under
the dDPP trades a negligible amount of fit 
for a markedly more separated and interpretable clustering. It also
produces more concentrated posteriors over partition complexity, with
$95\%$ highest-credible intervals (HCIs) as tight as or tighter than
the DP's. 

Figures~\ref{fig:mixing_500000} and~\ref{fig:mixing_measure_summary}
show the point estimate \eqref{Ghat}
under $w=5\times 10^5$ and $\gamma=10^3$. In the UMAP comparison of
Figure~\ref{fig:mixing_500000}, the dDPP solution (right) assigns
donors to fewer, spatially more coherent groups, with atoms
better spread out across diverse regions of the embedding.
In contrast, the estimate under the DP model
(left) clusters the atoms more tightly and yields a more fragmented
partition. Analogous figures for $w=7\times 10^5$ and $w=10^6$ appear
in Figure~6 of Section B.1 of the Supplementary Materials, and PCA
visualizations of the 
donor clusters for $w=5\times 10^5$ are shown in Figures~7--10. In
each case, the donors within a cluster closely resemble the typical
donors $\thstar{k}$ shown in
Figure~\ref{fig:mixing_measure_summary}.
Figure~\ref{fig:mixing_measure_summary}
overlays the dDPP atoms over a kernel density estimate of the
mixture model corresponding to $\hat{G}$ in \eqref{Ghat}.
The figure illustrates how the recovered atoms align with the dominant
modes of the observed data.
The corresponding figure for the DP atoms is given in
Figure~11 of Section B.1 of the Supplementary Materials.

\subsection{Human Epilepsy Project Data}
\label{subsec:hep}

The Human Epilepsy Project~\citep{French_HumanEpilepsyProject} (HEP) enrolled 448 individuals newly diagnosed  with focal epilepsy between 2012 and 2017 across 34 clinical centers  worldwide. Eligible participants were between ages 12 and 60 at diagnosis  and enrolled within four months of initiating medical treatment. Participants recorded daily seizure counts via an electronic seizure diary  over a three-year follow-up period. After excluding individuals with no  diary entries, no properly tracked days, or no medication data, the final 
analysis cohort consisted of 407 individuals. For the illustrative purpose of this example, we only cluster the seizure data (not using treatment and baseline covariates). For participant $i$, the daily record reports $x_{i,t} \in \{0, 0.5, 1\}$ for $t = 1, \ldots, T_i$,  where 0 = no seizure, 1 = seizure, and 0.5 = missing (imputed). Since $T_i$ varies 
across participants and patients are enrolled at varying times, the data are unaligned. For meaningful clustering, we map the data for each patient to a distribution of “reads”, where reads are defined as length $W$ subsequences as follows. We extract $T_i - W + 1$ consecutive windows of width $W>1$:
$
\mathbf{x}_{i,k} = (x_{i,k}, x_{i,k+1}, \ldots, x_{i,k+W-1}) \in \{0, 0.5, 1\}^W, 
\quad k = 1, \ldots, T_i - W + 1.
$
Each participant is then represented as an empirical distribution over these reads:
$
\hat{F}_i = \frac{1}{T_i - W + 1} \sum_{k=1}^{T_i - W + 1} \delta_{\mathbf{x}_{i,k}} \in \mathcal{P}(\{0, 0.5, 1\}^W),
$
where $\delta_{\mathbf{x}_{i,k}}$ is a Dirac mass at
$\mathbf{x}_{i,k}$ and  $\mathcal{P}(\{0, 0.5, 1\}^W)$ denotes the
space of probability measures on $\{0, 0.5, 1\}^W$.  This places all
participants in a common space regardless of follow-up length. 
We cluster $F_i$, $i=1,\ldots,N$, using the proposed dDPP model for
random partitions with distribution-valued data.

\begin{table}[!t]
\centering
\caption{HEP data: Same as Table~\ref{tab:combined_all} for HEP data.}
\label{tab:combined_hep}
\resizebox{\textwidth}{!}{%
\begin{tabular}{ll cc cc cc cc}
\toprule
 & & \multicolumn{2}{c}{\textbf{Clusters }$k$} & \multicolumn{2}{c}{\textbf{Average  Cluster Size}} & \multicolumn{2}{c}{\textbf{Log-generalized likelihood}} & \multicolumn{2}{c}{\textbf{Repulsion}} \\
\cmidrule(lr){3-4}\cmidrule(lr){5-6}\cmidrule(lr){7-8}\cmidrule(lr){9-10}
\textbf{Model} & $w$ & \textbf{Point} & $\mathbb{E}$ and [HCI]& \textbf{Point} & $\mathbb{E}$ and [HCI]& \textbf{Point} & $\mathbb{E}$ and [HCI]& \textbf{Point} & $\mathbb{E}$ and [HCI]\\
\midrule
\multirow{3}{*}{DP}
  & $100$ & 9  & 8.00 [5.00, 11.00]   & 45.22 & 53.19 [37.00, 81.40] & $-2.12$ & $-2.11$ [$-2.13, -2.09$] & $-52.79$  & $-45.66$  [$-82.02, -11.16$] \\
  & $200$ & 14 & 14.06 [11.00, 17.00] & 29.07 & 29.40 [23.94, 37.00] & $-2.85$ & $-2.87$ [$-2.89, -2.85$] & $-115.78$ & $-114.48$ [$-161.27, -67.52$] \\
  & $500$ & 27 & 26.72 [22.00, 31.00] & 15.07 & 15.35 [12.72, 17.70] & $-4.13$ & $-4.14$ [$-4.18, -4.10$] & $-286.84$ & $-276.72$ [$-352.94, -216.60$] \\
\midrule
\multirow{3}{*}{dDPP ($\gamma=10$)}
  & $100$ & 6  & 6.22 [6.00, 7.00]    & 67.83 & 65.86 [58.14, 67.83] & $-2.10$ & $-2.10$ [$-2.12, -2.09$] & $-13.73$  & $-18.57$  [$-41.43, -13.38$] \\
  & $200$ & 12 & 11.87 [11.00, 13.00] & 33.92 & 34.43 [31.31, 37.00] & $-2.86$ & $-2.87$ [$-2.88, -2.85$] & $-84.43$  & $-81.27$  [$-106.24, -57.26$] \\
  & $500$ & 23 & 23.25 [21.00, 26.00] & 17.70 & 17.56 [15.65, 19.38] & $-4.12$ & $-4.13$ [$-4.17, -4.10$] & $-224.18$ & $-227.01$ [$-259.23, -189.43$] \\
\midrule
\multirow{3}{*}{dDPP ($\gamma=100$)}
  & $100$ & 6  & 6.28 [5.00, 8.00]    & 67.83 & 65.45 [50.88, 81.40] & $-2.10$ & $-2.11$ [$-2.13, -2.10$] & $-13.47$  & $-17.39$  [$-43.43, -11.61$] \\
  & $200$ & 13 & 12.23 [11.00, 14.00] & 31.31 & 33.48 [29.07, 37.00] & $-2.84$ & $-2.86$ [$-2.87, -2.84$] & $-109.33$ & $-91.45$  [$-114.51, -57.46$] \\
  & $500$ & 23 & 24.91 [22.00, 27.00] & 17.70 & 16.39 [15.07, 18.50] & $-4.16$ & $-4.14$ [$-4.19, -4.11$] & $-223.72$ & $-243.78$ [$-292.54, -216.96$] \\
\midrule
\multirow{3}{*}{dDPP ($\gamma=1000$)}
  & $100$ & 7  & 7.89 [6.00, 10.00]   & 58.14 & 53.18 [40.70, 67.83] & $-2.09$ & $-2.10$ [$-2.13, -2.09$] & $-20.24$  & $-41.56$  [$-77.90, -13.67$] \\
  & $200$ & 13 & 12.97 [11.00, 15.00] & 31.31 & 31.65 [27.13, 37.00] & $-2.84$ & $-2.86$ [$-2.88, -2.84$] & $-88.59$  & $-100.51$ [$-143.17, -62.84$] \\
  & $500$ & 26 & 24.49 [21.00, 28.00] & 15.65 & 16.74 [14.54, 19.38] & $-4.12$ & $-4.16$ [$-4.21, -4.12$] & $-258.44$ & $-246.23$ [$-304.46, -186.18$] \\
\bottomrule
\end{tabular}%
}
\end{table}

\begin{figure}[!t]
    \centering
    \begin{tabular}{cc}
        \includegraphics[width=0.43\linewidth]{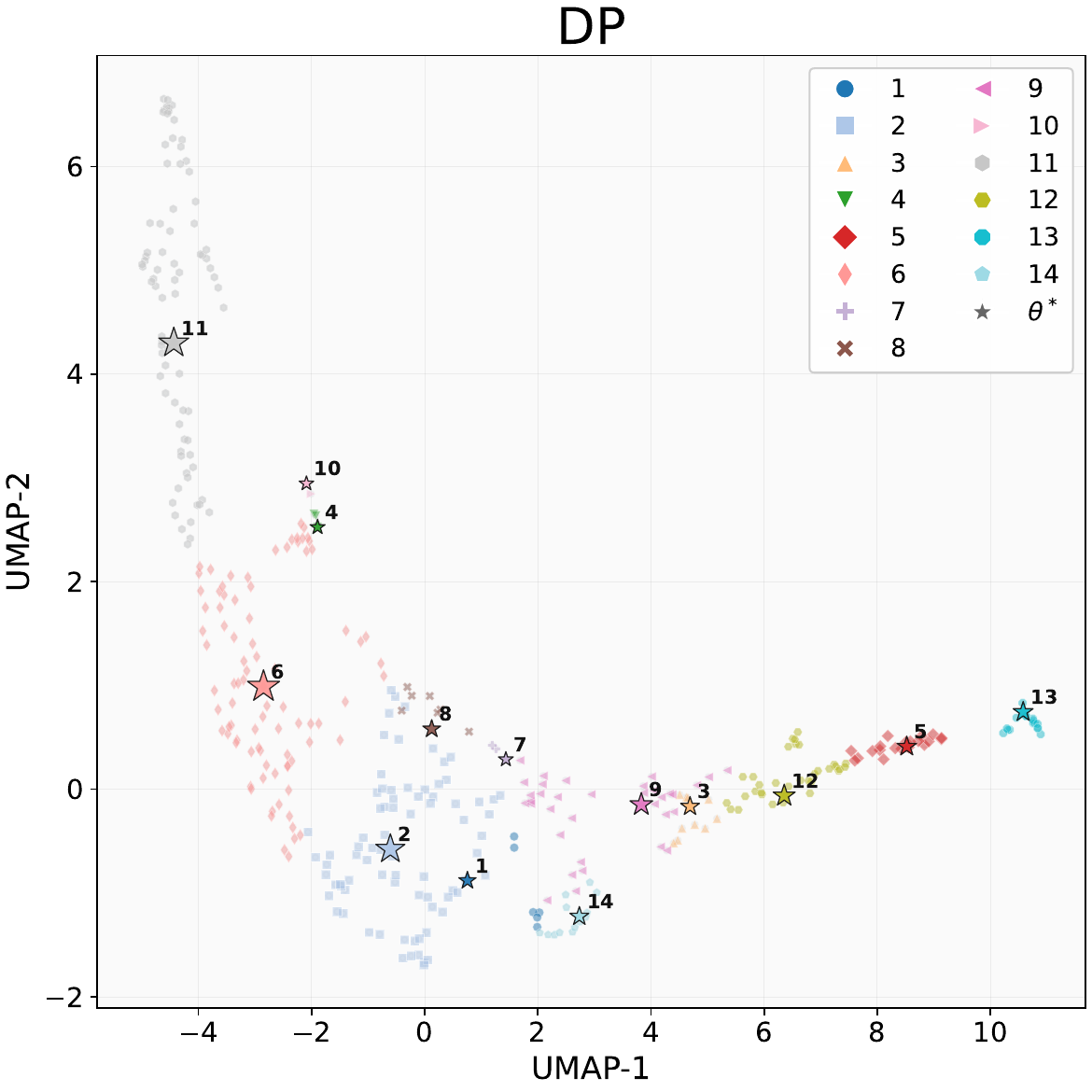} &  \includegraphics[width=0.43\linewidth]{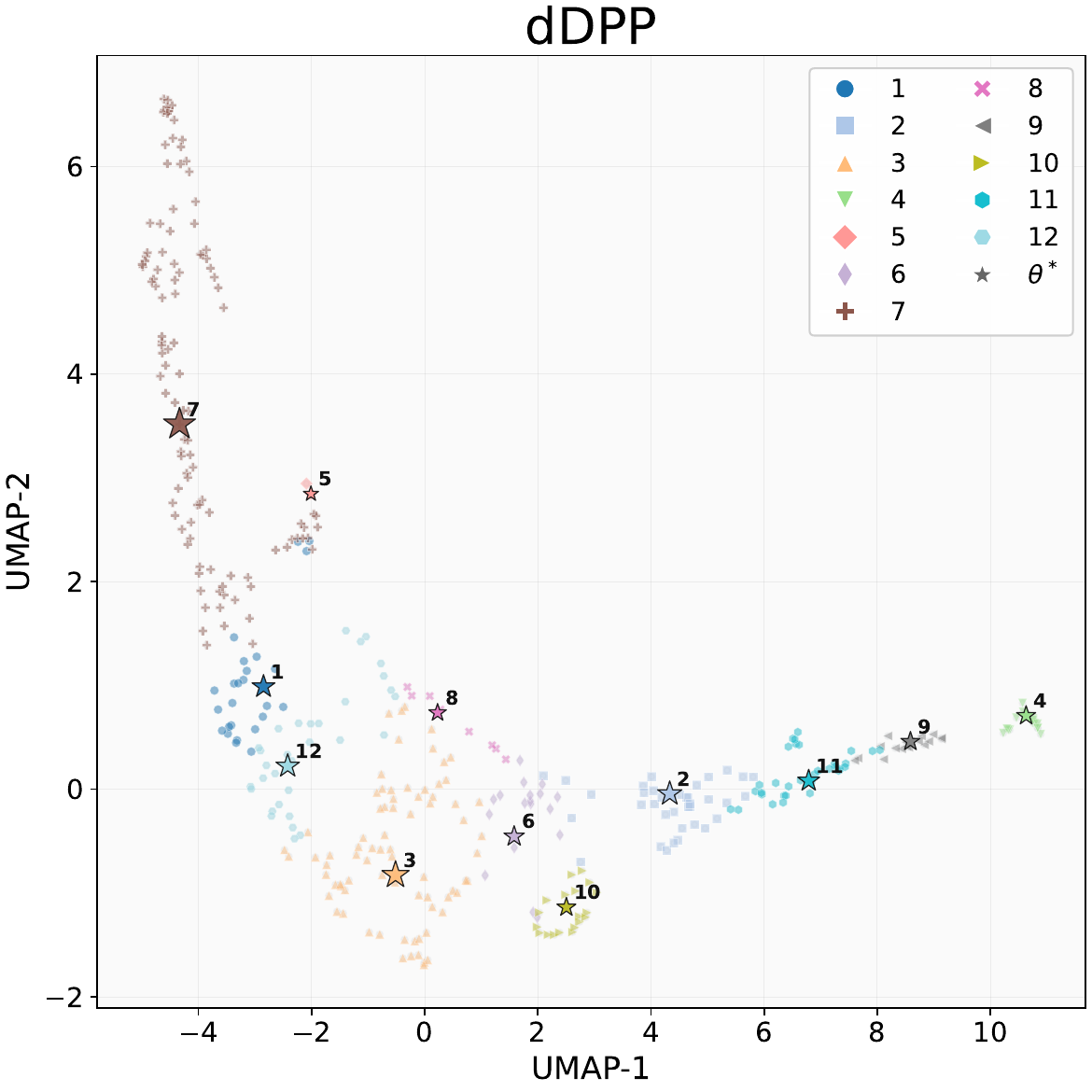}
    \end{tabular}
    \caption{HEP data.
      Same as Figure~\ref{fig:mixing_500000}  with $w=200,\gamma=10$, for the HEP data.}
    \label{fig:mixing_200_HEP}
\end{figure}

\begin{figure}[!t]
    \centering
    \includegraphics[width=1\linewidth]{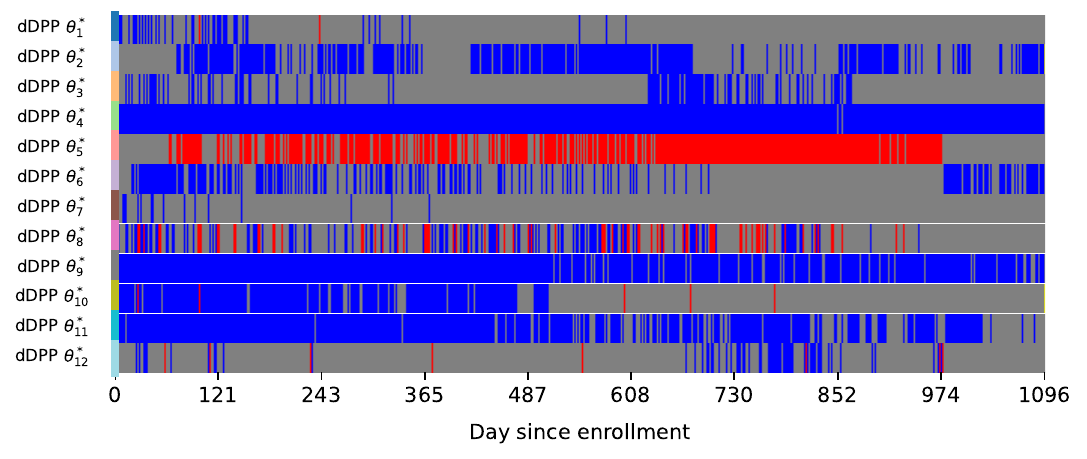} 
    \caption{HEP data.
      The atoms of the point estimate \eqref{Ghat} of the mixing
      measure under the dDPP model ($w=200,\gamma=10$).
      The vertical axis reports the $K=12$ estimated clusters.
      The horizontal axis are days of followup.
      The plot shows $\thstar{k}$.
      Blue represents
      0, red represents 1, and grey represents 0.5; yellow indicates
      padding where applicable (patients with fewer check-ups).} 
    \label{fig:mixing_measure_summary_hep}
\end{figure}

\begin{figure}[!t]
    \centering
    \includegraphics[width=1\linewidth]{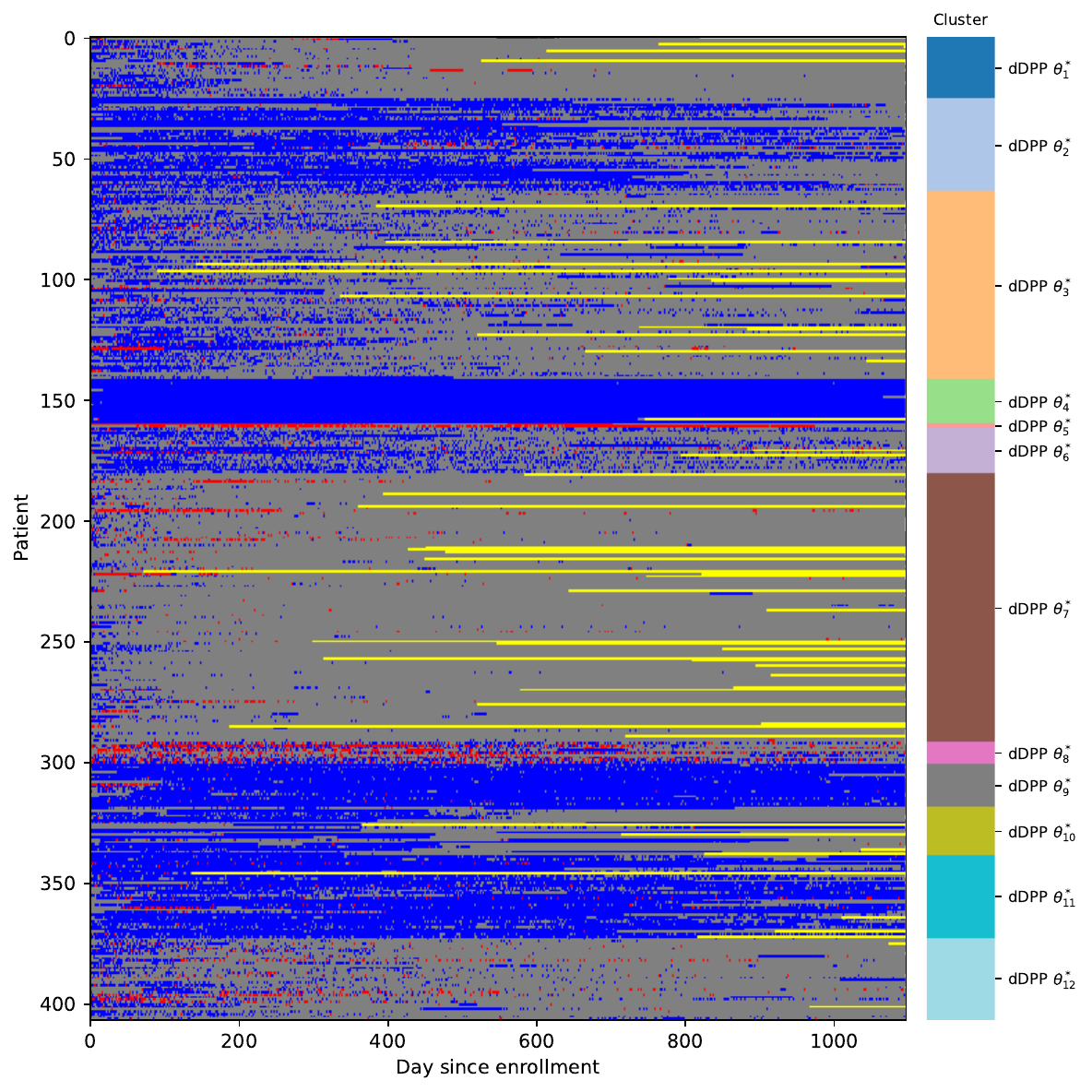}
    \caption{HEP data.
      Visualization of all patients with corresponding clusters label.
      Similar to Figure \ref{fig:mixing_measure_summary_hep}, with
      observed patients $i=1,\ldots,N$ on the vertical axis, and
      adding color coding 
      for cluster membership (on the right side) and yellow indicates
      padding for patients with $T_i < T_{\max}$.}
    \label{fig:HEP_clusters}
\end{figure}

The results on the HEP cohort in Table~\ref{tab:combined_hep} (with $W=7$) reproduce the pattern observed on the single-cell data. Across all choices of hyperparameter $w$, dDPP recovers fewer clusters than the DP baseline, with correspondingly larger mean cluster sizes, and the recovered atoms are markedly better separated, as reflected by the substantially higher repulsion score at every weight. The repulsion hyperparameter $\gamma$ governs the strength of this effect. This separation again comes at essentially no cost to fit. In particular, the log-likelihood is statistically indistinguishable between the two models throughout. Inference under the dDPP model yields more concentrated posteriors over partition complexity, with $95\%$ HCIs for the number of clusters that are consistently tighter than the DP's. In other words, inference under the dDPP model trades a negligible amount of likelihood for a more separated and more interpretable clustering of the seizure-diary distributions.

We again compare the point estimates \eqref{Ghat} under the DP and dDPP models. In the UMAP comparison of Figure~\ref{fig:mixing_200_HEP} ($w=200,\gamma=10$), the dDPP solution (right) assigns participants to fewer, more spatially coherent groups, with atoms distributed across distinct regions of the embedding, whereas the DP solution (left) clusters its atoms more tightly and yields a more fragmented partition. The same phenomenon appears for $w=100$ and $w=500$ in Figure~12 of Section B.2 of the Supplementary Materials. Figure~\ref{fig:mixing_measure_summary_hep} visualizes the recovered dDPP atoms as representative seizure-window patterns, highlighting the distinct temporal profiles each cluster captures; analogous figures for both the DP and dDPP models with $w=100$, $w=200,\gamma=10$, and $w=500$ are given in Figures~13--15 of Section B.2 of the Supplementary Materials. Finally, Figure~\ref{fig:HEP_clusters} displays the full cohort under the induced cluster labels ($w=200,\gamma=10$), confirming that the partition organizes participants into coherent, well-separated groups.

\section{Conclusion}
\label{sec:conclusion}

We introduced the distributional determinantal point process (dDPP)
as a repulsive point process with distribution-valued atoms with
sliced Wasserstein kernels to define the L-ensemble kernel.
While the use of the generalized likelihood in the hierarchical mixture
model construction keeps inference computationally tractable, the same
could be seen as compromising the nature of the approach as principled
model-based inference.
See, for example, \citet{mcalinn2026genBayes} for a recent discussion.
Another limitation is the use of a finite discrete base measure $P_0$.

Useful generalizations could extend the model to multiple dependent
random partitions, as it arises in many applications.
The use of common atoms (in this case, distribution-valued atoms) is
natural in model \eqref{eq:likelihood} and \eqref{eq:mixing_measure},
as it already separates the prior on atoms $\thtilde{h}$ and
corresponding marks $s_h$. The latter could vary across dependent
random partitions and the earlier could be shared. Moreover, additional theoretical studies of the proposed model can also be investigated such as clustering consistency.

\section{Data Availability Statement}\label{data-availability-statement}

Single-cell data is available as described
in~\citet{yazar2022single}. Human Epilepsy Project data is available
by application (see \citet{French_HumanEpilepsyProject}).


\phantomsection\label{supplementary-material}
\bigskip






\clearpage
  \bibliography{bibliography}
\clearpage

\appendix

\begin{center}
    \huge{Supplementary for  ``Distributional Determinantal Point Process  for Repulsive Clustering of  Distributions"}
\end{center}

\section{Proofs}

\subsection{Proof of Theorem 1}

 We define
$T_L: \mathbb{L}_2(\Theta,P_0)\to \mathbb{L}_2(\Theta,P_0)$ be the  integral operator of $L$:
\begin{align}\label{TL}
  (T_L f)(\nu) = \int_{\Theta} L(\nu,\nu')\,f(\nu')\mathrm{d}P_0(\nu'),
\end{align}
where $\mathbb{L}_2(\Theta, P_0)
= \left\{ f : \Theta \to \Re \mid \int_{\Theta} |f(\nu)|^2 \, \mathrm{d} P_0(\nu) < \infty
\right\}.$
We need to show two properties to prove that dDPP is a well-defined point process (i) $L(\nu,\nu')$ is positive definite, (ii) 
 $T_L$ to be trace-class on $\mathbb{L}_2(\Theta,P_0)$ 
which leads to  $\sum_{k=0}^{\infty} \frac{1}{k!}
\int_{\Theta^k}
\det\big( L(\nu_i, \nu_j) \big)_{i,j=1}^k
\, \mathrm{d} P_0(\nu_1)\ldots \mathrm{d} P_0(\nu_k)=\det(I+T_L)$ to be finite.

\textbf{Positive definiteness.} The kernel is symmetric due to the symmetry of SW distance. We only need to prove that it is positive definite. Let $\Theta \subset \mathcal{P}_2(\Re^d)$ and define
$
\Psi[\nu](v,t)
=
F^{-1}_{P_v \sharp \nu}(t),
$
for $(v,t) \in \Sbb^{d-1} \times [0,1]$.
From the quantile representation of SW distance, we have:
\begin{align}
\SW_2^2(\nu,\nu')
=
\|\Psi(\nu) - \Psi(\nu')\|_{\mathbb{L}_2(\Sbb^{d-1} \times [0,1])}^2.
\end{align}
For any $\nu_1,\dots,\nu_k \in \Theta$ and $c_1,\dots,c_k \in \Re$, we need to show
\begin{align}
\sum_{i=1}^k \sum_{j=1}^k c_i c_j L(\nu_i,\nu_j) > 0.
\end{align}
Define $f_i = \Psi(\nu_i) \in \mathbb{L}_2(\Sbb^{d-1} \times [0,1])$, we then have 
$
L(\nu_i,\nu_j)
=
\exp\big(-\gamma \|f_i - f_j\|^2\big).
$
Using the identity
\begin{align}
\|f_i - f_j\|^2
=
\|f_i\|^2 + \|f_j\|^2 - 2\langle f_i, f_j \rangle,
\end{align}
we obtain
\begin{align}
L(\nu_i,\nu_j)
=
\exp(-\gamma \|f_i\|^2)
\exp(-\gamma \|f_j\|^2)
\exp\big(2\gamma \langle f_i, f_j \rangle\big).
\end{align}
Thus,
\begin{align}
\sum_{i,j=1}^k c_i c_j L(\nu_i,\nu_j)
=
\sum_{i,j=1}^k
\big(c_i e^{-\gamma \|f_i\|^2}\big)
\big(c_j e^{-\gamma \|f_j\|^2}\big)
\exp\big(2\gamma \langle f_i, f_j \rangle\big).
\end{align}
Let $a_i = c_i e^{-\gamma \|f_i\|^2}$,  it suffices to show
\begin{align}
\sum_{i=1}^k \sum_{j=1}^k a_i a_j
\exp\big(2\gamma \langle f_i, f_j \rangle\big)
> 0.
\end{align}
We observe that the kernel
\begin{align}
\exp\big(2\gamma \langle f_i, f_j \rangle\big)
\end{align}
is positive definite on a Hilbert space. Using the power series expansion, we have
$
\exp\big(2\gamma \langle f_i, f_j \rangle\big)
=
\sum_{m=0}^\infty \frac{(2\gamma)^m}{m!}
\langle f_i, f_j \rangle^m,
$
and each kernel $\langle f_i, f_j \rangle^m$ is positive definite. Since nonnegative linear combinations of  positive definite kernels are positive definite, we have
\begin{align}
\sum_{i,j=1}^k a_i a_j \exp\big(2\gamma \langle f_i, f_j \rangle\big) > 0,
\end{align}
which implies
\begin{align}
\sum_{i,j=1}^k c_i c_j L(\nu_i,\nu_j) > 0.
\end{align}
Hence $L$ is a positive definite kernel.

\textbf{(ii) Trace-class.} We prove that $T_L$ is trace-class via Mercer's theorem. To apply Mercer's theorem~\citep{steinwart2012mercer}, we need $\Theta$ to be compact (given by Assumption 1) and $L$ to be continuous on $\Theta \times \Theta$. We establish continuity as follows. The map $(\nu,\nu') \to \SW_2^2(\nu,\nu')$ is continuous on $\Theta \times \Theta$, since for any sequences $\nu_n \to \nu$ and $\nu_n' \to \nu'$, we have
\begin{align}
|\SW_2^2(\nu_n,\nu_n') - \SW_2^2(\nu,\nu')|
&\leq
|\SW_2(\nu_n,\nu_n') - \SW_2(\nu,\nu')|
\big(\SW_2(\nu_n,\nu_n') + \SW_2(\nu,\nu')\big) \\
&\leq
C\big(\SW_2(\nu_n,\nu) + \SW_2(\nu_n',\nu')\big)
\to 0,
\end{align}
where the second inequality uses the triangle inequality for $\SW_2$, the convergence to zero follows from~\citep{nadjahi2020statistical}, and $C < \infty$ is the finite diameter of $\Theta$ under $\SW_2$, which follows from compactness. Since $x \mapsto \exp(-\gamma x)$ is continuous, the kernel
\begin{align}
L(\nu,\nu') = \exp(-\gamma\,\SW_2^2(\nu,\nu'))
\end{align}
is continuous on $\Theta \times \Theta$.

Therefore, since $P_0$ is a Borel probability measure on $\Theta$, Mercer's theorem applies and $T_L$ admits an orthonormal eigenexpansion with non-negative eigenvalues $\{\lambda_i\}_{i=1}^\infty$, and its trace satisfies
\begin{align}
\mathrm{Tr}(T_L) = \int_\Theta L(\nu,\nu)\, \mathrm{d} P_0(\nu).
\end{align}
Since $\SW_2(\nu,\nu)=0$ for all $\nu \in \Theta$, we have $L(\nu,\nu)=1$, and hence
\begin{align}
\mathrm{Tr}(T_L) = \int_\Theta 1\, \mathrm{d}P_0(\nu) = 1.
\end{align}
Since $L$ is positive definite (shown in part (i)), all eigenvalues satisfy $\lambda_i \geq 0$, and therefore
\begin{align}
\sum_{i=1}^\infty \lambda_i = \mathrm{Tr}(T_L) = 1 < \infty,
\end{align}
which by definition establishes that $T_L$ is trace-class. Consequently, the Fredholm determinant admits the spectral representation
\begin{align}
\det(I + T_L) = \prod_{i=1}^\infty (1 + \lambda_i).
\end{align}
Using the inequality $\log(1+\lambda_i) \le \lambda_i$ for $\lambda_i \ge 0$, we obtain
\begin{align}
\log\det(I+T_L)
= \sum_{i=1}^\infty \log(1 + \lambda_i)
\le \sum_{i=1}^\infty \lambda_i
= \mathrm{Tr}(T_L)
= 1 < \infty,
\end{align}
and hence $\det(I + T_L) \leq e < \infty$, which completes the proof.

\subsection{Proof of Theorem 2}

Since the map $x \to e^{-\gamma x}$ is differentiable with derivative $-\gamma e^{-\gamma x}$, the mean value
theorem yields
\begin{align}
    |\hat{L}_{ij} - L_{ij}|
    = \gamma\, e^{-\gamma\xi_{ij}}
    \left|\SW_2^2\left(\hat{\nu}^{(m)}_i,\hat{\nu}^{(m)}_j\right)
    - \SW_2^2(\thbase{i},\thbase{j})\right|
\end{align}
for some $\xi_{ij}$ lying between
$\SW_2^2(\hat{\nu}^{(m)}_i,\hat{\nu}^{(m)}_j)$ and $\SW_2^2(\thbase{i},\thbase{j})$.
Since $e^{-\gamma\xi_{ij}} \leq 1$, we have
\begin{align}
\label{eq:mvt}
    |\hat{L}_{ij} - L_{ij}|
    \leq \gamma
    \left|\SW_2^2\left(\hat{\nu}^{(m)}_i,\hat{\nu}^{(m)}_j\right)
    - \SW_2^2(\thbase{i},\thbase{j})\right|.
\end{align}
It therefore suffices to bound $|\SW_2^2(\hat{\nu}^{(m)}_i,\hat{\nu}^{(m)}_j) - \SW_2^2(\thbase{i},\thbase{j})|$
directly.
Using the closed-form expression of $\SW_2^2$ and Jensen's inequality (twice):
\begin{align}
    &\left|\SW_2^2\left(\hat{\nu}^{(m)}_i,\hat{\nu}^{(m)}_j\right)
    - \SW_2^2\left(\thbase{i},\thbase{j}\right)\right| \nonumber \\
    &\quad= \left|\mathbb{E}_{v \sim \mathcal{U}(\Sbb^{d-1})}
    \left[\int_0^1\left(
    \left(F_{P_v\sharp\hat{\nu}^{(m)}_i}^{-1}(t)
    - F_{P_v\sharp\hat{\nu}^{(m)}_j}^{-1}(t)\right)^2
    -
    \left(F_{P_v\sharp \thbase{i}}^{-1}(t)
    - F_{P_v\sharp \thbase{j}}^{-1}(t)\right)^2
    \right)\mathrm{d}t\right]\right|
    \nonumber \\
    &\quad\leq \mathbb{E}_{v \sim \mathcal{U}(\Sbb^{d-1})}
    \left[\int_0^1
    \left|
    \left(F_{P_v\sharp\hat{\nu}^{(m)}_i}^{-1}(t)
    - F_{P_v\sharp\hat{\nu}^{(m)}_j}^{-1}(t)\right)^2
    -
    \left(F_{P_v\sharp \thbase{i}}^{-1}(t)
    - F_{P_v\sharp \thbase{j}}^{-1}(t)\right)^2
    \right|\mathrm{d}t\right].
\label{eq:sw2_triangle}
\end{align}
For each fixed $v \in \Sbb^{d-1}$ and $t \in [0,1]$, we define
\begin{align}
    p_v(t) := F_{P_v\sharp\hat{\nu}^{(m)}_i}^{-1}(t)
    - F_{P_v\sharp\hat{\nu}^{(m)}_j}^{-1}(t),
    \qquad
    q_v(t) := F_{P_v\sharp \thbase{i}}^{-1}(t)
    - F_{P_v\sharp \thbase{j}}^{-1}(t).
\end{align}
Both quantities are bounded in $[-2R, 2R]$ since $\mathrm{supp}(\thbase{i}) \subset B(0,R)$ implies $P_v\sharp \thbase{i}$ is supported on $[-R,R]$ for all $v$. Applying $|p_v(t)^2 - q_v(t)^2| = |p_v(t) - q_v(t)||p_v(t) +
q_v(t)| \leq 4R|p_v(t) - q_v(t)|$ and the triangle inequality
$|p_v(t) - q_v(t)| \leq |F_{P_v\sharp\hat{\nu}^{(m)}_i}^{-1}(t) -
F_{P_v\sharp \thbase{i}}^{-1}(t)| + |F_{P_v\sharp\hat{\nu}^{(m)}_j}^{-1}(t)
- F_{P_v\sharp \thbase{j}}^{-1}(t)|$, we have:
\begin{align}
    &\int_0^1 |p_v(t)^2 - q_v(t)^2|\,\mathrm{d}t \nonumber \\
    &\quad\leq 4R\left(
    \int_0^1\left|F_{P_v\sharp\hat{\nu}^{(m)}_i}^{-1}(t)
    - F_{P_v\sharp \thbase{i}}^{-1}(t)\right|\mathrm{d}t
    + \int_0^1\left|F_{P_v\sharp\hat{\nu}^{(m)}_j}^{-1}(t)
    - F_{P_v\sharp \thbase{j}}^{-1}(t)\right|\mathrm{d}t
    \right) \nonumber \\
    &\quad= 4R\left(
    W_1\left(P_v\sharp\hat{\nu}^{(m)}_i, P_v\sharp \thbase{i}\right)
    + W_1\left(P_v\sharp\hat{\nu}^{(m)}_j, P_v\sharp \thbase{j}\right)
    \right),
\label{eq:pq_bound}
\end{align}
where we used the quantile representation $W_1(\mu,\nu) = \int_0^1
|F_\mu^{-1}(t) - F_\nu^{-1}(t)|\,\mathrm{d}t$. For measures on $[-R,R]$,
$W_1(\mu,\nu) = \int_\Re|F_\mu(x) - F_\nu(x)|\,dx \leq
2R\|F_\mu - F_\nu\|_\infty$. Therefore~\eqref{eq:pq_bound} gives:
\begin{align}
   \int_0^1 |p_v(t)^2 - q_v(t)^2|\,\mathrm{d}t
    \leq 8R^2\left(
    \left\|F_{P_v\sharp\hat{\nu}^{(m)}_i} -
    F_{P_v\sharp \thbase{i}}\right\|_\infty
    + \left\|F_{P_v\sharp\hat{\nu}^{(m)}_j} -
    F_{P_v\sharp \thbase{j}}\right\|_\infty
    \right).
\label{eq:w2sq_cdf_linear}
\end{align}
Substituting~\eqref{eq:w2sq_cdf_linear} into~\eqref{eq:sw2_triangle} and
defining, for each atom $i \in [Q]$, the projection-averaged error is
\begin{align}
\label{eq:Di_def}
    D_i := \mathbb{E}_{v \sim \mathcal{U}(\Sbb^{d-1})}
    \left[\left\|F_{P_v\sharp\hat{\nu}^{(m)}_i} - F_{P_v\sharp \thbase{i}}\right\|_\infty\right].
\end{align}
we the have
\begin{align}
\label{eq:sw2sq_cdf_final}
    \left|\SW_2^2\left(\hat{\nu}^{(m)}_i,\hat{\nu}^{(m)}_j\right)
    - \SW_2^2\left(\thbase{i},\thbase{j}\right)\right|
    \leq 8R^2\left(D_i + D_j\right).
\end{align}
 Next, we bound
each $D_i$ in high probability through a Dvoretzky--Kiefer--Wolfowitz (DKW) expectation bound applied pointwise in $v$, followed by a
bounded-differences concentration inequality. For each fixed $v \in \Sbb^{d-1}$, the projected samples
$\{v^\top x_t^{(i)}\}_{t=1}^m$ are i.i.d.\ from $P_v\sharp \thbase{i}$, and
$F_{P_v\sharp\hat{\nu}^{(m)}_i}$ is their empirical CDF. The DKW inequality
$\mathbb{P}(\|F_{P_v\sharp\hat{\nu}^{(m)}_i} - F_{P_v\sharp \thbase{i}}\|_\infty > \epsilon)
\leq 2e^{-2m\epsilon^2}$ gives
\begin{align}
    \mathbb{E}\left[\left\|F_{P_v\sharp\hat{\nu}^{(m)}_i} - F_{P_v\sharp \thbase{i}}\right\|_\infty\right]
    = \int_0^\infty \mathbb{P}\left(\|\cdot\|_\infty > \epsilon\right)d\epsilon
    \leq \int_0^\infty 2e^{-2m\epsilon^2}\,d\epsilon
    = \sqrt{\frac{\pi}{2m}},
\end{align}
uniformly in $v$. By Fubini's theorem, we have
\begin{align}
\label{eq:dkw_expectation}
    \mathbb{E}[D_i]
    = \mathbb{E}_v\,\mathbb{E}\left[\left\|F_{P_v\sharp\hat{\nu}^{(m)}_i} - F_{P_v\sharp \thbase{i}}\right\|_\infty\right]
    \leq \sqrt{\frac{\pi}{2m}}.
\end{align}

Viewing $D_i$ as a function of the $m$ i.i.d.\ samples $\{x_t^{(i)}\}_{t=1}^m$,
replacing a single sample changes each empirical CDF
$F_{P_v\sharp\hat{\nu}^{(m)}_i}$ by at most $1/m$ in sup-norm, hence changes
$\|F_{P_v\sharp\hat{\nu}^{(m)}_i} - F_{P_v\sharp \thbase{i}}\|_\infty$ and its
$\mathbb{E}_v$-average $D_i$ by at most $1/m$. McDiarmid's inequality therefore
gives $\mathbb{P}(D_i - \mathbb{E}[D_i] \geq s) \leq \exp(-2ms^2)$, so with
probability at least $1 - \delta_0$,
\begin{align}
\label{eq:dkw_bound}
    D_i
    \leq \mathbb{E}[D_i] + \sqrt{\frac{\log(1/\delta_0)}{2m}}
    \leq \sqrt{\frac{\pi}{2m}} + \sqrt{\frac{\log(1/\delta_0)}{2m}}.
\end{align}
Combining~\eqref{eq:sw2sq_cdf_final},~\eqref{eq:dkw_bound} (for both $i$ and
$j$) and~\eqref{eq:mvt}, with probability at least $1 - 2\delta_0$:
\begin{align}
\label{eq:entry_bound}
    |\hat{L}_{ij} - L_{ij}|
    \leq 16\gamma R^2\left(\sqrt{\frac{\pi}{2m}} + \sqrt{\frac{\log(1/\delta_0)}{2m}}\right).
\end{align}
There are $Q(Q-1) \leq Q^2$ off-diagonal pairs $(i,j)$ (the diagonal entries
satisfy $\hat{L}_{ii} = L_{ii} = 1$). Setting $\delta_0 = \delta/(2Q^2)$ and
applying a union bound over all pairs, with probability at least $1 - \delta$,
we have for all $(i,j) \in [Q]^2$:
\begin{align}
    |\hat{L}_{ij} - L_{ij}|
    \leq 16\gamma R^2\left(\sqrt{\frac{\pi}{2m}} + \sqrt{\frac{\log(2Q^2/\delta)}{2m}}\right).
\end{align}
Taking the maximum and applying $\|\cdot\|_2 \leq \|\cdot\|_F \leq
Q\|\cdot\|_{\max}$, together with $\sqrt{a} + \sqrt{b} \leq \sqrt{2}\sqrt{a+b}$, we have
\begin{align}
\label{eq:L_spectral}
    \|\hat{\mathbf{L}} - \mathbf{L}\|_2
    \leq \frac{16\gamma R^2 Q}{\sqrt{2m}}\left(\sqrt{\pi} + \sqrt{\log(2Q^2/\delta)}\right)
    \leq \frac{16\gamma R^2 Q}{\sqrt{m}}\sqrt{\pi + \log(2Q^2/\delta)},
\end{align}
establishing the bound for $\mathbf{L}$.

\textbf{Propagation to $\hat{\mathbf{K}}$ via Resolvent Identity.}
We propagate through the map $\mathbf{K} = \mathbf{L}(I_Q + \mathbf{L})^{-1}$.
For any two matrices $A, B \succ -I$, the resolvent identity gives
\begin{align}
    A(I+A)^{-1} - B(I+B)^{-1} = (I+A)^{-1}(A - B)(I+B)^{-1}.
\end{align}
Applying this with $A = \hat{\mathbf{L}}$ and $B = \mathbf{L}$, we have
\begin{align}
\label{eq:resolvent}
    \hat{\mathbf{K}} - \mathbf{K}
    = (I + \hat{\mathbf{L}})^{-1}
    (\hat{\mathbf{L}} - \mathbf{L})
    (I + \mathbf{L})^{-1}.
\end{align}
Taking spectral norms and applying submultiplicativity, we have
\begin{align}
    \|\hat{\mathbf{K}} - \mathbf{K}\|_2
    \leq
    \|(I + \hat{\mathbf{L}})^{-1}\|_2
    \|\hat{\mathbf{L}} - \mathbf{L}\|_2
    \|(I + \mathbf{L})^{-1}\|_2.
\end{align}
Since $\mathbf{L} \succ 0$, we have $\|(I + \mathbf{L})^{-1}\|_2 = (1 +
\lambda_{\min}(\mathbf{L}))^{-1} =: \kappa < 1$, and since
$\hat{\mathbf{L}} \succ 0$, we have $\|(I + \hat{\mathbf{L}})^{-1}\|_2 \leq 1$.
Substituting and applying~\eqref{eq:L_spectral}:
\begin{align}
    \|\hat{\mathbf{K}} - \mathbf{K}\|_2
    \leq \frac{16\kappa\gamma R^2 Q}{\sqrt{m}}\sqrt{\pi + \log(2Q^2/\delta)}.
\end{align}
This completes the proof of Theorem~\ref{thm:sample_complexity}.

\subsection{Proof of Corollary 1}

 We denote $\mathbf{A}$ as either $\mathbf{L}_S$ or $\mathbf{K}_S$, and write $\hat{\mathbf{A}} = \mathbf{A} + \mathbf{E}$ where $\mathbf{E}= \hat{\mathbf{A}} - \mathbf{A}$. We denote the rows of $\mathbf{A}$ and
$\mathbf{E}$ by $\mathbf{a}_1,\ldots,\mathbf{a}_k$ and
$\mathbf{e}_1,\ldots,\mathbf{e}_k$ respectively.

Since the determinant is linear in each row separately, applying linearity to each of the $k$ rows of $\mathbf{A} + \mathbf{E}$
produces $2^k$ terms, one for each subset $T \subseteq [k]$ recording which rows are taken from $\mathbf{E}$:
\begin{align}
    \det(\mathbf{A} + \mathbf{E})
    = \sum_{T \subseteq [k]} \det(\mathbf{M}_T),
\end{align}
where $\mathbf{M}_T$ is the $k \times k$ matrix with rows
\begin{align}
    (\mathbf{M}_T)_a =
    \begin{cases}
        \mathbf{e}_a & a \in T, \\
        \mathbf{a}_a & a \notin T.
    \end{cases}
\end{align}
The term $T = \emptyset$ gives $\det(\mathbf{M}_\emptyset) =\det(\mathbf{A})$. Subtracting and grouping by $\ell = |T|$, we have
\begin{align}
    \det(\mathbf{A} + \mathbf{E}) - \det(\mathbf{A})
    = \sum_{\ell=1}^{k}
    \sum_{\substack{T \subseteq [k] \\ |T| = \ell}}
    \det(\mathbf{M}_T).
\end{align}
Each $\mathbf{M}_T$ with $|T| = \ell$ has exactly $\ell$ rows from $\mathbf{E}$ and $k - \ell$ rows from $\mathbf{A}$. We apply
Hadamard's inequality $|\det(\mathbf{M})| \leq \prod_{a=1}^k
\|\mathbf{m}_a\|_2$ to each term:
\begin{align}
    |\det(\mathbf{M}_T)|
    \leq \prod_{a \in T} \|\mathbf{e}_a\|_2
     \prod_{a \notin T} \|\mathbf{a}_a\|_2
    \leq \|\mathbf{E}\|_2^\ell  \|\mathbf{A}\|_2^{k-\ell}.
\end{align}
Since there are $\binom{k}{\ell}$ subsets $T$ of size $\ell$, taking
absolute values and summing over all $\ell$:
\begin{align}
    \left|\det(\hat{\mathbf{A}}) - \det(\mathbf{A})\right|
    \leq \sum_{\ell=1}^{k} \binom{k}{\ell}
    \|\mathbf{E}\|_2^\ell  \|\mathbf{A}\|_2^{k-\ell}
    = \left(\|\mathbf{A}\|_2 + \|\mathbf{E}\|_2\right)^k
    - \|\mathbf{A}\|_2^k.
\label{eq:det_binom}
\end{align}
Factoring out $\|\mathbf{A}\|_2^k$ and applying $(1+x)^k - 1 \leq
kx\,e^{(k-1)x}$ for $x \geq 0$ with $x = \|\mathbf{E}\|_2 /
\|\mathbf{A}\|_2$, we have:
\begin{align}
\label{eq:det_perturb}
    \left|\det(\hat{\mathbf{A}}) - \det(\mathbf{A})\right|
    \leq k\, e^{(k-1)\|\mathbf{E}\|_2/\|\mathbf{A}\|_2}
     \|\mathbf{A}\|_2^{k-1}  \|\mathbf{E}\|_2.
\end{align}

\textbf{Bound for $\det(\hat{\mathbf{L}}_S)$.} Applying the inequality $b^k - a^k \leq k(b-a)b^{k-1}$ for $0 \leq a
\leq b$ to~\eqref{eq:det_binom} with $a = \|\mathbf{L}_S\|_2$ and $b =
\|\mathbf{L}_S\|_2 + \|\mathbf{E}\|_2$, we have:
\begin{align}
\label{eq:det_L_intermediate}
    \left|\det(\hat{\mathbf{L}}_S) - \det(\mathbf{L}_S)\right|
    \leq k  \|\mathbf{E}\|_2 
    \left(\|\mathbf{L}_S\|_2 + \|\mathbf{E}\|_2\right)^{k-1}.
\end{align}
Since all entries of $\mathbf{L}_S$ and $\hat{\mathbf{L}}_S$ satisfy $L_{ij}, \hat{L}_{ij} \in (0,1]$ with unit diagonal, we have
$\|\mathbf{L}_S\|_2 \leq \|\mathbf{L}_S\|_F \leq k$ and$\|\hat{\mathbf{L}}_S\|_2 \leq k$. Consequently, we have
$\|\mathbf{E}\|_2 = \|\hat{\mathbf{L}}_S - \mathbf{L}_S\|_2 \leq\|\hat{\mathbf{L}}_S\|_2 + \|\mathbf{L}_S\|_2 \leq 2k$ and therefore
$\|\mathbf{L}_S\|_2 + \|\mathbf{E}\|_2 \leq 3k$. Together with $\|\mathbf{E}\|_2 = \|\hat{\mathbf{L}}_S - \mathbf{L}_S\|_2 \leq
\|\hat{\mathbf{L}} - \mathbf{L}\|_2$, we have:
\begin{align}
    \left|\det(\hat{\mathbf{L}}_S) - \det(\mathbf{L}_S)\right|
    \leq k(3k)^{k-1}  \|\hat{\mathbf{L}} - \mathbf{L}\|_2.
\end{align}
Substituting the bound on $\|\hat{\mathbf{L}} - \mathbf{L}\|_2$ from
Theorem~\ref{thm:sample_complexity}, with probability at least $1 -
\delta$:
\begin{align}
    \left|\det(\hat{\mathbf{L}}_S) - \det(\mathbf{L}_S)\right|
    \leq
    \frac{16\, k(3k)^{k-1} \gamma R^2 Q}{\sqrt{m}}
    \sqrt{\pi + \log(2Q^2/\delta)},
\end{align}
establishing~\eqref{eq:det_L_bound}.

\textbf{Bound for $\det(\hat{\mathbf{K}}_S)$.}
Since all eigenvalues of $\mathbf{K}$ are of the form $\lambda_\ell/(1+\lambda_\ell) \in (0,1)$, we have
$\|\mathbf{K}_S\|_2 \leq 1$ and similarly $\|\hat{\mathbf{K}}_S\|_2 \leq 1$ (as $\hat{\mathbf{K}} =
\hat{\mathbf{L}}(I+\hat{\mathbf{L}})^{-1}$ has the same eigenvalue
form). Applying $b^k - a^k \leq k(b-a)b^{k-1}$ to~\eqref{eq:det_binom} with $a = \|\mathbf{K}_S\|_2$ and $b = \|\mathbf{K}_S\|_2 +
\|\mathbf{E}\|_2$, and using $\|\mathbf{E}\|_2 = \|\hat{\mathbf{K}}_S -
\mathbf{K}_S\|_2 \leq \|\hat{\mathbf{K}}_S\|_2 + \|\mathbf{K}_S\|_2
\leq 2$ so that $\|\mathbf{K}_S\|_2 + \|\mathbf{E}\|_2 \leq 3$:
\begin{align}
    \left|\det(\hat{\mathbf{K}}_S) - \det(\mathbf{K}_S)\right|
    \leq k  \|\mathbf{E}\|_2 
    \left(\|\mathbf{K}_S\|_2 + \|\mathbf{E}\|_2\right)^{k-1}
    \leq k\,3^{k-1}  \|\hat{\mathbf{K}} - \mathbf{K}\|_2,
\end{align}
where we used $\|\mathbf{E}\|_2 = \|\hat{\mathbf{K}}_S - \mathbf{K}_S\|_2
\leq \|\hat{\mathbf{K}} - \mathbf{K}\|_2$. Substituting the
bound on $\|\hat{\mathbf{K}} - \mathbf{K}\|_2$ from
Theorem~\ref{thm:sample_complexity}, with probability at least $1 -
\delta$:
\begin{align}
    \left|\det(\hat{\mathbf{K}}_S) - \det(\mathbf{K}_S)\right|
    \leq
    \frac{16\, k\,3^{k-1}\, \kappa\gamma R^2 Q}{\sqrt{m}}
    \sqrt{\pi + \log(2Q^2/\delta)},
\end{align}
establishing~\eqref{eq:det_K_bound}.

\subsection{Proof of Theorem 3}

We first define the individual log generalized likelihood as follows:
\begin{align}
    r(F, \pi) := -\log\sum_{q=1}^Q \pi_q \exp\left(-w\, \SW_2^2(F, \thbase{q})\right).
\end{align}
We then define the population risk $R(\pi) = \mathbb{E}_{F\sim \Gtrue}[r(F,\pi)]$ and the empirical
risk
$\hat{R}_N(\pi) = \frac{1}{N}\sum_{i=1}^N r(F_i,\pi).$

\textbf{Bounding the population risk and its continuity.}
We first note that assumption~\eqref{eq:moment_assumption} implies
\begin{align}
\label{eq:second_moment}
    \mathbb{E}_{F\sim \Gtrue}\left[\SW_2^2(F,\thbase{q})\right] < \infty
    \quad\text{for each } q,
\end{align}
since $w\,\SW_2^2(F,\thbase{q}) \leq \exp(w\sum_{q'} \SW_2^2(F,\thbase{q'}))$,
and the right-hand side is integrable by assumption.

Since $\sum_q \pi_q\,\exp(-w\,\SW_2^2(F,\thbase{q})) \leq \sum_q \pi_q = 1$, we have $r(F,\pi) \geq 0$. For the upper bound, since
$\sum_q \pi_q\,\exp(-w\,\SW_2^2(F,\thbase{q}))
\geq \exp(-w\max_q \SW_2^2(F,\thbase{q}))$, we have
\begin{align}
   r(F,\pi)
    \leq w\max_q \SW_2^2(F,\thbase{q})
    \leq w\sum_{q=1}^Q \SW_2^2(F,\thbase{q}).
\end{align}
Taking expectations and using~\eqref{eq:second_moment}, we have
\begin{align}
    R(\pi)
    \leq w\sum_{q=1}^Q
         \mathbb{E}_{F\sim \Gtrue}\left[\SW_2^2(F,\thbase{q})\right]
    < \infty
    \quad\text{for all } \pi\in\Delta^{Q-1}.
\end{align}

The continuity of $R$ on $\Delta^{Q-1}$ follows from continuity of $\pi\mapsto r(F,\pi)$ for each fixed $F$, together with the
dominated convergence theorem using the integrable dominating function $H(F) = w\sum_q \SW_2^2(F,\thbase{q})$ (integrable by~\eqref{eq:second_moment}).

Since $R$ is continuous on the compact set $\Delta^{Q-1}$, the set $\mathcal{M}^* := \argmin_{\pi \in \Delta^{Q-1}} R(\pi)$ is non-empty and compact. For every $\varepsilon > 0$, the set $\{\pi : \tilde{d}(\pi,\mathcal{M}^*) \geq \varepsilon\}$ is compact
and disjoint from $\mathcal{M}^*$, we define
\begin{align}
\label{eq:gap}
    \delta(\varepsilon)
    := \inf_{\pi\,:\tilde d(\pi,\mathcal{M}^*)>\varepsilon}
       \bigl[R(\pi) - R^*\bigr] > 0,
    \qquad R^* := \min_{\pi \in \Delta^{Q-1}} R(\pi).
\end{align}

\textbf{Uniform strong law of large numbers.}
Given a fixed $\pi \in \Delta^{Q-1}$, the random variables $r(F_1,\pi), r(F_2,\pi),\ldots$ are
i.i.d.\ with finite mean $R(\pi)$. By the strong law of large numbers, we have
\begin{align}
\label{eq:pointwise_slln}
    \hat{R}_N(\pi) \overset{a.s.}{\to} R(\pi).
\end{align}
We extend~\eqref{eq:pointwise_slln} to uniform convergence via a
finite-net argument. For any $\pi, \pi' \in \Delta^{Q-1}$ and any $F$,
setting $f_q(F) := \exp(-w\,\SW_2^2(F,\thbase{q})) \in (0,1]$, let
$A := \sum_q \pi_q f_q(F)$ and $B := \sum_q \pi'_q f_q(F)$, both in $(0,1]$.
Using the inequality $|\log(A/B)| \leq |A-B|/\min(A,B)$ (valid for $A,B > 0$):
\begin{align}
    |r(F,\pi) - r(F,\pi')|
    &= \left|\log\frac{B}{A}\right|
    \leq \frac{|A - B|}{\min(A, B)}
    = \frac{\left|\sum_q(\pi_q-\pi'_q)f_q(F)\right|}
              {\min(A,\,B)}.
\end{align}
Since $f_q(F) \leq 1$, the numerator satisfies
$|\sum_q(\pi_q-\pi'_q)f_q(F)| \leq \|\pi-\pi'\|_1$.
For the denominator,
$\min(A,B) \geq \min_q f_q(F) = \exp(-w\max_q \SW_2^2(F,\thbase{q}))$.
Therefore:
\begin{align}
    |r(F,\pi) - r(F,\pi')|
   &\leq \|\pi-\pi'\|_1  \exp\left(w\max_q \SW_2^2(F,\thbase{q})\right) \nonumber \\
    &\leq \|\pi-\pi'\|_1  \exp\left(w\sum_q \SW_2^2(F,\thbase{q})\right).\label{eq:modulus}
\end{align}
We define
$C_1 := \mathbb{E}_{\Gtrue}[\exp(w\sum_q \SW_2^2(F,\thbase{q}))]
< \infty$
by assumption~\eqref{eq:moment_assumption}.
Taking expectations in~\eqref{eq:modulus}, we have
\begin{align}
\label{eq:population_modulus}
    |R(\pi) - R(\pi')| \leq C_1\,\|\pi-\pi'\|_1.
\end{align}

Since $\Delta^{Q-1}$ is compact, for any $\eta > 0$ there exists a
finite $\eta$-net
$\mathcal{N}_\eta = \{\pi^{(1)},\ldots,\pi^{(M_\eta)}\}
\subset \Delta^{Q-1}$ such that for every $\pi \in \Delta^{Q-1}$
there exists some $\pi^{(j)} \in \mathcal{N}_\eta$ with
$\|\pi - \pi^{(j)}\|_1 \leq \eta$.
For any $\pi$, pick its nearest net point $\pi^{(j)}$, by the triangle inequality, we have
\begin{align}
\label{eq:three_terms}
    |\hat{R}_N(\pi) - R(\pi)|
    &\leq \underbrace{|\hat{R}_N(\pi) -
                       \hat{R}_N(\pi^{(j)})|}_{\text{Term 1}}
         +\underbrace{|\hat{R}_N(\pi^{(j)}) -
                       R(\pi^{(j)})|}_{\text{Term 2}}
         +\underbrace{|R(\pi^{(j)}) -
                       R(\pi)|}_{\text{Term 3}}.
\end{align}

\emph{Term 3.} By~\eqref{eq:population_modulus}: $|R(\pi^{(j)}) - R(\pi)| \leq C_1\,\eta$.

\emph{Term 1.} By~\eqref{eq:modulus}, averaging over $i$:
\begin{align}
|\hat{R}_N(\pi) - \hat{R}_N(\pi^{(j)})|
    \leq \eta\,\frac{1}{N}\sum_{i=1}^N
         \exp\left(w\sum_q \SW_2^2(F_i,\thbase{q})\right) =: \eta\, Z_N.
\end{align}
Since $\exp(w\sum_q \SW_2^2(F_i,\thbase{q}))$ are i.i.d.\ with finite mean
$C_1 < \infty$, the strong law of large numbers gives
$Z_N \xrightarrow{a.s.} C_1$.
Hence there exists an a.s.-finite random index $N_1$ such that for all
$N \geq N_1$, $Z_N \leq 2C_1$ almost surely, giving
$|\hat{R}_N(\pi) - \hat{R}_N(\pi^{(j)})| \leq 2C_1\,\eta$.

\emph{Term 2.} By the pointwise strong law of large numbers~\eqref{eq:pointwise_slln} applied at
each of the $M_\eta < \infty$ net points, we have
\begin{align}
    \max_{j=1,\ldots,M_\eta}
    |\hat{R}_N(\pi^{(j)}) - R(\pi^{(j)})|
    \xrightarrow{a.s.} 0.
\end{align}

Taking the supremum over $\pi \in \Delta^{Q-1}$ in~\eqref{eq:three_terms}
and applying the bounds above, for all $N \geq N_1$ almost surely:
\begin{align}
    \sup_{\pi\in\Delta^{Q-1}}|\hat{R}_N(\pi) - R(\pi)|
    \leq 3C_1\,\eta
         + \max_{j=1,\ldots,M_\eta}|\hat{R}_N(\pi^{(j)}) - R(\pi^{(j)})|.
\end{align}
Taking $\limsup_{N\to\infty}$ and using Term~2, we have
\begin{align}
    \limsup_{N\to\infty}
    \sup_{\pi\in\Delta^{Q-1}}|\hat{R}_N(\pi) - R(\pi)|
    \leq 3C_1\,\eta \quad\text{a.s.}
\end{align}
Since $\eta > 0$ was arbitrary, letting $\eta \to 0$ along a countable sequence gives
\begin{align}
\label{eq:uniform_slln}
    \sup_{\pi\in\Delta^{Q-1}}|\hat{R}_N(\pi) - R(\pi)|
    \overset{a.s.}{\to} 0.
\end{align}
Setting $\eta_\varepsilon := \delta(\varepsilon)/4 > 0$, there exists an
a.s.-finite random index $N_0$ such that for all $N \geq N_0$:
\begin{align}
\label{eq:uniform_bound}
    \sup_{\pi\in\Delta^{Q-1}}|\hat{R}_N(\pi) - R(\pi)|
    \leq \eta_\varepsilon.
\end{align}

\textbf{Full support of the induced prior on $\pi$.}
The dDPP draws a random configuration $\Phi\subseteq\Theta$, and conditionally on $\Phi=\{\thbase{q}\}_{q\in S}$ with $S\neq\emptyset$ the model \eqref{eq:mixing_measure} attaches i.i.d.\ $\mathrm{Gamma}(a,1)$ marks to the selected atoms and normalizes them, so that $\pi$ is supported on $S$.  Write
$\Pi$ for the resulting marginal law of $\pi$ on $\Delta^{Q-1}$.

For any nonempty $S\subseteq[Q]$ the J\'anossy density satisfies
\begin{align}
    j_{|S|}(S)
    \propto \det\bigl(L(\thbase{p},\thbase{q})\bigr)_{\thbase{p},\thbase{q}\in S} > 0,
\end{align}
since the atoms of $\Theta$ are pairwise distinct and the SW kernel is \emph{strictly} positive definite on distinct distributions. Hence $\Pr(\Phi = \{\thbase{q}\}_{q\in S})>0$ for every nonempty $S\subseteq[Q]$.
In particular the full configuration satisfies $\Pr(\Phi=\Theta)>0$.

Conditional on $\Phi=\Theta$, the weights $\pi_q = s_q/\sum_{q'}s_{q'}$ follow a $\mathrm{Dirichlet}(a,\ldots,a)$ distribution, whose Lebesgue density on the relative interior $\mathrm{ri}(\Delta^{Q-1})$ is strictly positive, and which assigns zero mass to the boundary $\partial\Delta^{Q-1}$.  Now let $U\subseteq\Delta^{Q-1}$ be nonempty and relatively open.  Since $\mathrm{ri}(\Delta^{Q-1})$ is dense in $\Delta^{Q-1}$, the set $U\cap\mathrm{ri}(\Delta^{Q-1})$ is nonempty since it is the intersection of two
relatively open sets it is also relatively open. Therefore  it has strictly positive $(Q-1)$-dimensional Lebesgue measure in the affine hull of $\Delta^{Q-1}$.  As the Dirichlet density is bounded away from zero on compact subsets of $\mathrm{ri}(\Delta^{Q-1})$, it follows that $\mathrm{Dir}_a\bigl(U\cap\mathrm{ri}(\Delta^{Q-1})\bigr)>0$, and hence
\begin{align}
    \Pi(U)
    \;\ge\; \Pr(\Phi=\Theta)\;
      \mathrm{Dir}_a\bigl(U\cap\mathrm{ri}(\Delta^{Q-1})\bigr)
    \;>\;0,
\end{align}
for every nonempty relatively open $U \subseteq \Delta^{Q-1}$.

We emphasize that this argument does not require $\mathcal{M}^*$ to be contained in $\mathrm{ri}(\Delta^{Q-1})$. The set that must carry prior mass is the neighbourhood $U^*_\rho$, not $\mathcal{M}^*$ itself.  Since
$U_\rho^* = \{\pi : \tilde{d}(\pi,\mathcal{M}^*) < \rho\}$ is nonempty and relatively open for every $\rho>0$, we conclude $\Pi(U^*_\rho)>0$, which is all that is used in the final step of the proof.

\textbf{Connection between the posterior and the empirical risk.}
The model specifies a likelihood for observation $F$ given $\pi$ as $p(F\mid\pi) \propto \sum_{q=1}^Q \pi_q \exp(-w\,\SW_2^2(F,\thbase{q}))$,
so that $-\log p(F\mid\pi) = r(F,\pi)$ up to an additive constant independent of $\pi$. With prior $\Pi$, the posterior satisfies
\begin{align}
\label{eq:posterior_formula}
    \Pi_N(A \mid F_1,\ldots,F_N)
    = \frac{\displaystyle\int_A
            \exp\bigl(-N\hat{R}_N(\pi)\bigr)\,\mathrm{d}\Pi(\pi)}
           {\displaystyle\int_{\Delta^{Q-1}}
            \exp\bigl(-N\hat{R}_N(\pi)\bigr)\,\mathrm{d}\Pi(\pi)},
\end{align}
for any measurable $A \subseteq \Delta^{Q-1}$.

\textbf{Bounding the posterior probability.}
We fix $\varepsilon > 0$ and let $A_\varepsilon = \{\pi : \tilde{d}(\pi,\mathcal{M}^*) > \varepsilon\}$.
We work on the a.s.-event $\{N \geq N_0\}$ where~\eqref{eq:uniform_bound} holds.

For any $\pi \in A_\varepsilon$,
applying~\eqref{eq:uniform_bound} and~\eqref{eq:gap}, we have
\begin{align}
    \hat{R}_N(\pi)
    \geq R(\pi) - \eta_\varepsilon
    \geq R^* + \delta(\varepsilon) - \eta_\varepsilon
    = R^* + 3\eta_\varepsilon,
\end{align}
where the last equality uses $\delta(\varepsilon) = 4\eta_\varepsilon$.
Therefore:
\begin{align}
\label{eq:numerator}
    \int_{A_\varepsilon} \exp\left(-N\hat{R}_N(\pi)\right) \mathrm{d}\Pi(\pi)
    \leq \exp\left(-N(R^*+3\eta_\varepsilon)\right).
\end{align}

We choose $\rho = \rho(\varepsilon) \in (0,\varepsilon)$ small enough so that
$U_\rho^* = \{\pi : \tilde{d}(\pi,\mathcal{M}^*) < \rho\} \subseteq A_\varepsilon^c$
(which holds for any $\rho < \varepsilon$ since $A_\varepsilon^c = \{d(\pi,\mathcal{M}^*)\leq\varepsilon\}$)
and, by continuity of $R$, we have
\begin{align}
\label{eq:r_neighborhood}
    \sup_{\pi \in U_\rho^*} R(\pi) \leq R^* + \eta_\varepsilon.
\end{align}
For any $\pi \in U_\rho^*$,
applying~\eqref{eq:uniform_bound} and~\eqref{eq:r_neighborhood} leads to
\begin{align}
    \hat{R}_N(\pi)
    \leq R(\pi) + \eta_\varepsilon
    \leq R^* + 2\eta_\varepsilon.
\end{align}
Therefore, we have
\begin{align}
\label{eq:denominator}
    \int_{\Delta^{Q-1}} \exp\left(-N\hat{R}_N(\pi)\right) \mathrm{d}\Pi(\pi)
    &\geq \int_{U_\rho^*} \exp\left(-N\hat{R}_N(\pi)\right) \mathrm{d}\Pi(\pi) \nonumber \\
    &\geq \Pi(U_\rho^*)  \exp\left(-N(R^*+2\eta_\varepsilon)\right).
\end{align}

Dividing~\eqref{eq:numerator} by~\eqref{eq:denominator} and
using~\eqref{eq:posterior_formula}, with
$C := 1/\Pi(U_\rho^*) \in (0,\infty)$ (finite by full support of the prior)
and $c(\varepsilon) := \eta_\varepsilon = \delta(\varepsilon)/4 > 0$
(positive by~\eqref{eq:gap}), we final have
\begin{align}
    \Pi_N\left(\tilde{d}(\pi,\mathcal{M}^*) > \varepsilon
    \Big| F_1,\ldots,F_N\right)
    \leq C \exp\left(-c(\varepsilon)\,N\right)
    \overset{a.s.}{\to} 0
    \quad\text{as }N\to\infty,
\end{align}
which completes the proof.

\section{Additional Experiments}
\label{sec:add_exp}

In this section we provide additional qualitative results that complement the main text. All figures below use the point estimates of the mixing measure obtained from the decision-theoretic summarization of Section~\ref{subsec:summarization}, and the DP and dDPP models are run under identical settings, same atom set $\Theta=\mathcal{S}$ and the same generalized likelihood scale $w$, so that any difference is attributable solely to the prior. 

\subsection{Single-Cell Data}
\label{subsec:add_single_cell}

\begin{figure}[!t]
    \centering
    \begin{tabular}{cc}
         \includegraphics[width=0.5\linewidth]{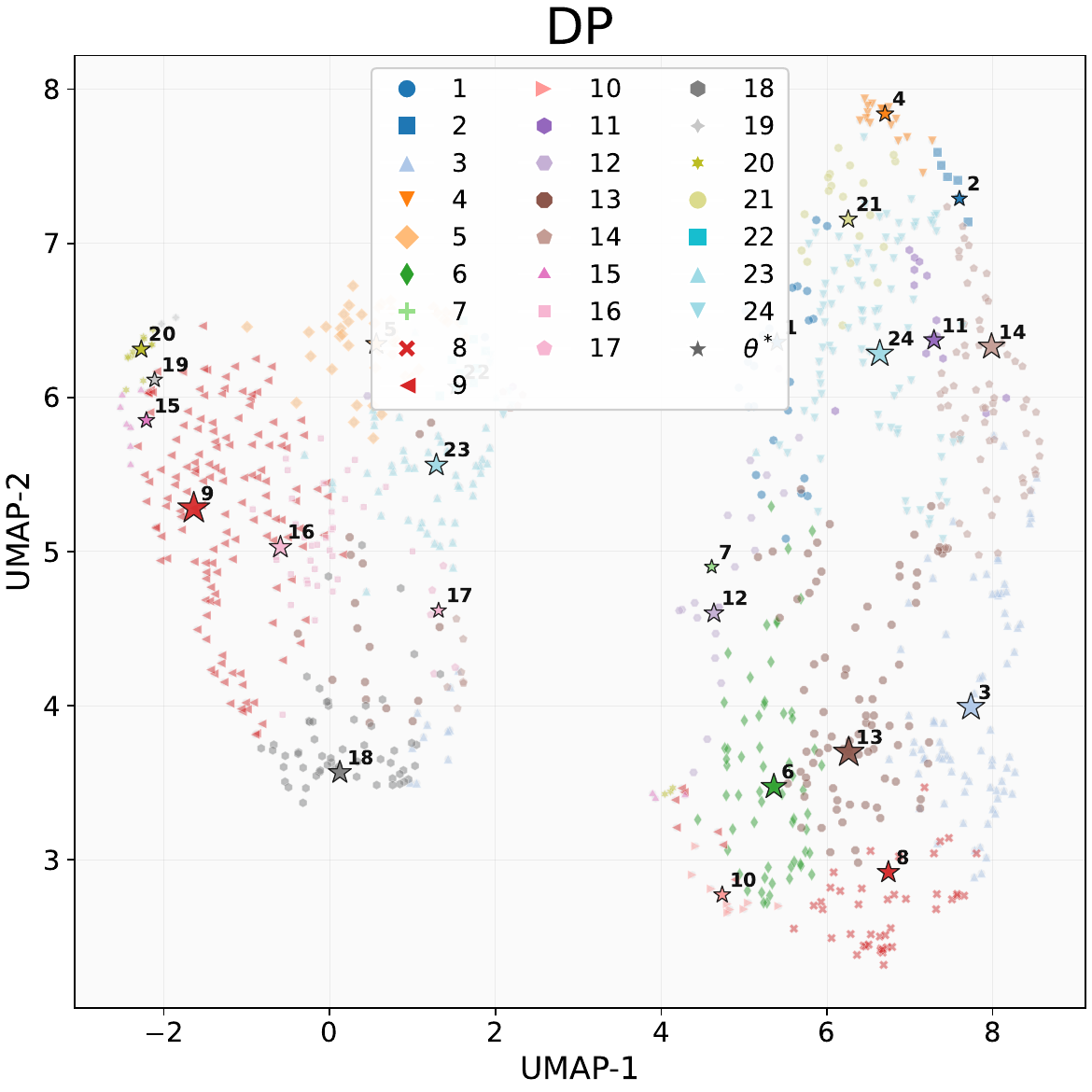} & \includegraphics[width=0.5\linewidth]{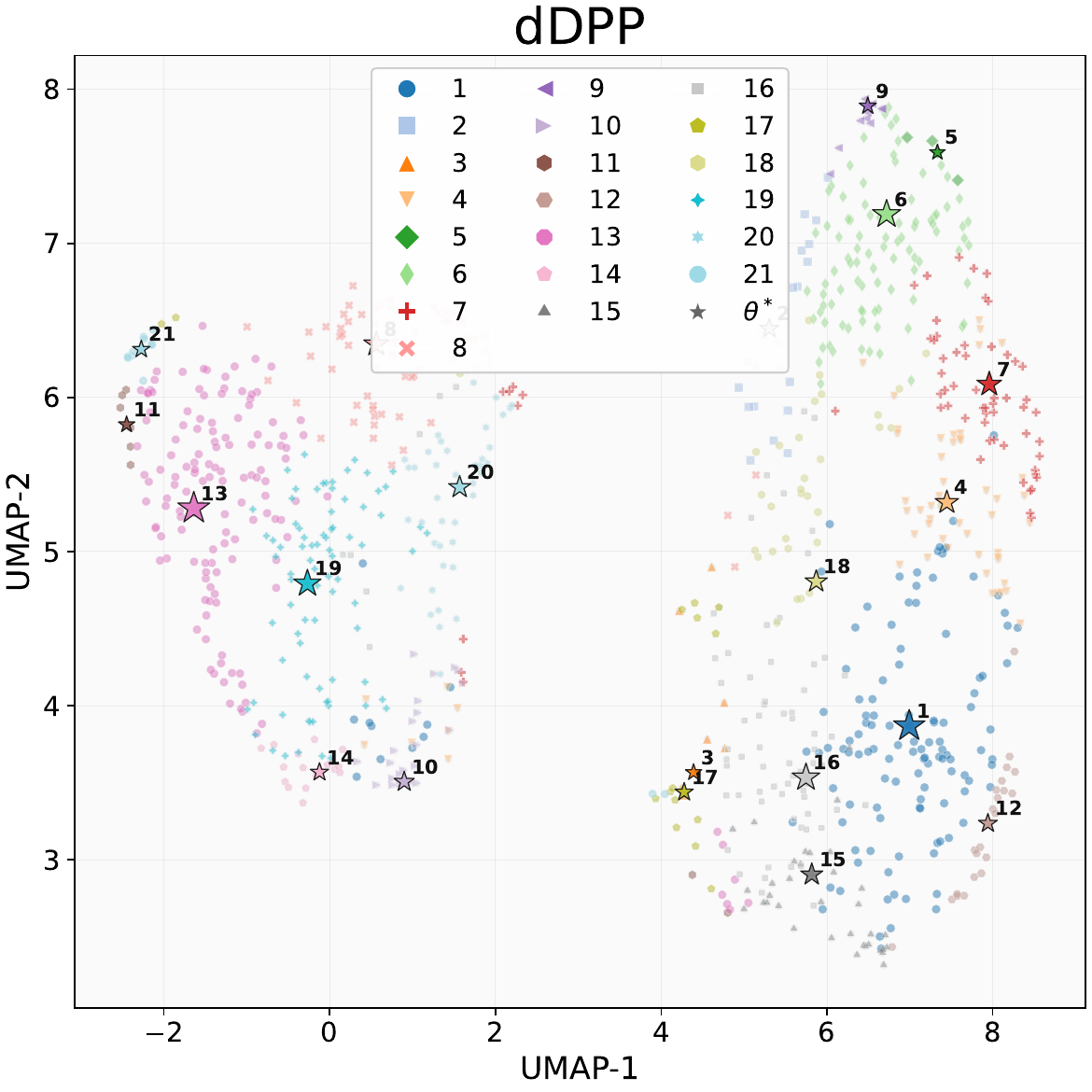}\\
        \includegraphics[width=0.5\linewidth]{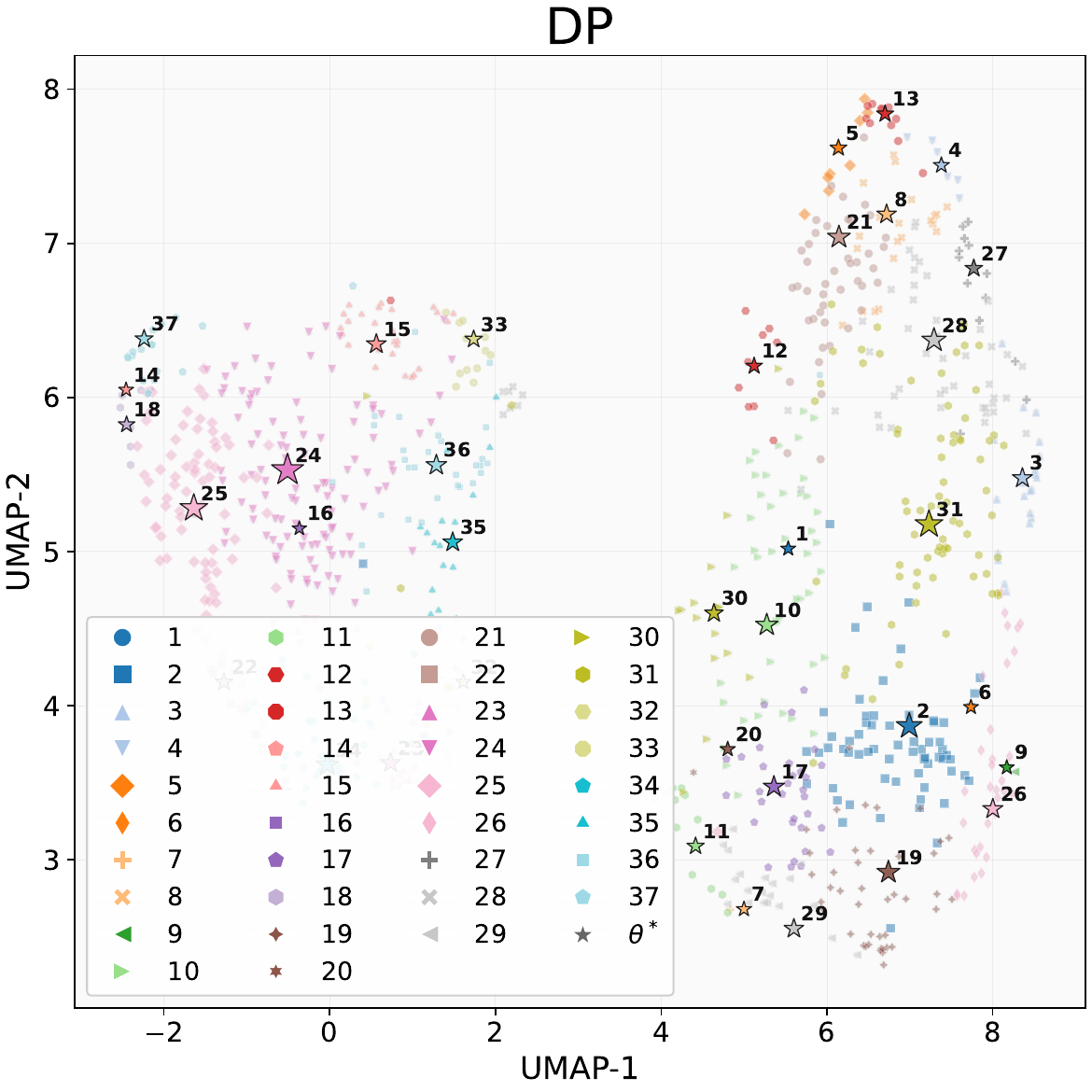} &  \includegraphics[width=0.5\linewidth]{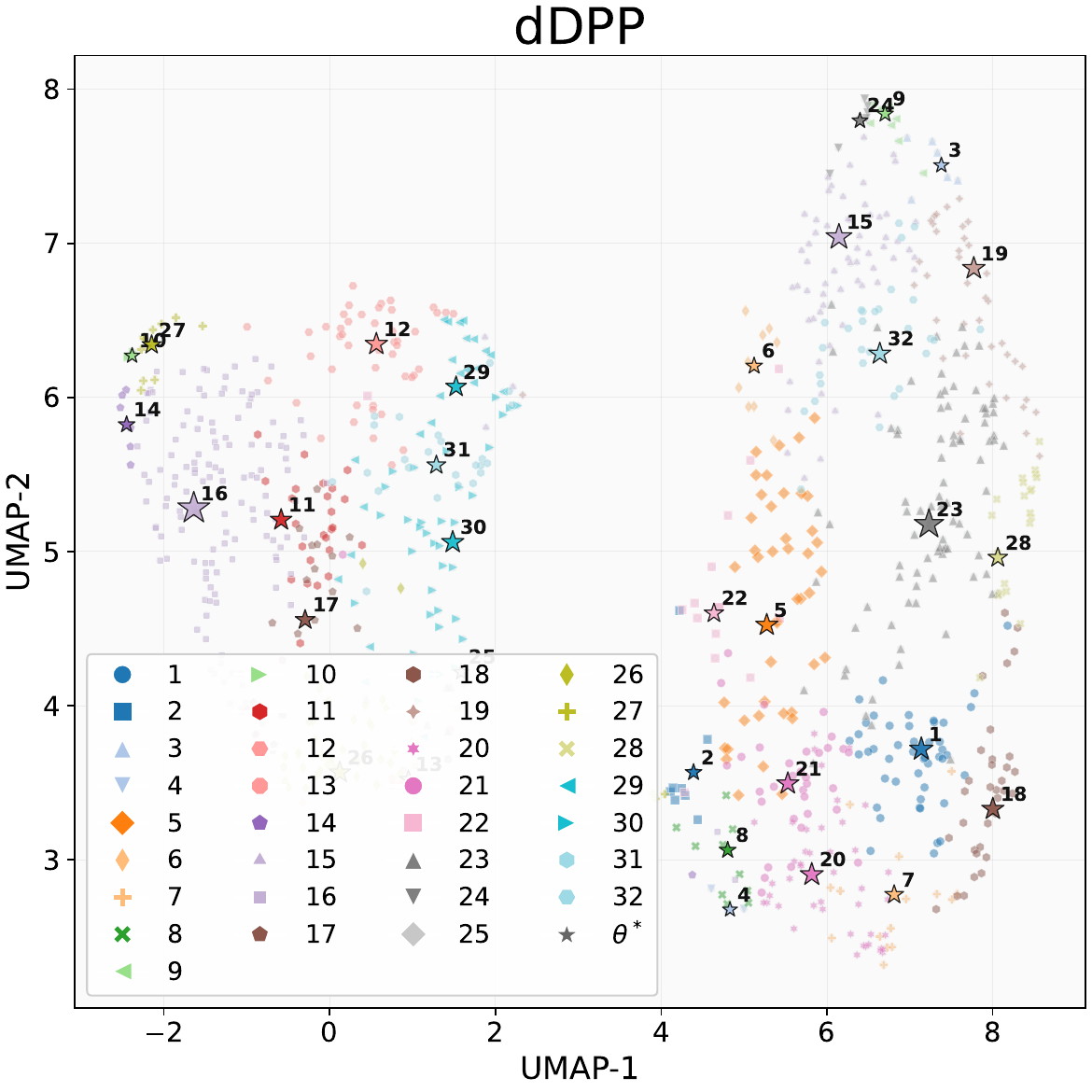}

    \end{tabular}
    \caption{Same as Figure~\ref{fig:mixing_500000} with $7 \times 10^5$ (first row) and  $w= 10^6$ (second row).}
    \label{fig:mixing_700000}
\end{figure}

Figure~\ref{fig:mixing_700000} repeats the UMAP comparison of Figure~\ref{fig:mixing_500000} for the two larger likelihood scales
$w = 7\times 10^5$ (first row) and $w = 10^6$ (second row). The qualitative contrast observed at $w = 5\times 10^5$ persists, and in fact sharpens, as $w$ increases: at every scale the dDPP atoms (right column) spread across distinct
regions of the embedding and induce fewer, more spatially coherent groups, whereas the DP atoms (left column) concentrate more tightly and yield a more fragmented partition. This is consistent with Table~\ref{tab:combined_all},
where the gap in cluster count between the two models widens with $w$: as the likelihood drives toward finer partitions, the repulsion in the dDPP prior increasingly suppresses redundant atoms.

\begin{figure}[!t]
    \centering
    \includegraphics[width=1\linewidth]{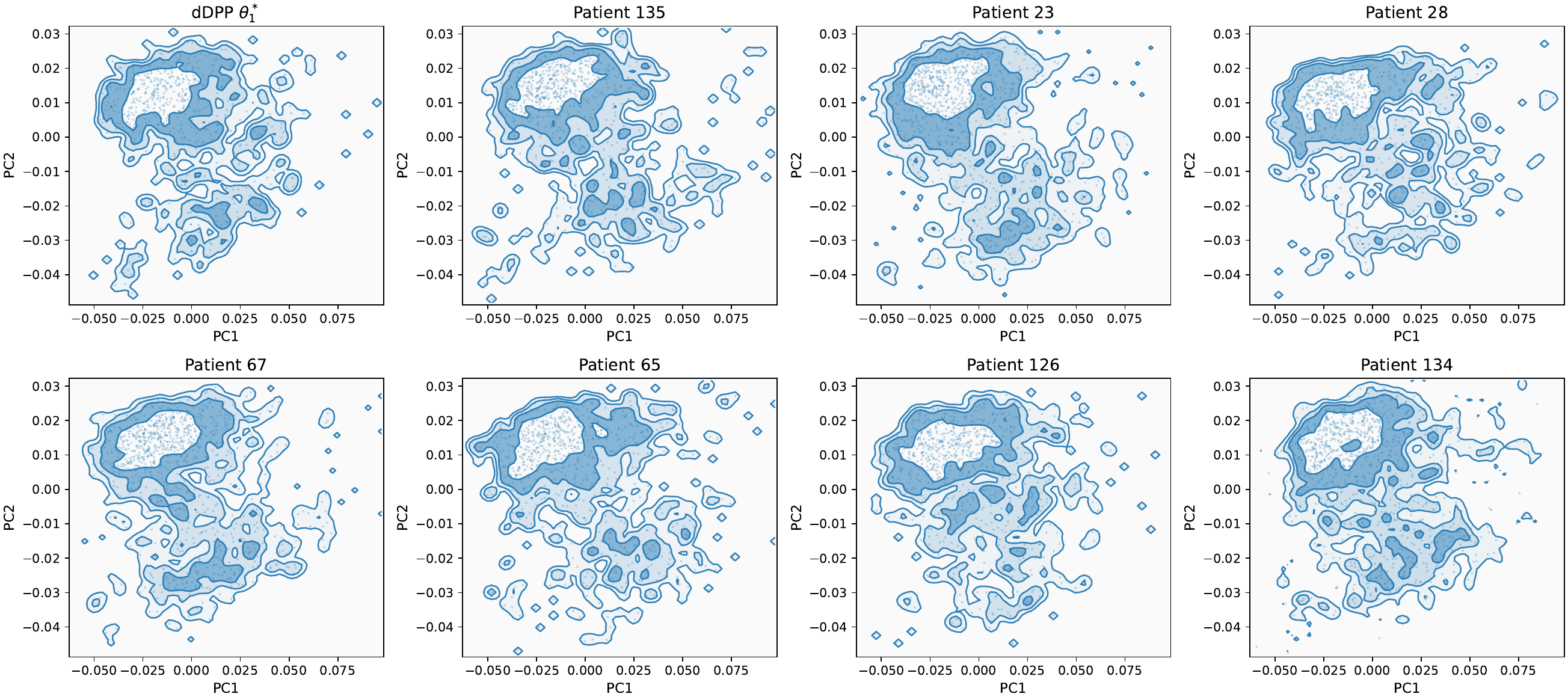} \\
    \includegraphics[width=1\linewidth]{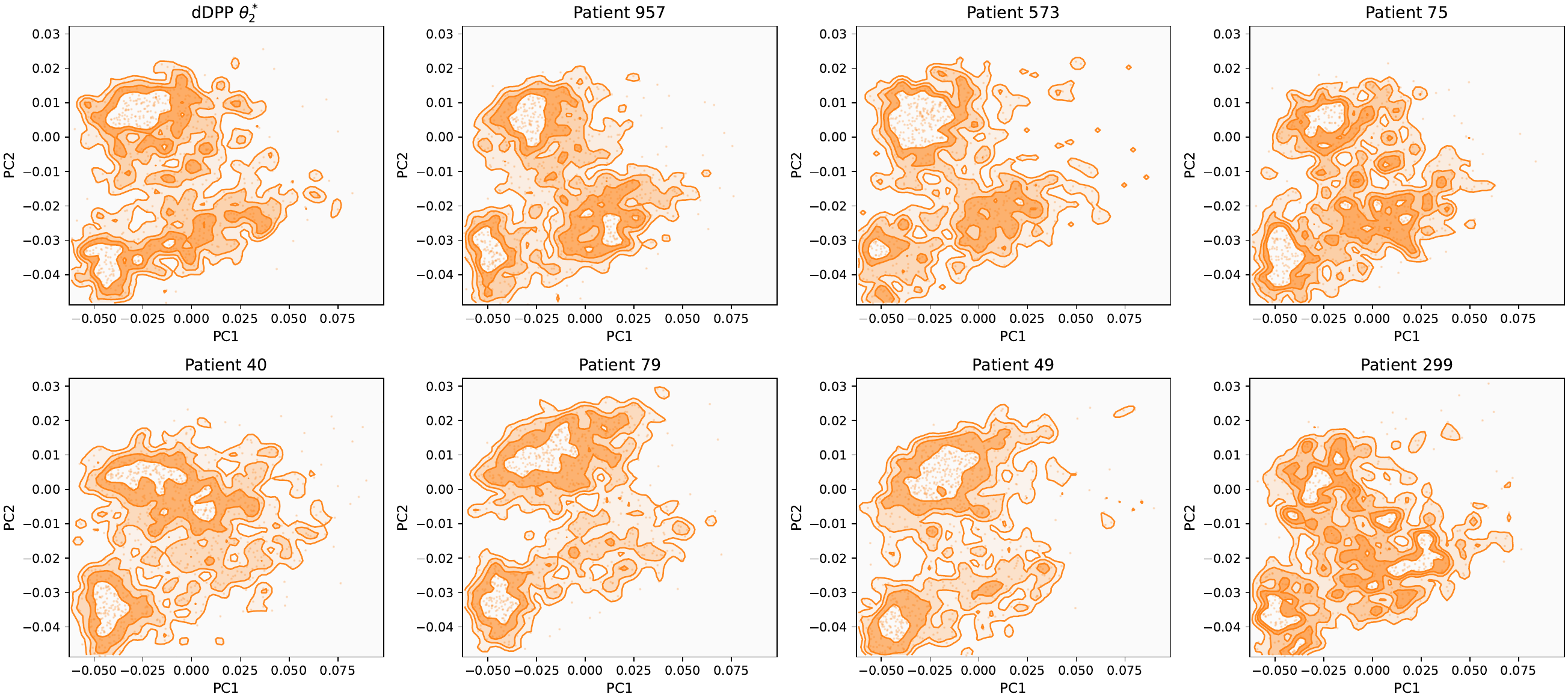} \\
    \includegraphics[width=1\linewidth]{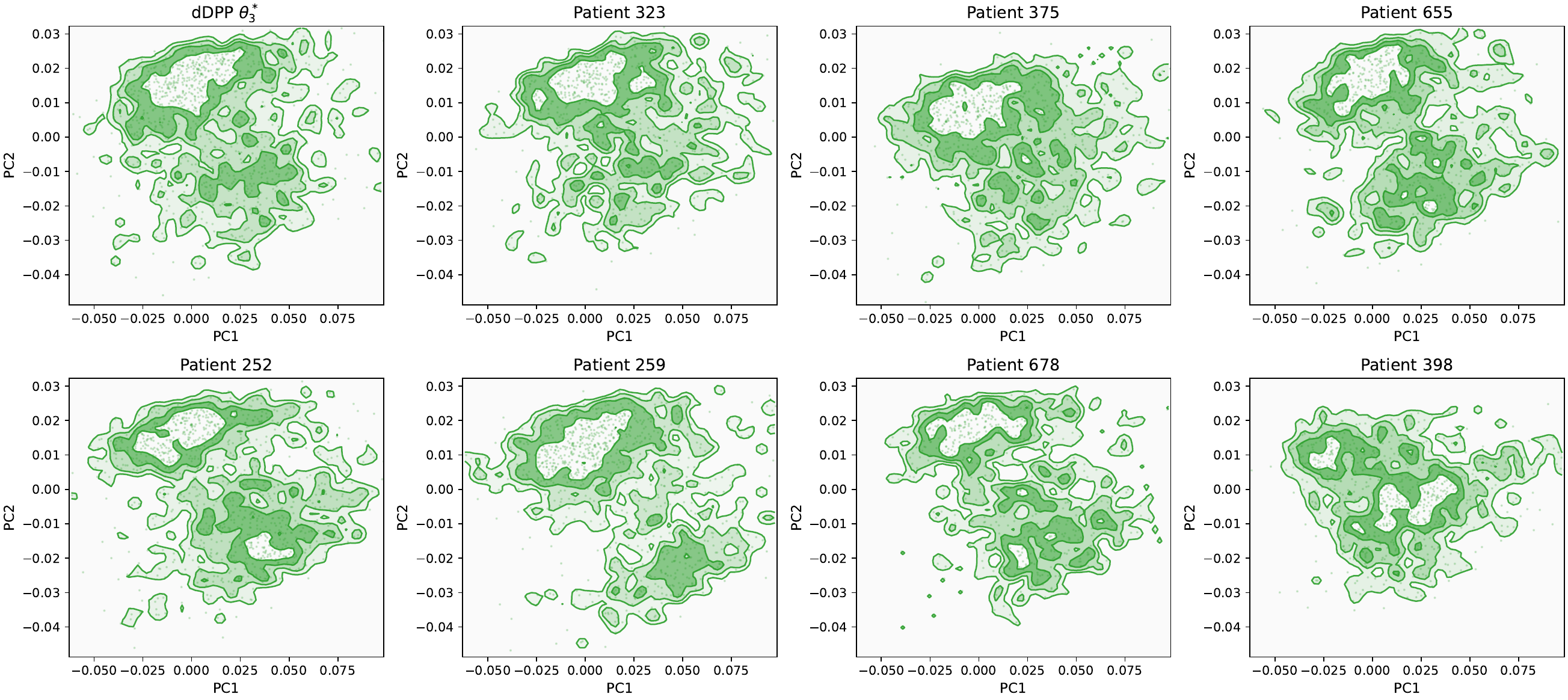}
    \caption{PCA visualization of some donors in clusters the dDPP model ($w=5\times10^5,\gamma=10^3$).}
    \label{fig:patients_by_cluster_1}
\end{figure}

\begin{figure}[!t]
    \centering
    \includegraphics[width=1\linewidth]{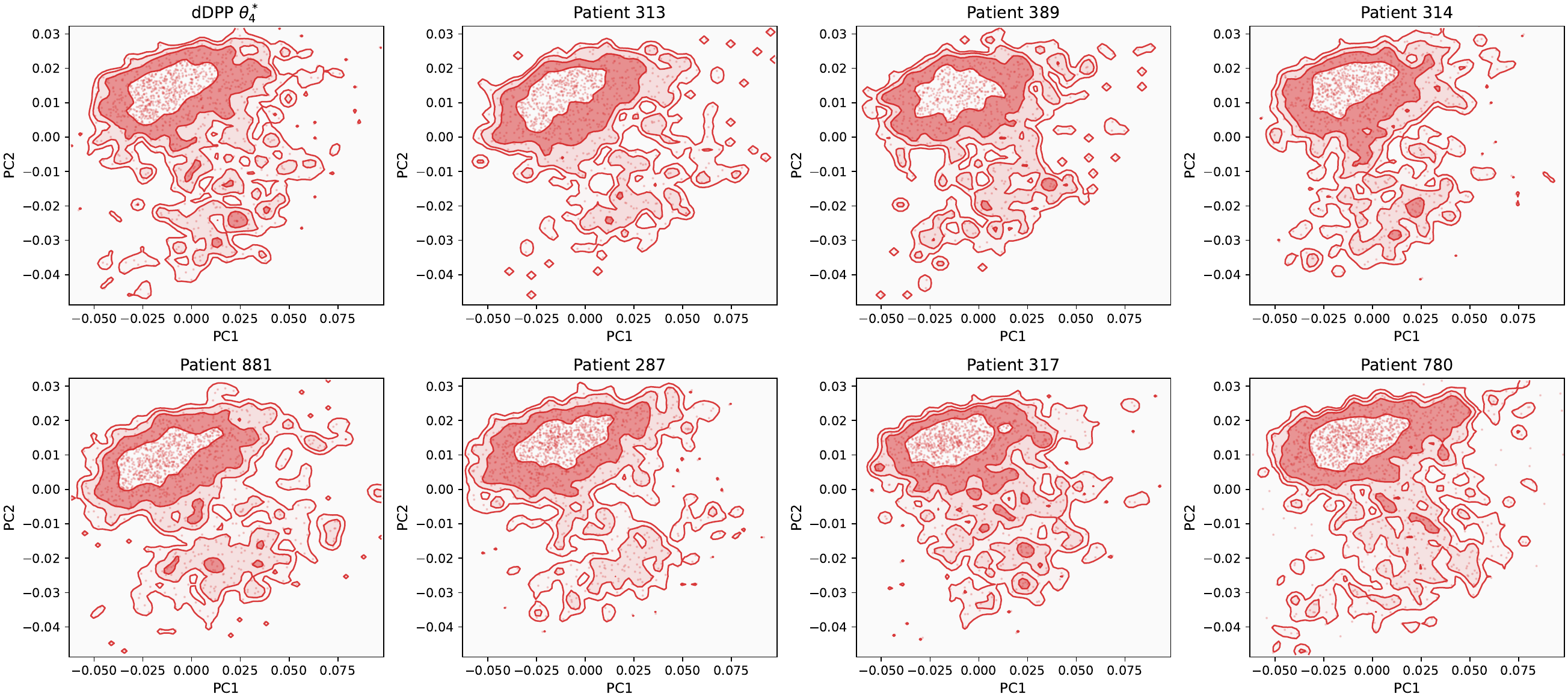} \\
    \includegraphics[width=1\linewidth]{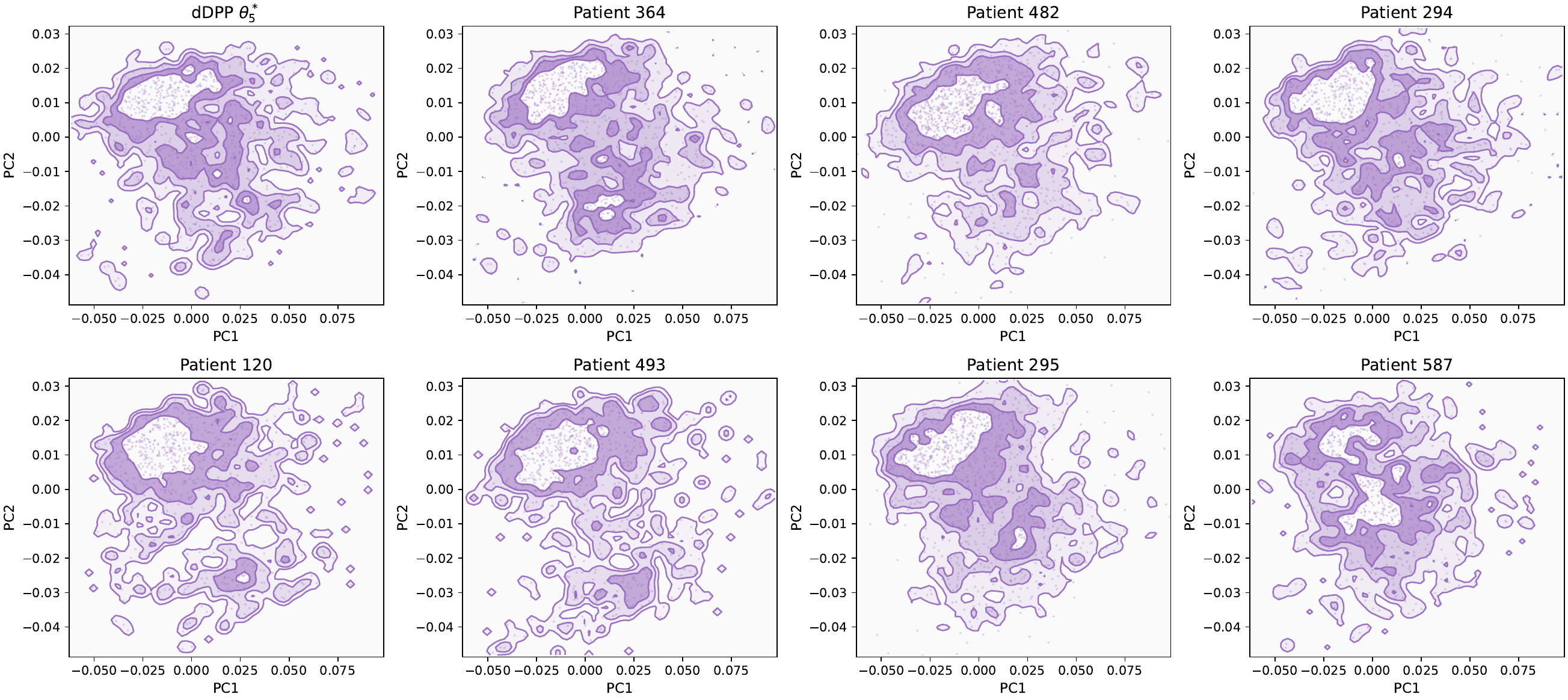} \\
    \includegraphics[width=1\linewidth]{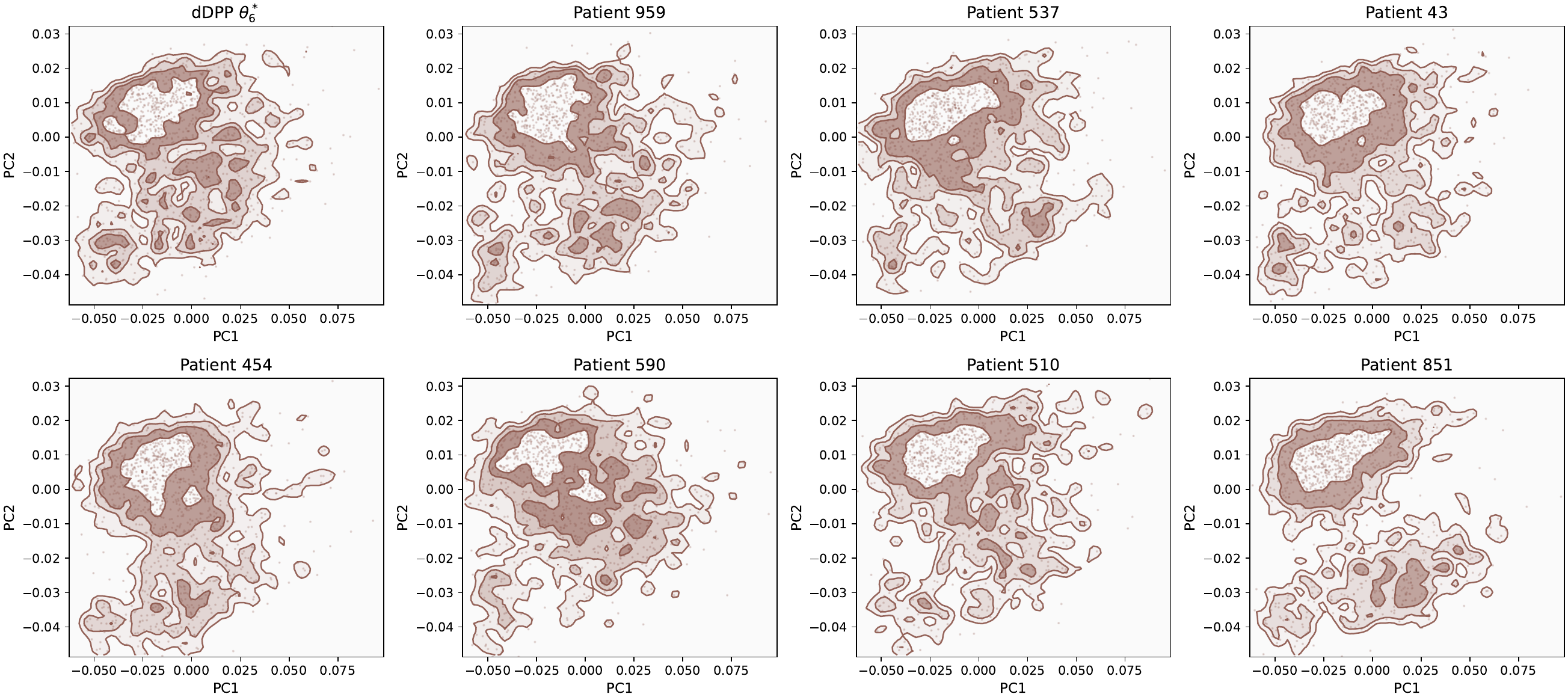} 
    \caption{PCA visualization of some donors in clusters the dDPP model ($w=5\times10^5,\gamma=10^3$) }
    \label{fig:patients_by_cluster_2}
\end{figure}

\begin{figure}[!t]
    \centering
    \includegraphics[width=1\linewidth]{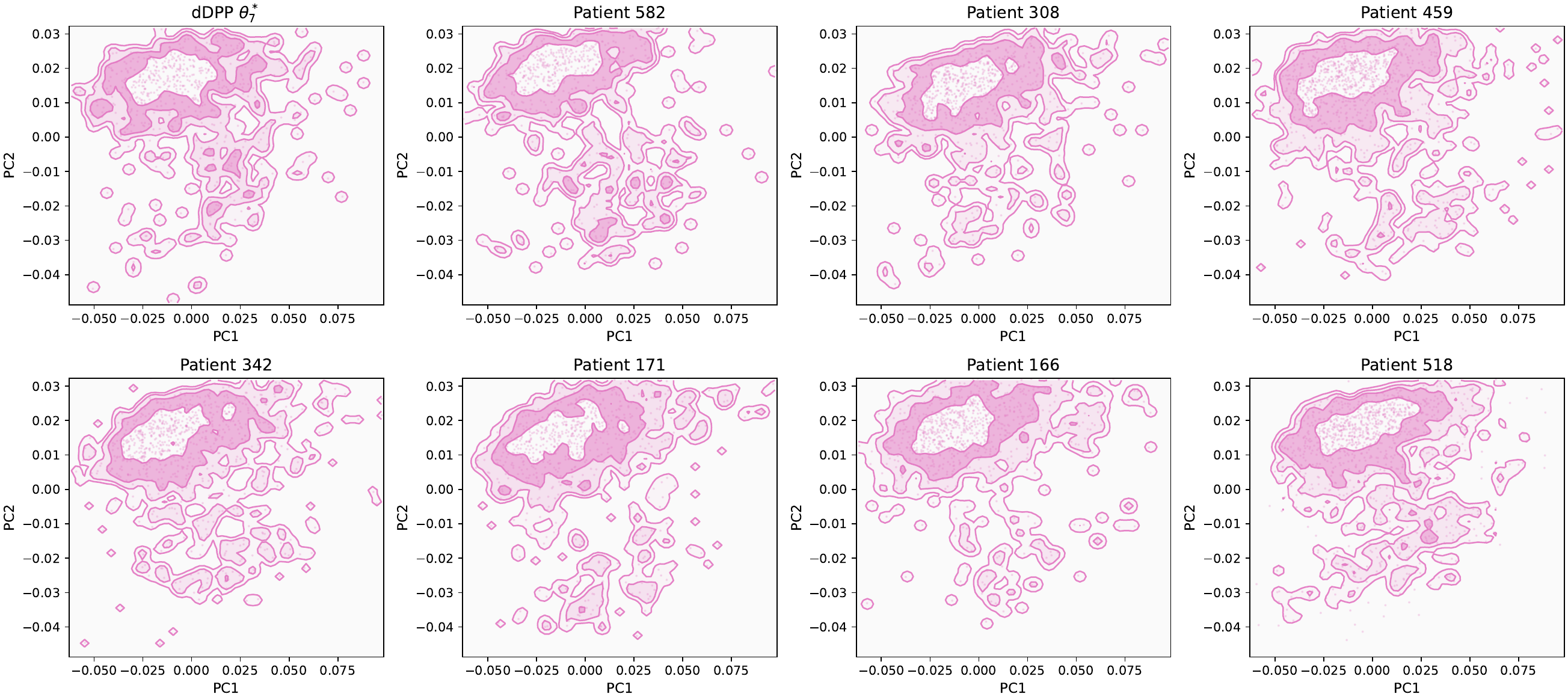} \\
    \includegraphics[width=1\linewidth]{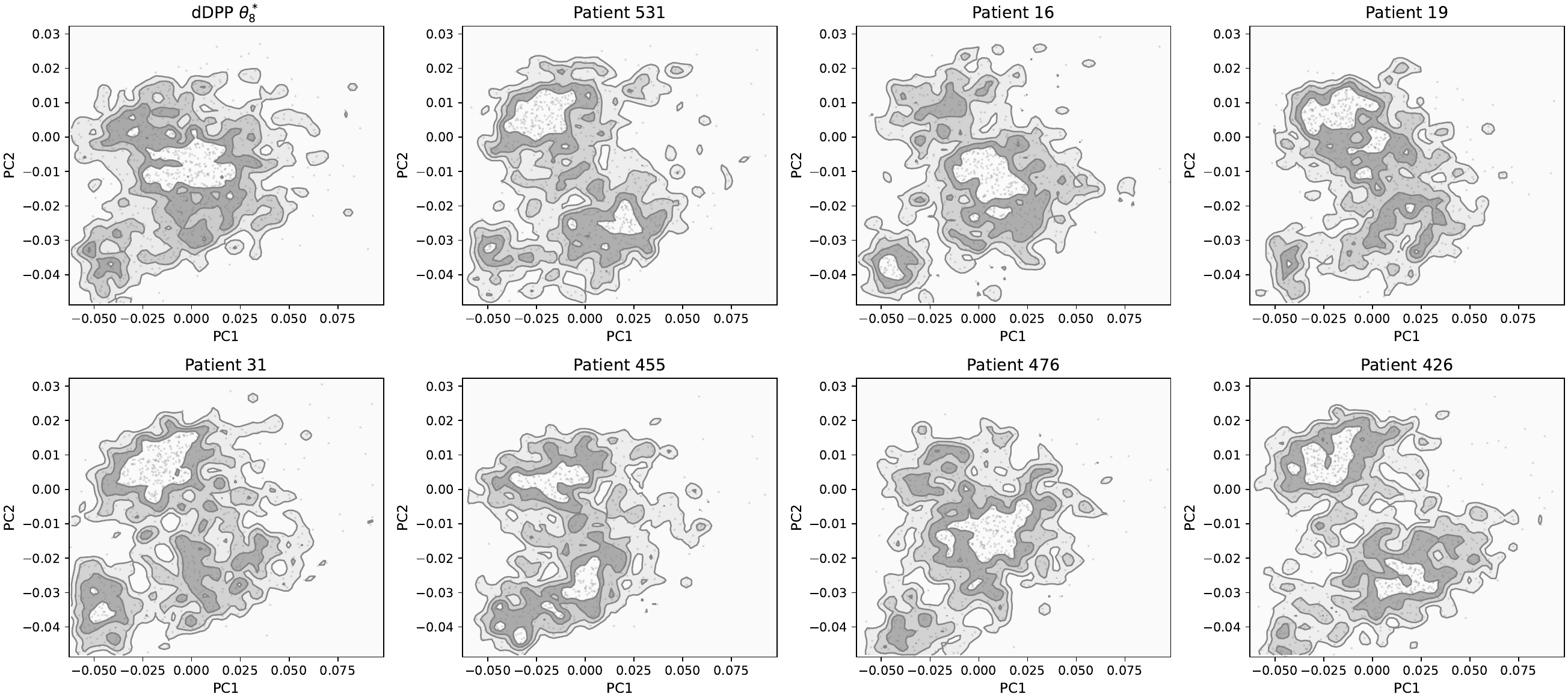} \\
    \includegraphics[width=1\linewidth]{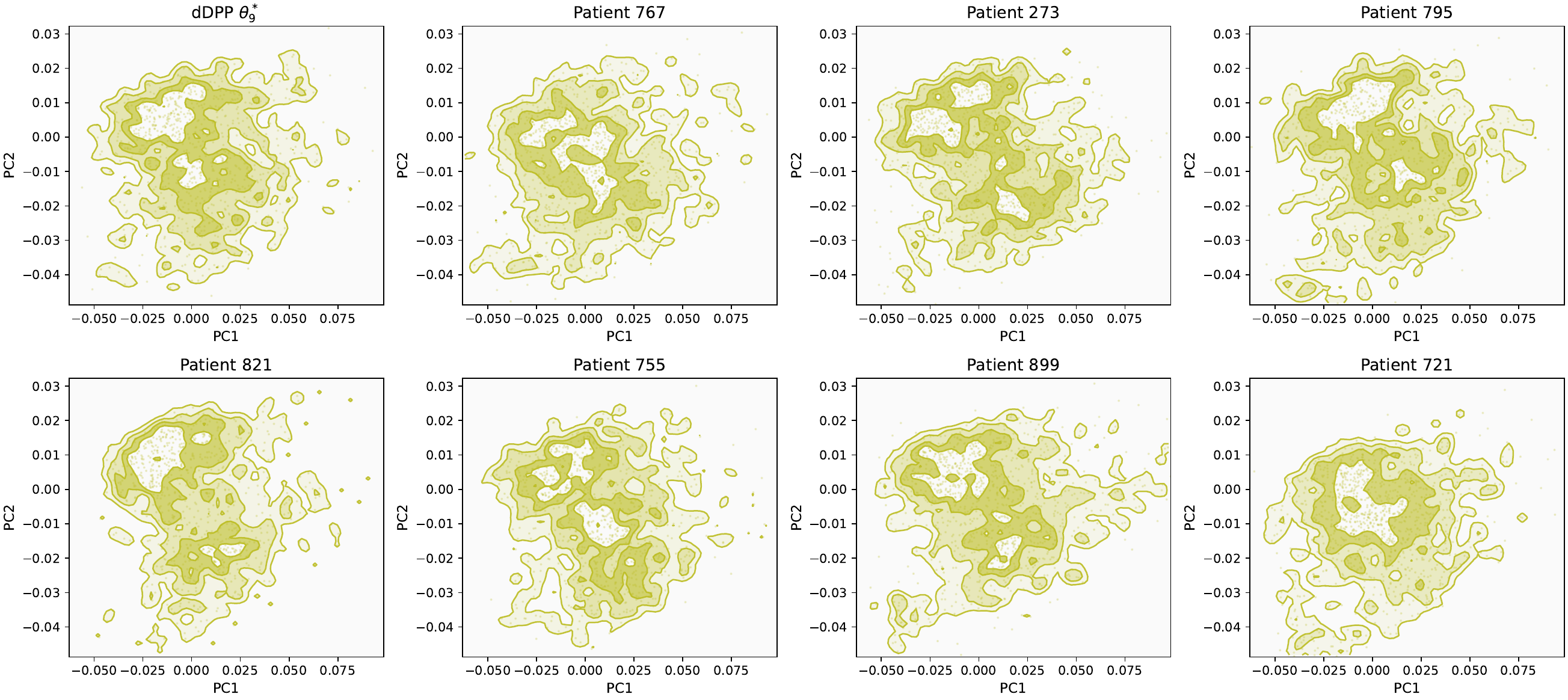} 
    \caption{PCA visualization of some donors in clusters the dDPP model ($w=5\times10^5,\gamma=10^3$)}
    \label{fig:patients_by_cluster_3}
\end{figure}

\begin{figure}[!t]
    \centering
    \includegraphics[width=1\linewidth]{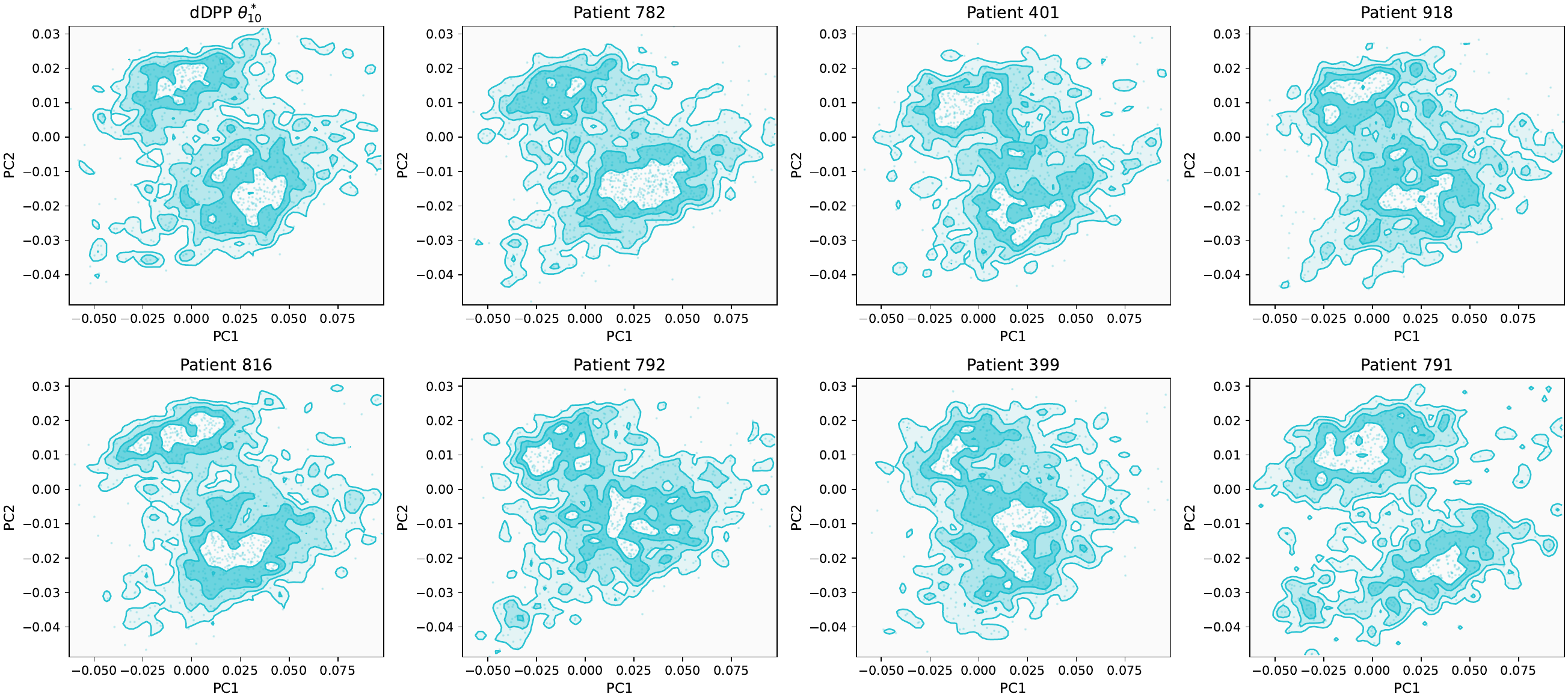} \\
    \includegraphics[width=1\linewidth]{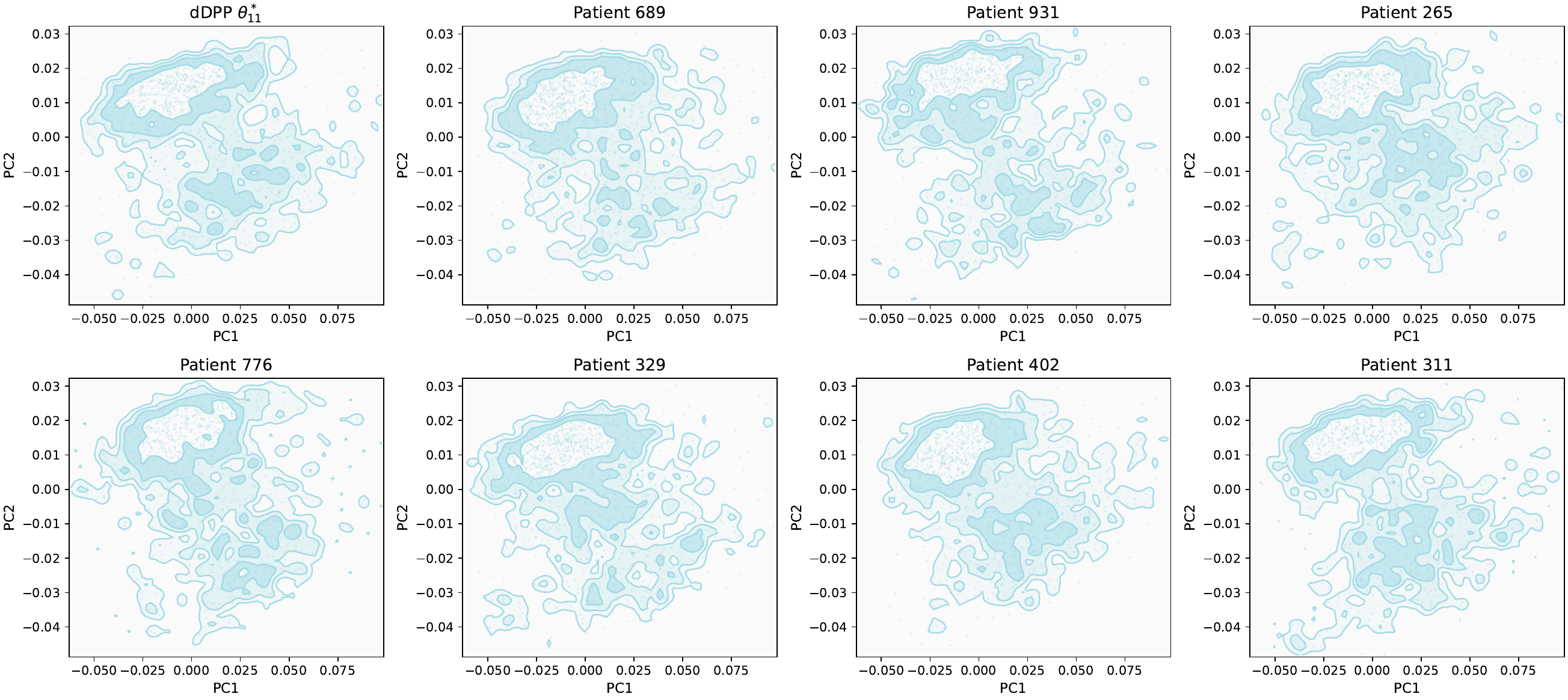} 
    \caption{PCA visualization of some donors in clusters the dDPP model ($w=5\times10^5,\gamma=10^3$)}
    \label{fig:patients_by_cluster_4}
\end{figure}

Figures~\ref{fig:patients_by_cluster_1}--\ref{fig:patients_by_cluster_4} display, for the dDPP solution at $w = 5\times 10^5$, the donor-level cell distributions in the $17$-dimensional PCA space for the individual
clusters. Within each cluster, the donors exhibit visibly similar distributional shapes, and they closely resemble the corresponding
representative atom shown in Figure~\ref{fig:mixing_measure_summary}. This indicates that the recovered atoms act as faithful prototypes for the donors assigned to them, and that the partition groups together donors with genuinely comparable gene-expression profiles rather than merely co-locating
them in the embedding.

\begin{figure}[!t]
    \centering
    \includegraphics[width=1\linewidth]{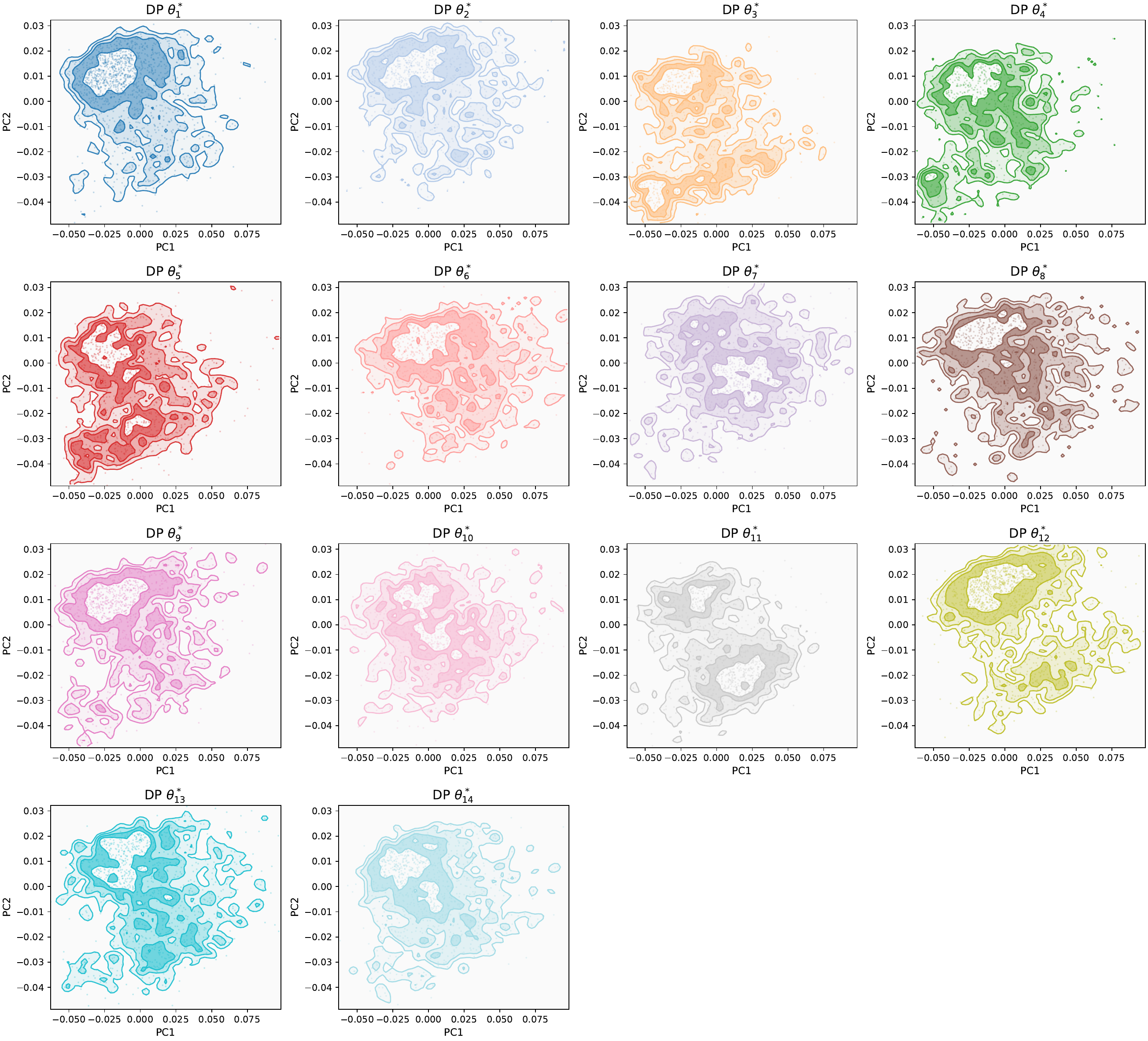}
    \caption{Same as Figure~\ref{fig:mixing_measure_summary} for the DP model.}
    \label{fig:mixing_measure_summary_dp}
\end{figure}

Figure~\ref{fig:mixing_measure_summary_dp} is the DP analogue of Figure~\ref{fig:mixing_measure_summary}, overlaying the recovered atoms on a KDE of the point-estimate mixing measure in the donor-level PCA space. Relative to the dDPP atoms, the DP atoms are placed closer together and several fall in overlapping high-density regions, so that distinct modes of the donor distribution are not as cleanly separated. This visual difference mirrors the higher (less negative) repulsion score attained by dDPP in
Table~\ref{tab:combined_all}.

\subsection{Human Epilepsy Project Data}
\label{subsec:add_hep}

\begin{figure}[!t]
    \centering
    \begin{tabular}{cc}
        \includegraphics[width=0.43\linewidth]{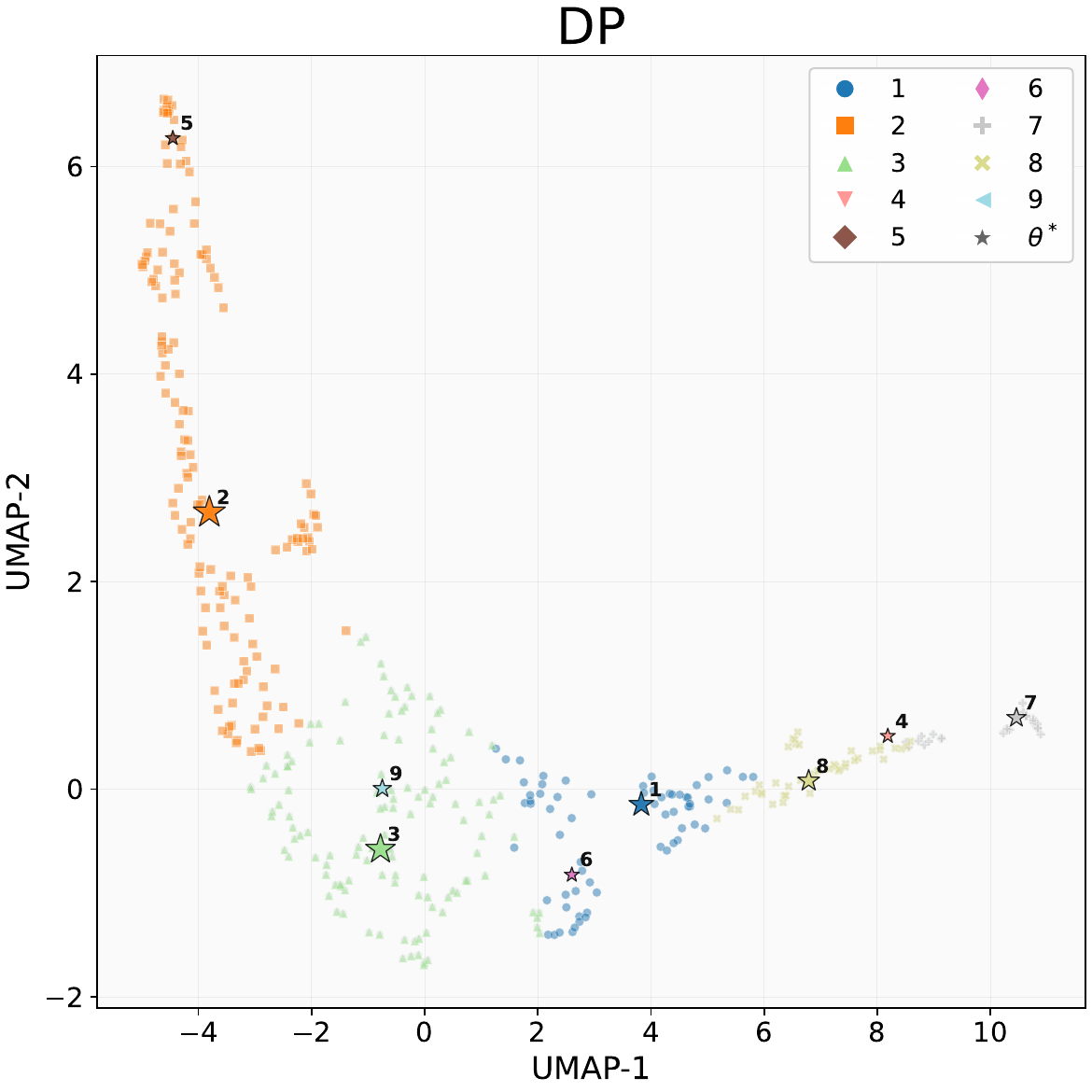} & \includegraphics[width=0.43\linewidth]{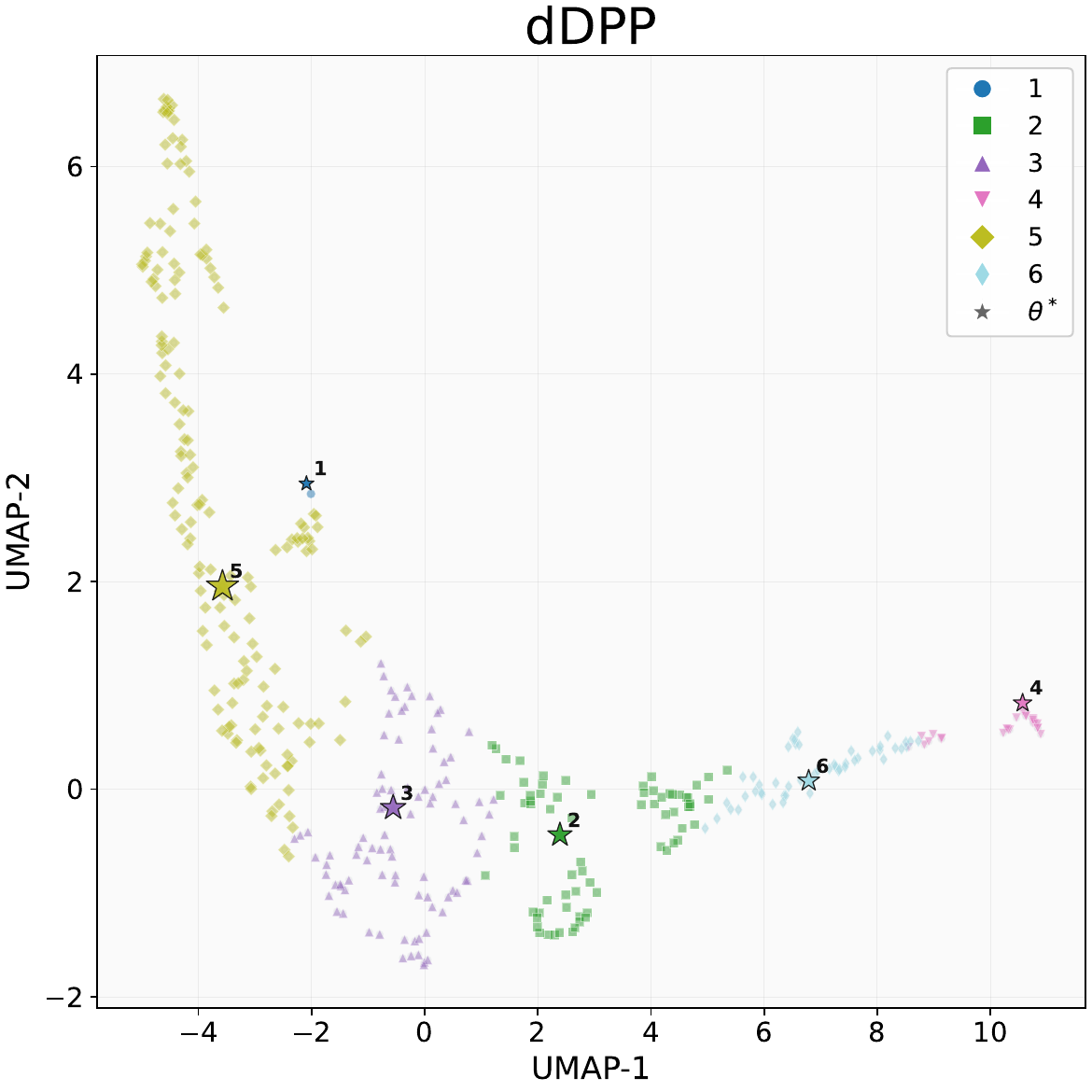} \\
        \includegraphics[width=0.43\linewidth]{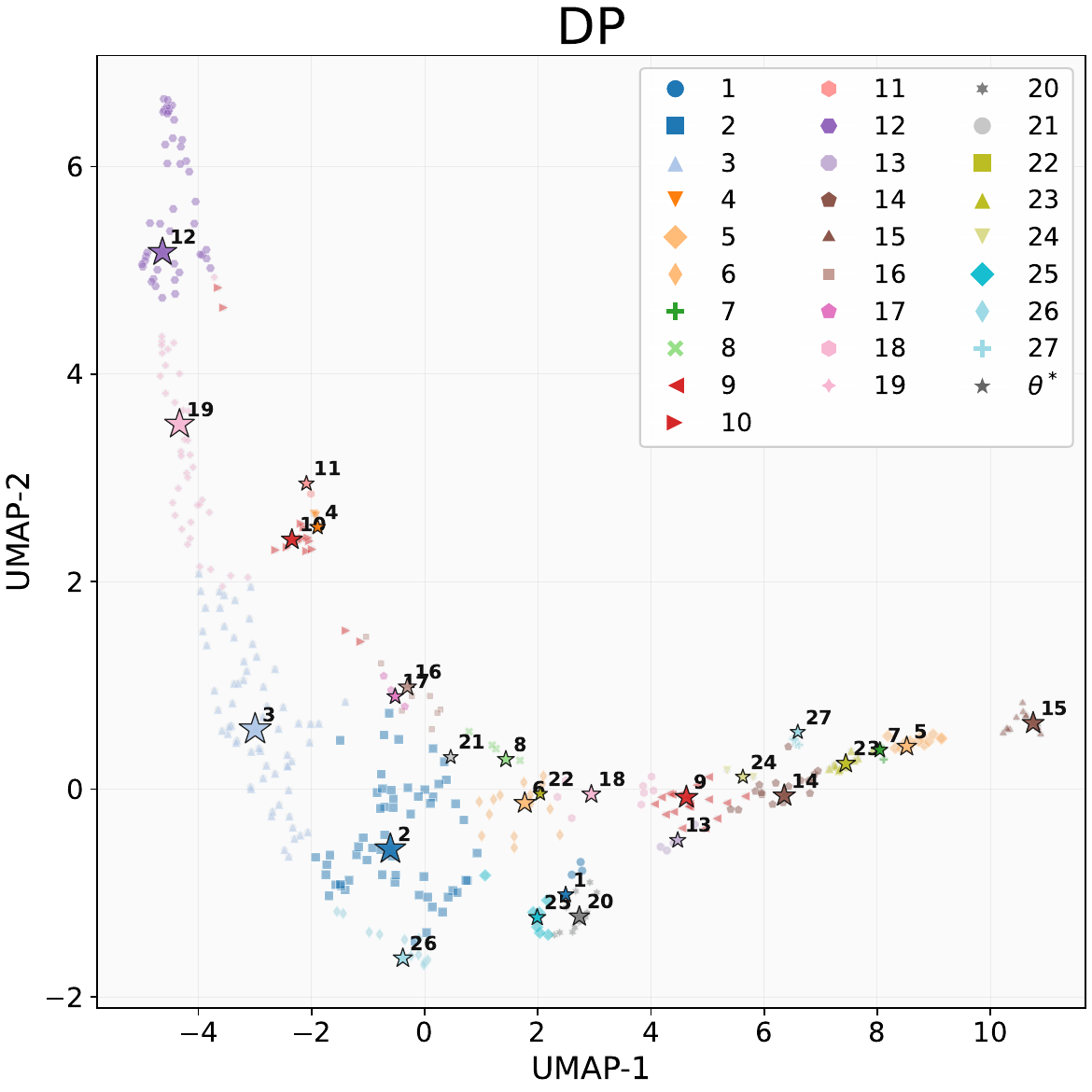} &  \includegraphics[width=0.43\linewidth]{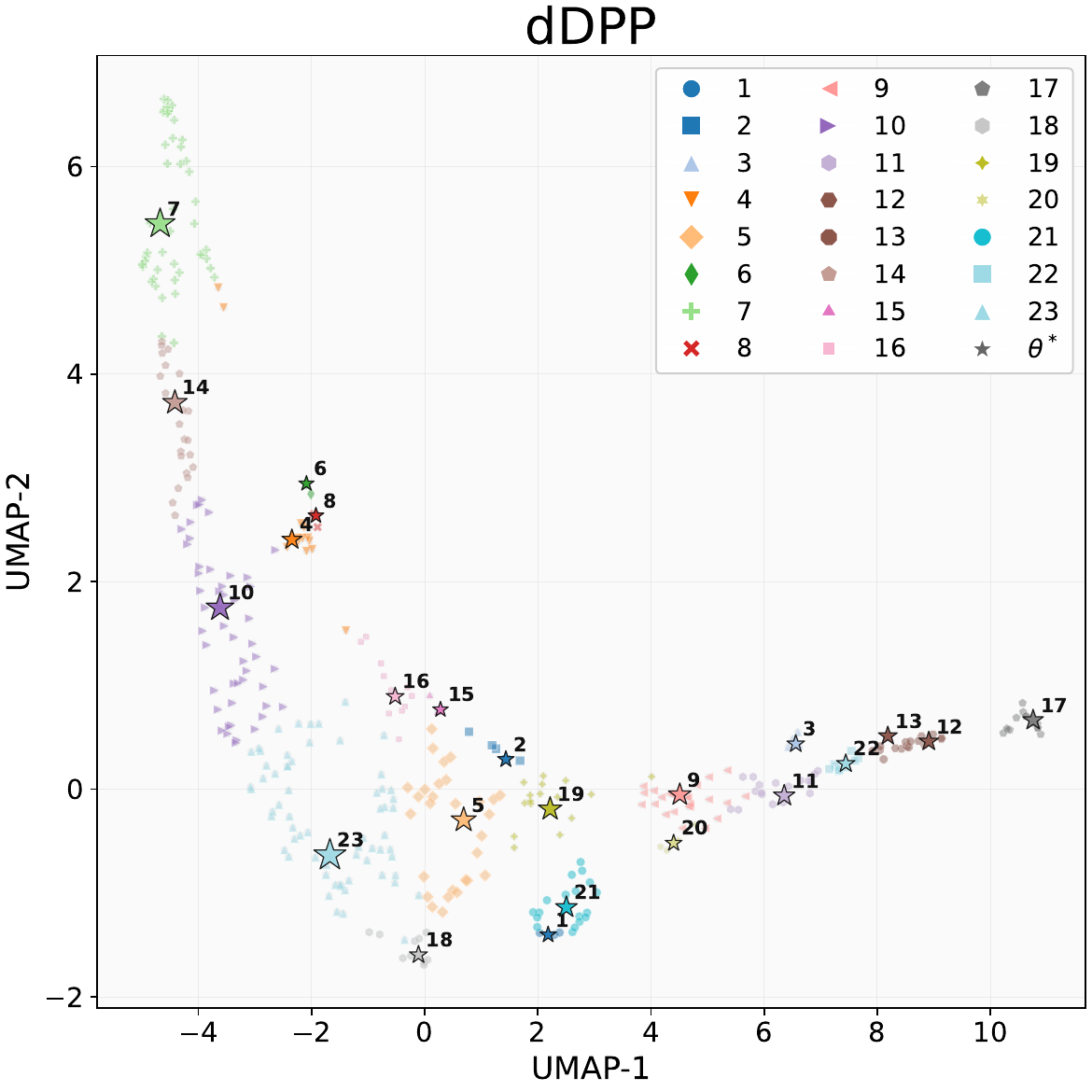}
    \end{tabular}
    \caption{Same as Figure~\ref{fig:mixing_200_HEP} with $w=100$ (first row) and $w=500$ (second row).}
    \label{fig:mixing_100_500_HEP}
\end{figure}

Figure~\ref{fig:mixing_100_500_HEP} repeats the UMAP comparison of Figure~\ref{fig:mixing_200_HEP} for $w = 100$ (first row) and $w = 500$
(second row). As on the single-cell data, the dDPP solution (right column) assigns participants to fewer, more spatially coherent groups whose atoms are distributed across distinct regions of the embedding, while the DP solution
(left column) clusters its atoms more tightly and produces a more fragmented partition. The effect is present at both the smaller and larger likelihood scales, confirming that the behavior reported at $w = 200$ in the main text is not an artifact of a particular choice of $w$.

\begin{figure}[!t]
    \centering
     \includegraphics[width=1\linewidth]{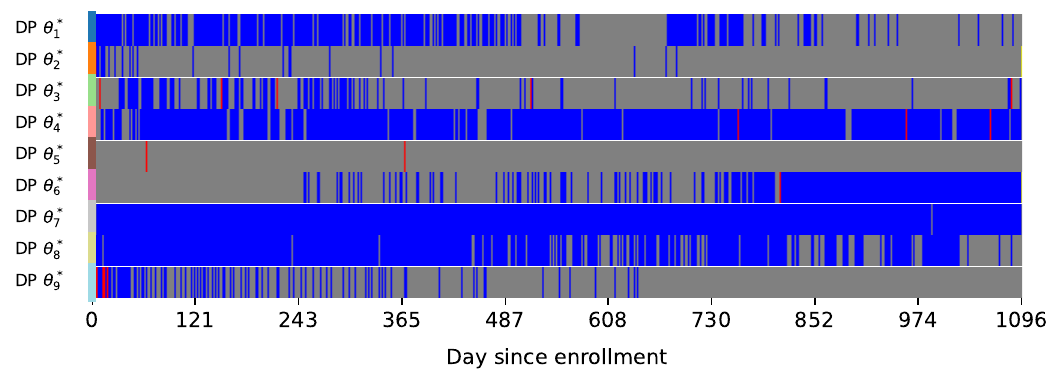} \\
    \includegraphics[width=1\linewidth]{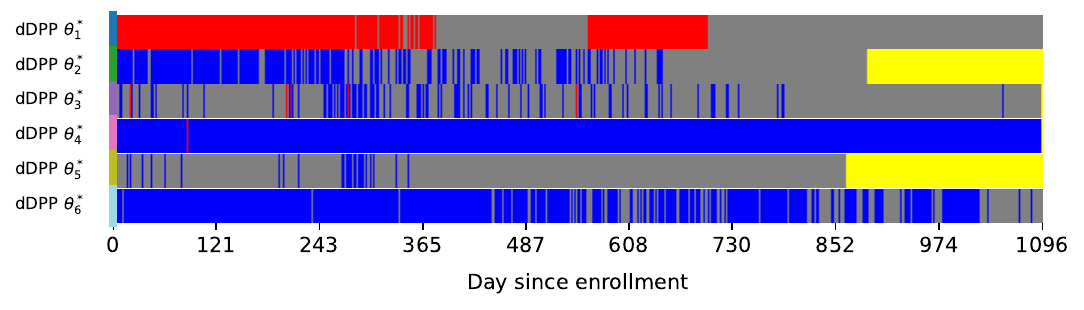} 
    \caption{Same as Figure~\ref{fig:mixing_measure_summary_hep} with $w=100$ for DP model (first row) and dDPP model (second row).}
    \label{fig:mixing_measure_summary_hep_100}
\end{figure}
\begin{figure}[!t]
    \centering
     \includegraphics[width=1\linewidth]{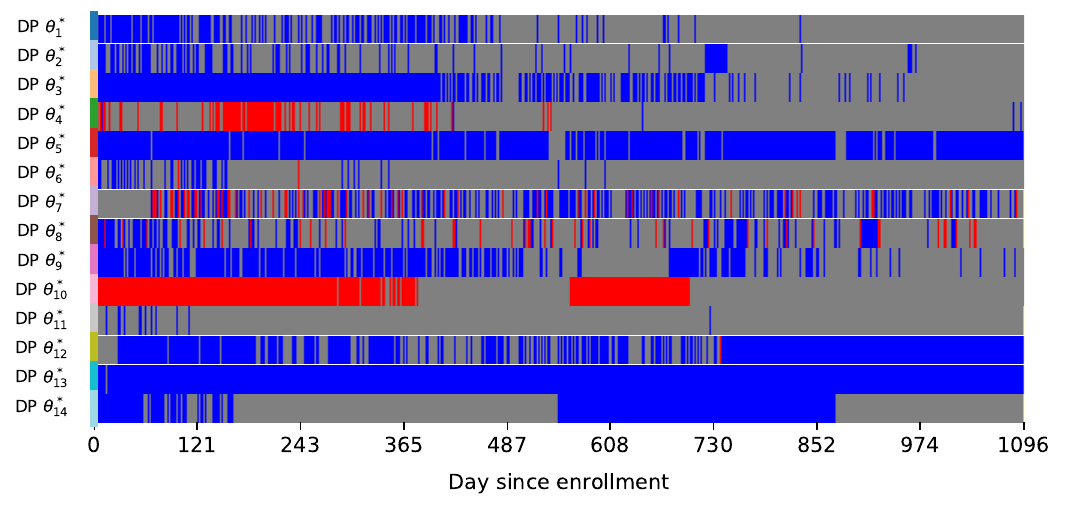}\\
    \includegraphics[width=1\linewidth]{figures/HEP/D2PP/d2pp_mixing_measure_iter_1442_typical_patients_10_200.pdf} 
    \caption{Same as Figure~\ref{fig:mixing_measure_summary_hep} with $w=200,\gamma=10$ for DP model (first row) and dDPP model (second row).}
    \label{fig:mixing_measure_summary_hep_200}
\end{figure}

\begin{figure}[!t]
    \centering
    \includegraphics[width=0.9\linewidth]{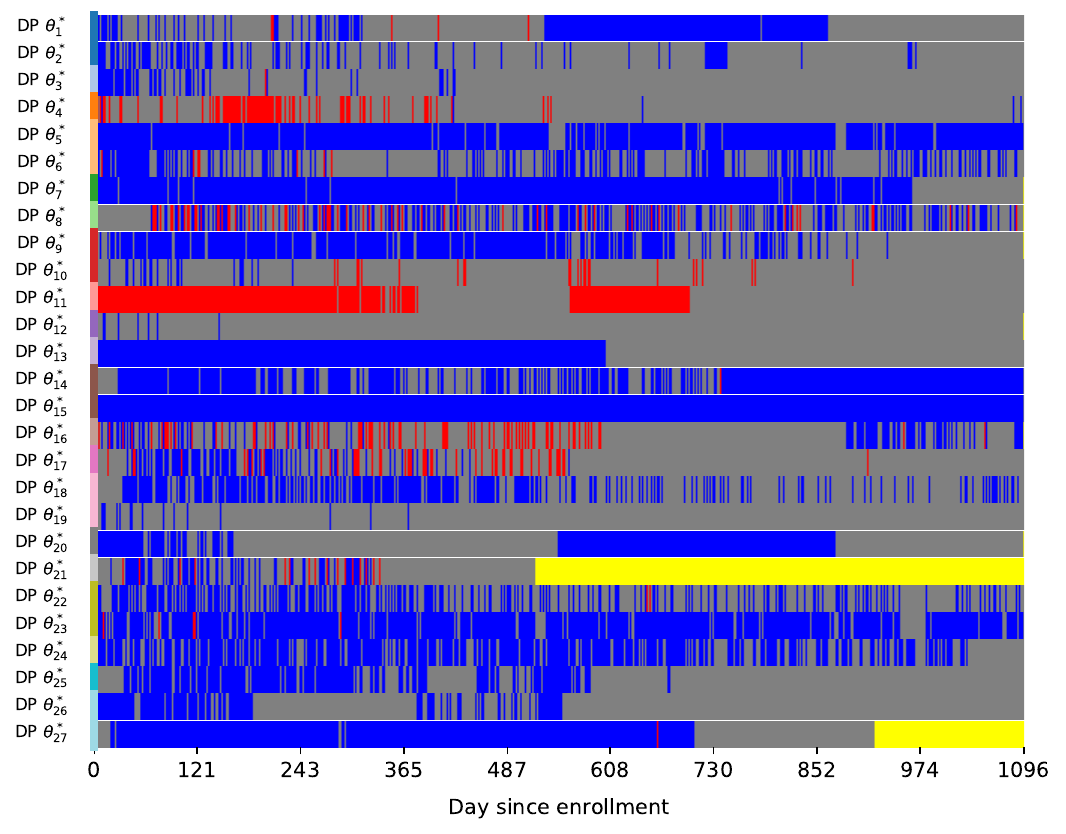} \\
      \includegraphics[width=0.9\linewidth]{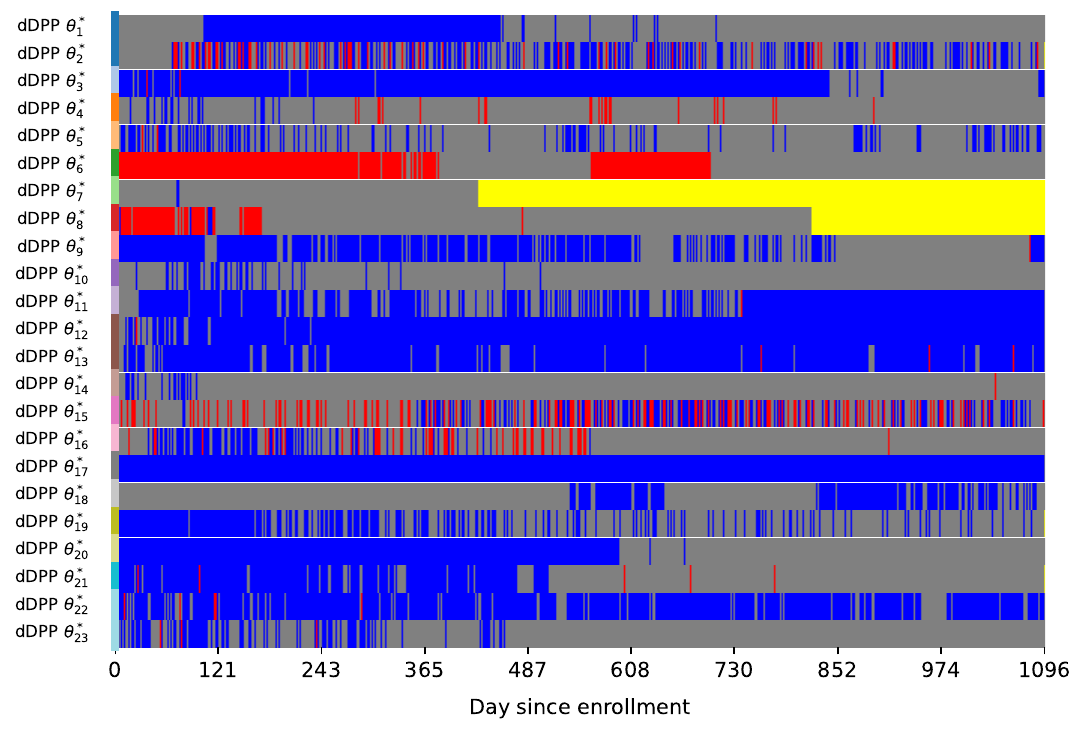} 
      \vspace{-1em}
    \caption{Same as Figure~\ref{fig:mixing_measure_summary_hep} with $w=500$ for DP model (first row) and dDPP model (second row).}
    \label{fig:mixing_measure_summary_hep_500}
\end{figure}

Figures~\ref{fig:mixing_measure_summary_hep_100}--\ref{fig:mixing_measure_summary_hep_500} visualize the recovered atoms as representative seizure-window patterns for both models at $w = 100$, $w = 200$, and $w = 500$ (DP in the first row, dDPP in the second row of each figure). Blue, red, and grey encode the daily states $0$, $1$, and $0.5$ respectively, with yellow indicating padding for participants with shorter follow-up. Across all three scales, the dDPP atoms capture more clearly differentiated temporal profiles, ranging from seizure-free or near seizure-free patterns to denser, more irregular patterns, whereas the DP atoms include more nearly duplicated profiles, again reflecting the weaker separation quantified by the repulsion column of
Table~\ref{tab:combined_hep}.

\end{document}